\documentclass[twocolumn,twocolappendix,12pt]{aastex7}
\usepackage{placeins,verbatim,amsmath}
\usepackage{xcolor}
\usepackage{multirow}
\usepackage{calc}
\usepackage{threeparttable}

\newcommand{\del}[1]{}
\newcommand{\red}[1]{{#1}}

\sfcode`A=1000 \sfcode`B=1000 \sfcode`C=1000 \sfcode`D=1000
\sfcode`E=1000 \sfcode`F=1000 \sfcode`G=1000 \sfcode`H=1000
\sfcode`I=1000 \sfcode`J=1000 \sfcode`K=1000 \sfcode`L=1000
\sfcode`M=1000 \sfcode`N=1000 \sfcode`O=1000 \sfcode`P=1000
\sfcode`Q=1000 \sfcode`R=1000 \sfcode`S=1000 \sfcode`T=1000
\sfcode`U=1000 \sfcode`V=1000 \sfcode`W=1000 \sfcode`X=1000
\sfcode`Y=1000 \sfcode`Z=1000

\newcommand{\Msol}{\hbox{M$_{\sun}$}}
\newcommand{\hii}{\hbox{\ion{H}{2}}}
\newcommand{\no}{\nodata}
\newcommand{\eg}{e.g.}
\newcommand{\ie}{i.e.}
\newcommand{\mmJy}{\hbox{$\mu$Jy}}

\newcommand{\zph}{\ensuremath{z_{\rm ph}}}
\newcommand{\zsp}{\ensuremath{z_{\rm sp}}}
\newcommand{\etal}{et~al.}

\newcommand{\degree}{{}$^\circ$}
\newcommand{\pkg}[1]{\texttt{#1}}

\defcitealias{Willner2023}{Paper~I}
\newcommand{\Pa}{\citetalias{Willner2023}}
\newcommand{\Pap}{\citepalias{Willner2023}}
\newcommand{\galfit}{\hbox{\tt GALFIT}}

\begin{document}

\title{PEARLS: JWST Counterparts of Micro-Jy Radio Sources in the NEP Time Domain Field. II. All Four Spokes}
\shorttitle{Radio Counterparts in the TDF}
\shortauthors{Willner, Gim, Polletta \etal}

\author[0000-0002-9895-5758]{S.\ P.\ Willner}
\affiliation{Center for Astrophysics \textbar\ Harvard \& Smithsonian, 60 Garden Street, Cambridge, MA, 02138, USA}
\email{swillner@cfa.harvard.edu}

\correspondingauthor{S.\ P.\ Willner}
\email{swillner@cfa.harvard.edu}

\author[0000-0003-1436-7658]{Hansung B.\ Gim} 
\affiliation{Eureka Scientific, 2452 Delmer Street, Suite 100, Oakland, CA 94602, USA}
\email{hansung.b.gim@gmail.com}

\author[0000-0001-7411-5386]{Maria del Carmen Polletta}
\affiliation{INAF--–Istituto di Astrofisica Spaziale e Fisica Cosmica Milano,  Via A.\ Corti 12, I-20133 Milano, Italy}
\email{maria.polletta@inaf.it}

\author[0009-0007-0782-0721]{Gibson B.\ Bowling}
\affiliation{School of Earth and Space Exploration, Arizona State University,
Tempe, AZ 85287-1404, USA}
\email{gbbowlin@asu.edu}

\author[0000-0003-3329-1337]{Seth H. Cohen} 
\affiliation{School of Earth and Space Exploration, Arizona State University, Tempe, AZ 85287-1404, USA}
\email{seth.cohen@asu.edu}

\author[0000-0001-8489-2349]{Vicente Estrada-Carpenter}
\affiliation{School of Earth and Space Exploration, Arizona State University, Tempe, AZ 85287-1404, USA}
\email{vestrad9@asu.edu}

\author[0000-0002-6610-2048]{Anton M. Koekemoer} 
\affiliation{Space Telescope Science Institute,
3700 San Martin Drive, Baltimore, MD 21218, USA}
\email{koekemoer@stsci.edu}

\author[0000-0003-3351-0878]{Rosalia O'Brien}
\affiliation{School of Earth and Space Exploration, Arizona State University, Tempe, AZ 85287-1404, USA}
\email{robrien5@asu.edu}

\author[0000-0001-9369-6921]{Alex Pigarelli}
\affiliation{School of Earth and Space Exploration, Arizona State University, Tempe, AZ 85287-1404, USA}
\affiliation{Beus Center for Cosmic Foundations, Arizona State University, Tempe, AZ 85287, USA}
\email{apigarel@asu.edu}

\author[0000-0001-9262-9997]{Christopher N.\ A.\ Willmer} 
\affiliation{Steward Observatory, University of Arizona, 933 N Cherry Ave, Tucson, AZ, 85721-0009, USA}
\email{cnawillmer@gmail.com}

\author[0000-0001-8156-6281]{Rogier A. Windhorst}
\affiliation{School of Earth and Space Exploration, Arizona State University, Tempe, AZ 85287-1404, USA}
\email{Rogier.Windhorst@gmail.com}

\author[0000-0003-1268-5230]{Rolf A. Jansen} 
\affiliation{School of Earth and Space Exploration, Arizona State University, Tempe, AZ 85287-1404, USA}
\email{rolfjansen.work@gmail.com}

\author[0000-0002-6150-833X]{Rafael {Ortiz~III}} 
\affiliation{School of Earth and Space Exploration, Arizona State University, Tempe, AZ 85287-1404, USA}
\email{rortizii@asu.edu}

\author[0000-0002-7265-7920]{Jake Summers} 
\affiliation{School of Earth and Space Exploration, Arizona State University, Tempe, AZ 85287-1404, USA}
\email{jakesummers7200@gmail.com}

\author[0000-0001-7363-6489]{William Cotton}
\affiliation{National Radio Astronomy Observatory (NRAO), 520 Edgemont Road, Charlottesville, VA 22903, USA}
\email{bcotton@nrao.edu}

\author[0000-0002-2115-1137]{Francesca Civano}
\affiliation{Goddard Space Flight Center, USA}
\email{francesca.m.civano@nasa.gov}

\author[0000-0001-9440-8872]{Norman A.\ Grogin}
\affiliation{Space Telescope Science Institute,
3700 San Martin Drive, Baltimore, MD 21218, USA}
\email{nagrogin@stsci.edu}

\author[0000-0002-2203-7889]{W.\ P.\ Maksym}
\affiliation{NASA Marshall Space Flight Center, Huntsville, AL 35812, USA}
\email{peter.maksym@gmail.com}

\author[0000-0002-5319-6620]{Payaswini Saikia}
\affiliation{Department of Astronomy, Yale University, PO Box 208101, New Haven, CT 06520-8101, USA}
\email{payaswini.ssc@gmail.com}

\author[0000-0001-6564-0517]{Ross M.\ Silver}
\affiliation{NASA Goddard Space Flight Center, Greenbelt, MD 20771, USA}
\affiliation{Southeastern Universities Research Association, Washington, DC 20005, USA}
\email{ross.m.silver@nasa.gov}

\author[0000-0002-7791-3671]{Xiurui Zhao}
\affiliation{Cahill Center for Astrophysics, California Institute of Technology, 1216 East California Boulevard, Pasadena, CA 91125, USA}
\email{xiurui.zhao.work@gmail.com}

\begin{abstract}

JWST/NIRCam observations in the North Ecliptic Pole Time Domain Field (TDF) identify 4.4~\micron\ counterparts for 206 of 211 radio sources with $S(\rm 3~GHz)\ga5$~\mmJy\ in a 65\,arcmin$^2$ field.  One of the remaining radio sources is likely to be a radio lobe of a nearby Seyfert galaxy, and the four radio sources without counterparts could be spurious. All but five counterparts are brighter than magnitude 23.5~AB at 4.4~\micron. A simple position match with radius 0\farcs3 would have identified 198 of the counterparts but only in a 4.4~\micron\ catalog created with aggressive deblending of multiple peaks 
within an object's brightness distribution 
into distinct catalog sources. The properties of the radio-host galaxies are mostly consistent with those found in \Pa: the median redshift is \red{1.15}, and the radio emission, calculated taking into account the non-linear dependence of radio luminosity on star-formation rate, is consistent with a star formation origin in $\sim$79\% of the sample. For the other $\sim$21\%, the radio flux could come from star formation hidden behind dust or from an active galactic nucleus. One difference from other studies of radio-source counterparts is that 66\% of the radio hosts show at least one indication of an AGN's presence. The presence of AGN and of hidden star formation could be elucidated by monitoring for source variability, and the TDF is the field most suited to such observations.

\end{abstract}



\section{Introduction} \label{s:intro}

The history of energy generation in the Universe includes both accretion, especially onto supermassive black holes (SMBHs), and star formation.  Early radio surveys {with the NSF's Karl G.\ Jansky Very Large Array (VLA)\footnote{The National Radio Astronomy Observatory is a facility of the National Science Foundation operated under cooperative agreement by Associated Universities, Inc.}} \citep[\eg, FIRST, NVSS;][]{Becker94,Condon98} explored mainly luminous radio galaxies, powered by accretion onto SMBHs, because such sources could be detected at enormous distances 
\citep[\eg,][]{Chambers1996,Seymour2007,Miley2008,Saxena2019}. When radio surveys began to reach flux densities near 1~mJy, a significant fraction of radio sources turned out to be starburst galaxies \citep[\eg,][]{Windhorst1985}.  
Modern, far deeper, radio surveys probe both types of energy generation and complement optical studies of cosmic star formation and accretion
\citep[\eg,][and references therein]{Smolcic2017data,Algera2020,Tompkins2023}. 
Broad multi-wavelength (and multi-messenger) coverage leads to better understanding of sources'  underlying physical processes.

A limitation of radio-continuum surveys is that they do not provide redshifts and therefore distances.  Redshifts, ideally spectroscopic but at least photometric, can come from identifying visible or infrared counterparts of the radio sources, but finding counterparts is not always as easy as a simple positional match.  Radio-source position uncertainties sometimes encompass multiple possible counterparts, and for powerful double-lobe radio sources, the host galaxy is between the lobes rather than coincident with either of them. The problem is even worse when only a single lobe is visible, i.e., when one radio lobe is much fainter than the other.  Despite the difficulties, an early success came when \citet{Spinrad1985} provided nearly complete identifications for the extragalactic sources in the 3CR, a survey complete to 8~Jy at 178~MHz. At flux densities of 10--1000~\mmJy, the optical identification fraction has been lower, about $\sim$75--85\%  in early {Hubble Space Telescope} (HST) imaging surveys \citep[\eg,][]{Fomalont2006, Russell2008, Windhorst1995}. More recent, deeper surveys  have identification completeness above 90\% \citep[\eg,][]{Bonzini2012,Smolcic2017id,Owen2018,Algera2020,Kondapally2021}, even with radio flux densities reaching $<$10~\mmJy\ levels.

To reach very high completeness levels, observations {at wavelengths $\ga$2\,\micron} are essential because passive stellar populations are intrinsically red, and star-forming galaxies can be reddened by dust.  High redshift increases the need for infrared (IR) observations because light observed at visible wavelengths originated as rest-frame ultraviolet (UV) emission, which is often absent or severely reddened.  The stellar emission from galaxies typically peaks near 1.6~\micron\ in the rest frame \citep{Sawicki02}, and observing longwards of that means the increasing intrinsic flux density towards shorter wavelengths somewhat cancels the effect of redshift (\ie, a ``negative K correction''). Another benefit of longer wavelengths is that dust extinction is much smaller. 
As an example, \citet{Strazzullo2010} found visible-band ($gri$) counterparts for $\ge$83\% of a sample of $S(\rm 1.4~GHz) > 13.5$~\mmJy\ (5$\sigma$) radio sources. Adding \textit{JHK} data raised the identification rate to $\ge$90\%, and adding IRAC 3.6 and 4.5\,\micron\ data (reaching mags $\sim$24.3) raised it to $\ge$93\%. Sources added at the longer wavelengths were mostly at $z>1.3$.  Another example is the VLA-COSMOS 3~GHz radio survey, which reached rms sensitivity of 2.3~\mmJy.  \citet{Smolcic2017id} used the field's deep HST data and additional multi-band data to achieve a 90\% identification rate.  Adding IRAC 3.6~\micron\ data raised the completeness rate to 92.4\%.  More dramatically, \citet{Willner2012} used 4.5~\micron\ IRAC observations to identify 100\% of radio sources in a  1.4~GHz survey \citep{Ivison2007} in the Extended Groth Strip, though the radio survey depth was only 50~\mmJy.  More recently, \citet{Cotton2018} found 4.5\,\micron\ counterparts for 98\% of 3~GHz sources reaching 4~\mmJy.

An initial attempt \citep[][hereafter Paper~1]{Willner2023} to identify counterparts of 5~\mmJy\ radio sources in JWST/NIRCam images succeeded at least 97\% and plausibly 100\% of the time.  However, the sample contained only 64 objects because of the limited NIRCam area. This paper reports results for an area four times larger. Section~\ref{s:obs} describes the radio survey, the JWST observations used for the identifications, and the additional data used.  Section~\ref{s:match} describes the matching procedure and results.  Section~\ref{s:disc} gives the source and sample characteristics, and Section~\ref{s:summ} summarizes the results, including recommendations for future matching studies. Distances are based on flat $\Lambda$CDM cosmology with $H_0=67.66$~km\,s$^{-1}$\,Mpc$^{-1}$ and $\Omega _{\rm M} =0.310$ \citep{Planck2018}.  All magnitudes are in the AB system \citep{Oke1983}.

\section{Observations} \label{s:obs}

\subsection{JWST observations}
\label{s:jwst}

The JWST Prime Extragalactic Areas for Reionization and Lensing Science (PEARLS) GTO program \citep{Windhorst2023} includes a NIRCam survey of the Time Domain Field (TDF)  near the North Ecliptic Pole (NEP). The field is within JWST's continuous viewing zone, making observations possible at any time. The observational design and layout of the original JWST TDF were
described by \citet{Jansen2018}, and Jansen \etal\ (in prep.)\ will give full details of the observations (PID 2738, PIs R.\ Windhorst and H.\ Hammel) and data.
Briefly, the TDF area consists of four ``spokes,'' each subtending an intended area $2\farcm3\times 7\farcm4$ at orientation differences of 90\arcdeg. Figure~1 of \citet{Willmer2023} depicts the field layout and location. NIRCam observed the spokes in succession at time intervals of about three months with filters F090W, F115W, F150W, F200W, F277W, F356W, F410M, and F444W.  
The four short-wavelength (SW) filters covered identical areas, as did the four long-wavelength (LW) filters, but the SW area is about 1\farcs5--2\arcsec\ larger in all dimensions than the LW.
The telescope image quality is diffraction-limited \citep{Rigby2023}, but the pixels undersample the point-spread function (PSF) in some filters.\footnote{\url{https://jwst-docs.stsci.edu/jwst-near-infrared-camera/nircam-performance/nircam-point-spread-functions}} 
Achieved image FWHMs range from 60 to 160 milliarcseconds (mas) at the wavelengths observed
\citep[][their Table~1]{Windhorst2023}. 
NIRCam point-source sensitivities range from 28.1 to 29.1\,mag (5$\sigma$); values for each filter and further details were given by \citet[][their {Table~2}]{Windhorst2023}.

Each NIRCam observation also included parallel NIRISS imaging and slitless spectroscopy in the inner half of each spoke. Because the NIRISS observations were done as coordinated parallels, each TDF spoke was observed by NIRISS six months before or after the NIRCam observation of that spoke, \ie,  Spoke~1 NIRCam observations coincided with Spoke~3 NIRISS observations, and vice versa. The NIRISS direct imaging was done in the F200W filter, and the slitless spectroscopy is described in Section~\ref{s:niriss}.

All images were calibrated using version 1.11.3 of the JWST pipeline\footnote{\url{https://github.com/spacetelescope/jwst}} {\citep{Bushouse2023}} with reference files specified by the context file jwst\_1100.pmap. Wisp features in the NIRCam SW images were removed post-pipeline as described by \citet{Robotham2023}.  Final mosaics in all filters have pixel scales of 60 mas/pixel.  All the mosaics were aligned to the GAIA-DR3 \citep{gaia3}\footnote{\url{https://www.cosmos.esa.int/web/gaia/}} astrometric reference frame. 

\subsection{Radio observations}
\label{s:radio}

\citet[][their Appendix~A]{Hyun2023} described the 3~GHz radio observations used here. The observations used the VLA A and B arrays (44 and 4 hours, respectively) to cover a single field
centered on the $z=1.4429$ \citep{Lyke2020} quasar ICRF J172314.1+654746 about 1\arcmin\ south of the reddish \citep[Fig.~1 of][]{Willmer2023} star at the intersection of the southern and eastern TDF spokes.  The VLA field is 12\arcmin\ in radius, at which distance the 3~GHz primary-beam response falls to 0.1. All sources with JWST coverage (Section~\ref{s:jwst}) are within 8\farcm82 of the beam center, where the primary-beam response is $\ge$0.35.  The angular resolution is 0\farcs7 FWHM, and the rms noise near the beam center is 1~$\mu$Jy\,beam$^{-1}$. Away from the beam center, the noise grows as the reciprocal of the primary beam response.  \citet{Hyun2023} extracted 756 $\ge$5$\sigma$ sources from the image, and that list (their Table~3) is the input source list for this paper. Random position uncertainties for the faintest unresolved sources are about 70~mas in each coordinate, while a few resolved sources have uncertainties  $\ga$250~mas \citep{Hyun2023}. Based on the source matches (Section~\ref{s:match}), systematic  differences between the VLA and NIRCam coordinate systems are negligible.

\subsection{HST imaging}
\label{s:ground}

The NEP TDF was imaged with HST ACS/WFC in F606W and F435W and with WFC3/UVIS in F275W as part of programs GO~15278 (PI: R.~Jansen) and TREASUREHUNT (GO~16252, 16793; PIs: R.~Jansen \& N.~Grogin). The respective 3\,$\sigma$ limiting magnitudes are  $\simeq$29.0, 28.2, and 27.3~mag
(R.~Jansen et al.\ {in prep.}; \citealt{OBrien2024}). Coverage of the field was gradually built up over time between 2017 October 1 and 2022 October 31 using HST CVZ opportunities that allowed  $\sim$9\,263--10\,375~s per visit in each of F606W and F435W and 
$\sim$20\,350--21\,500~s per visit in F275W. The ACS/WFC images cover essentially the entire TDF to a radius of $\sim$7\farcm5 with a total area of $\sim$194~arcmin$^2$ \citep[Fig.~1 of][]{OBrien2024}, while the WFC3/UVIS images provide non-contiguous coverage of this field.
R.~Jansen et al. ({in prep.}) will describe the data reduction in detail, but in brief  
\citep{OBrien2024}, pipeline-processed data were retrieved from MAST and post-processed to remove satellite trails and to fit and subtract the background column-by-column. The images were then
astrometrically calibrated to the Gaia/DR3 \citep{gaia3}  reference grid.  For the present work, the HST images were drizzled, with cosmic-ray rejection, onto the same 30~mas pixel grid as the NIRCam mosaics rather than the original \citet{OBrien2024} grid.  This made matched-image photometry straightforward.

\subsection{Ground-based Spectroscopic redshifts}
The principal spectroscopy came from BINOSPEC
\citep{Fabricant2019} at the MMT Observatory. Willmer \etal\ (in prep.)\ will give full details, but in brief, the observations were obtained in 2018--2019.\footnote{Eight new redshifts are also included.} Because this was before JWST's launch, targets were primarily selected based on VLA and Chandra detections. The instrument setup used the
270\,lines/mm grating to achieve spectral coverage from 
$\sim$4000\,\AA\ to  $\sim$9000\,\AA\ (depending on the source position within the mask) with a typical dispersion of 1.30\,\AA/pixel and
resolution $R\approx 1340$. Typical exposure times were 90~minutes (six 900~s exposures). The redshifts were measured using an adaptation of the DEEP2 pipeline \citep{Newman2013}, which used a combination of template cross-correlation and emission-line fits. All redshifts were visually validated and assigned a quality code $Q$ ranging from 1 (failed) to 4 (definitive redshift based on 2 or more spectral lines).  This work uses redshifts with $Q\ge3$ (good redshift but minor doubt).  Redshifts with $Q=2$ (uncertain) are {used for comparison with photometric redshifts (Section~\ref{s:fits})} but not used for any analysis. Typical redshift uncertainties are $\delta z \approx 0.0005$.

The field was also observed by Hectospec \citep{Fabricant2005} at the
MMT on 2022 June 9 and 2023 May 21 (PIs F.~Civano and X.~Zhao).
Hectospec is fiber-fed with each fiber subtending 1\farcs5. The observations used a 270\,lines/mm grating to achieve spectral coverage from 3650 to 9200\,\AA\ with typical dispersion of 1.2\,\AA/pixel and spectral resolution 6\,\AA.  The galaxies targeted were primarily NuSTAR and XMM-Newton X-ray sources, and redshifts were derived with the cross-correlation program {\sc xcsao} \citep{Kurtz1998} from one-dimensional spectra, which are all that Hectospec produces.
The Hectospec observations added four redshifts (indicated in Table~\ref{t:ID}), promoted two
redshifts from Binospec $Q=3$ to Hectospec $Q=4$, and confirmed 11 other Binospec
redshifts. Altogether, the ground-based spectroscopy gave 38 $Q=4$ and 18 $Q=3$ redshifts (though one of the $Q=3$ redshifts is for ID~99, which is outside the NIRCam sample defined in Section~\ref{s:process}).

[In this preprint, Tables~1, 3, and 4 are given in full at the end of
the document.  In the published paper, only short versions will be
printed.  Machine-readable tables are in the arXiv source directory.]

\begin{figure*}
{\bf ~~~~3~GHz\hfill ~~F090W \hfill F115W \hfill F150W \hfill F200W \hfill F277W \hfill F356W \hfill F444W~~~~}\\
\includegraphics[width=\linewidth]{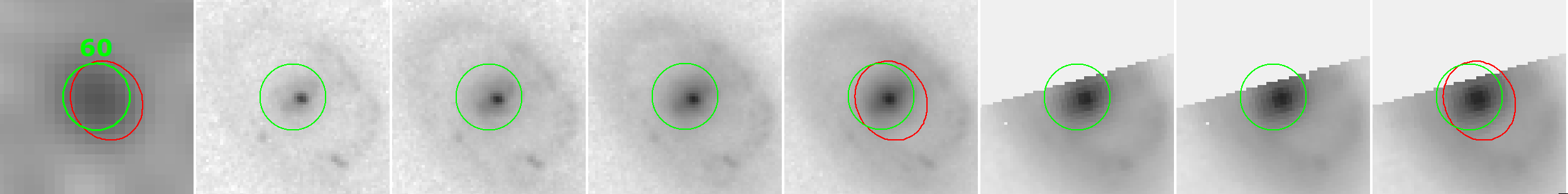}\\
\includegraphics[width=\linewidth]{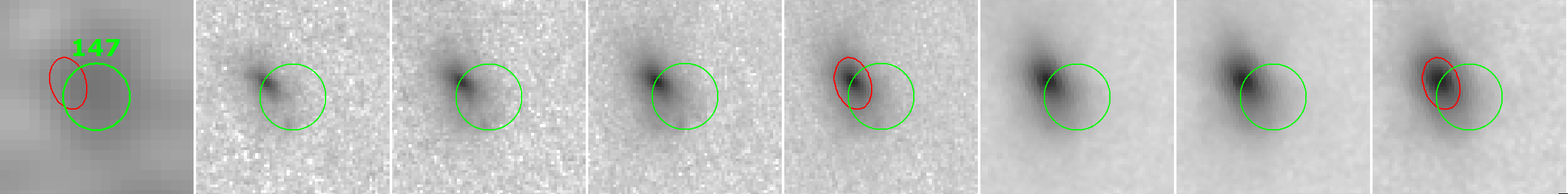}\\
\includegraphics[width=\linewidth]{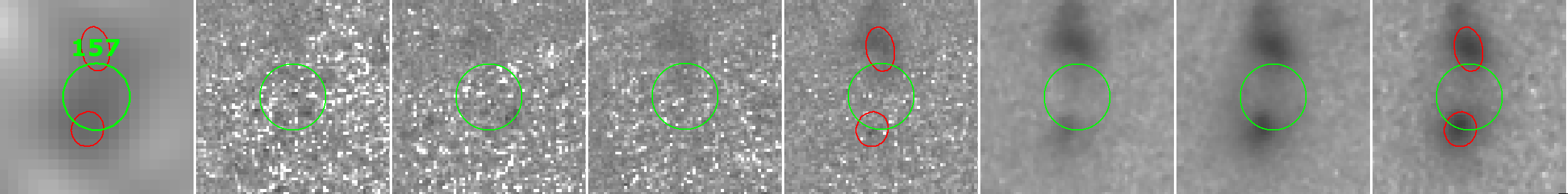}\\
\includegraphics[width=\linewidth]{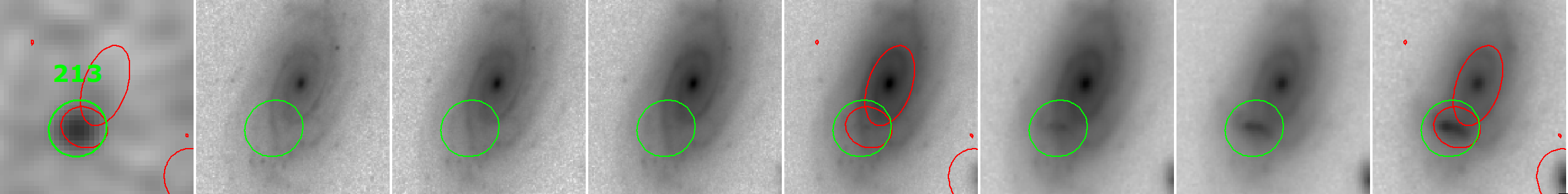}\\
\includegraphics[width=\linewidth]{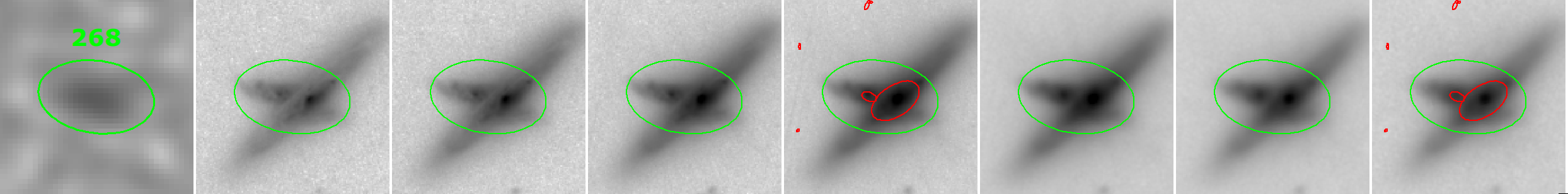}\\
\includegraphics[width=\linewidth]{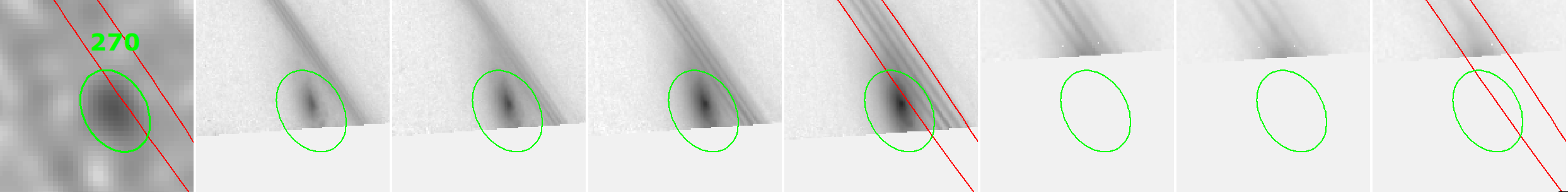}\\
\includegraphics[width=\linewidth]{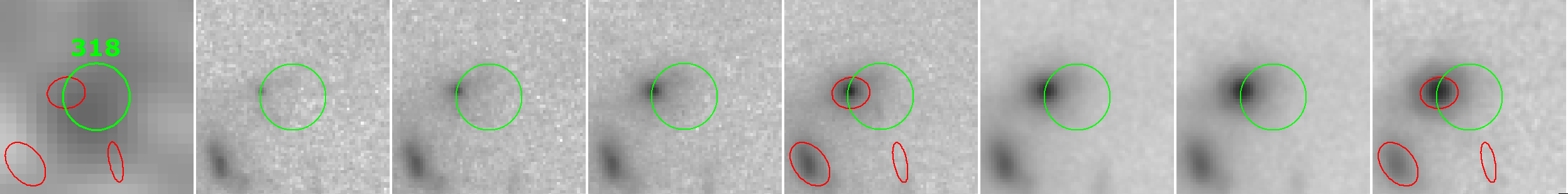}\\
\caption{Negative images of sources that illustrate some of the difficulties of position matching.  The leftmost panels show the 3\,GHz radio image with the source ID indicated. Other panels show the NIRCam images in order of wavelength as labeled at the top. (The F410M image is not shown.) Each thumbnail is 2\arcsec\ on a side except 4\arcsec\ for IDs~213/268/270.  North is up, east to the left. Radio images have greyscale $-3$ to +25\,\mmJy\,beam$^{-1}$ 
and are before primary-beam correction to give nearly constant noise.  Near-infrared (NIR) images have different greyscales for each source, but all images for a given source have the same greyscale to indicate color information.
Green circles show the radio positions \citep{Hyun2023} and, for extended radio sources, the elliptical radio sizes and orientations.  For point sources, the green circles are 0\farcs7 in diameter, the 3~GHz beam size.  Red ellipses show F444W source dimensions from SExtractor. MAG\_AUTO photometry ellipses are about 2.5 times larger than the ellipses shown.}
\label{f:match}
\end{figure*}

\begin{figure*}
{\bf ~~~~3~GHz\hfill ~~F090W \hfill F115W \hfill F150W \hfill F200W \hfill F277W \hfill F356W \hfill F444W~~~~}\\
\includegraphics[width=\linewidth]{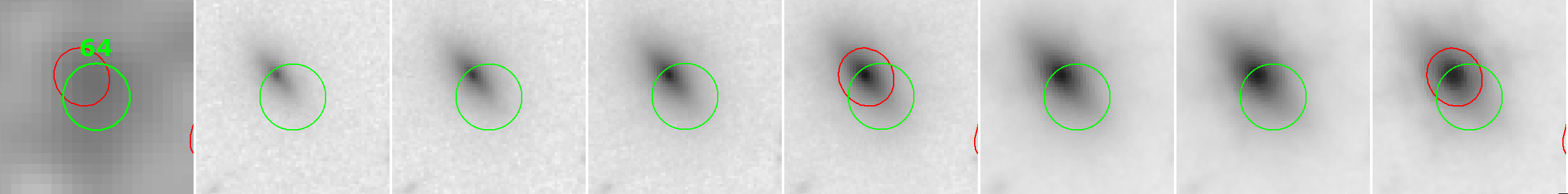}\\
\includegraphics[width=\linewidth]{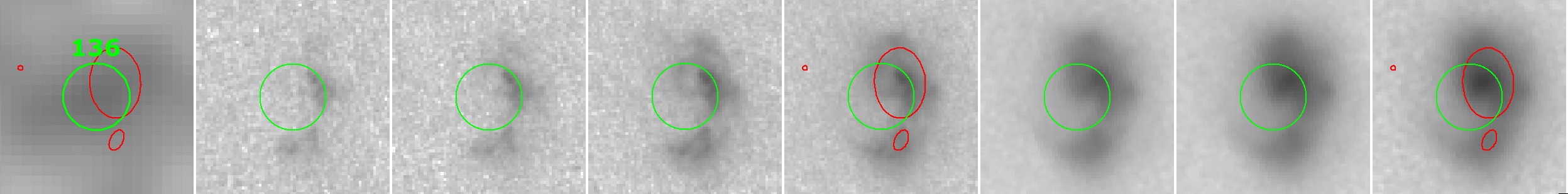}\\
\includegraphics[width=\linewidth]{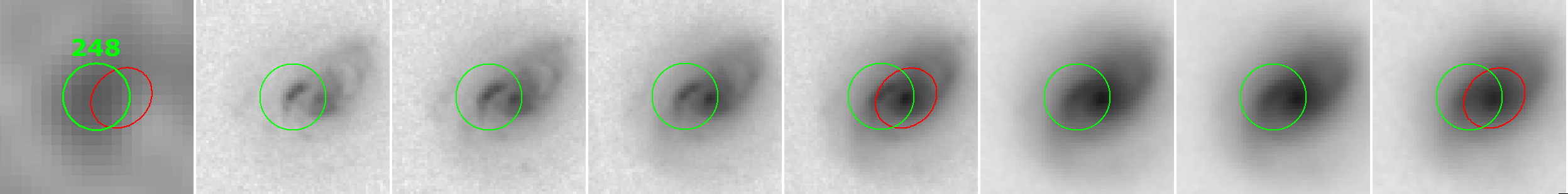}\\
\includegraphics[width=\linewidth]{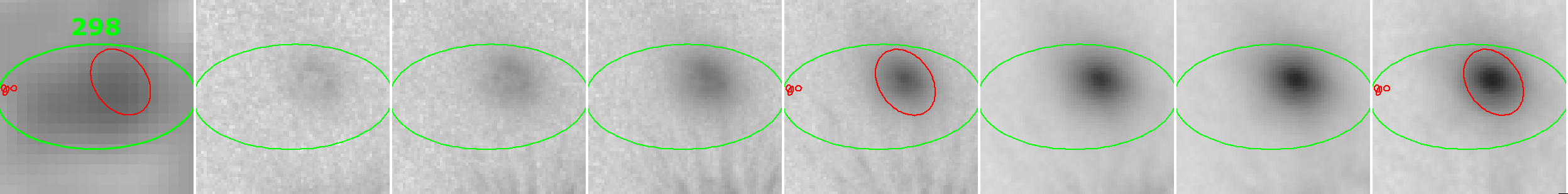}\\
\includegraphics[width=\linewidth]{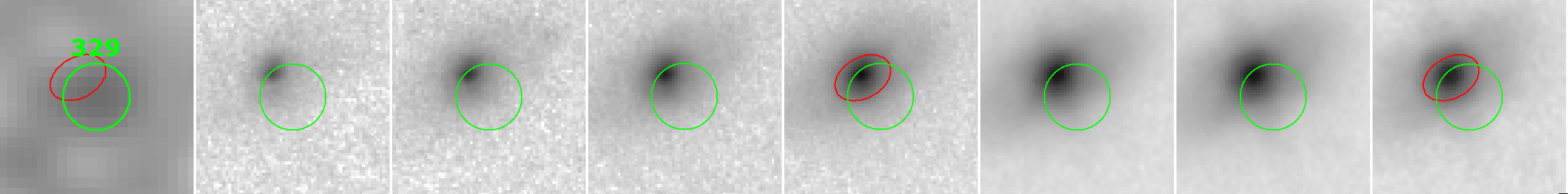}\\
\includegraphics[width=\linewidth]{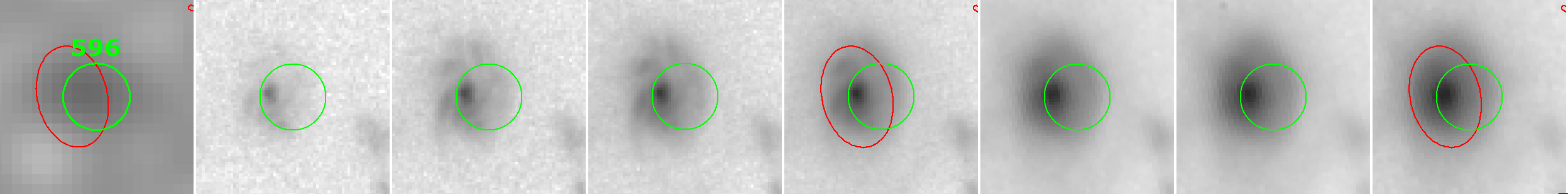}
\caption{Negative images of six sources for which the automated position search found a counterpart within 0\farcs3 but not within 0\farcs24, i.e., the sources for which the automated matches are most doubtful. All thumbnails are 2\arcsec\ on a side, and other details are as in Figure~1.}

\label{f:bigsep}
\end{figure*}

\begin{figure*}
{\bf ~~~~3~GHz\hfill ~~F090W \hfill F115W \hfill F150W \hfill F200W \hfill F277W \hfill F356W \hfill F444W~~~~}\\
\includegraphics[width=\linewidth]{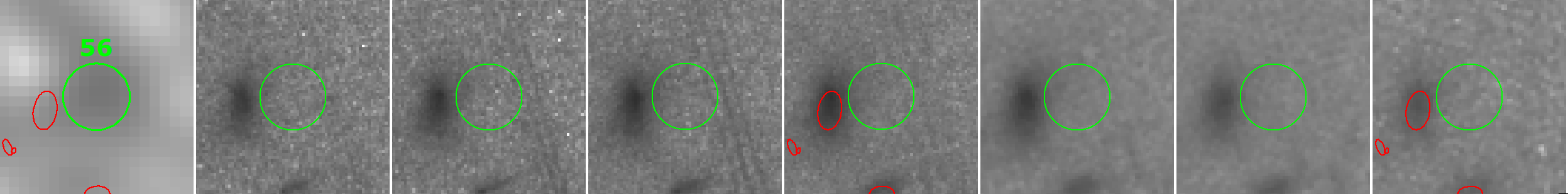}\\
\includegraphics[width=\linewidth]{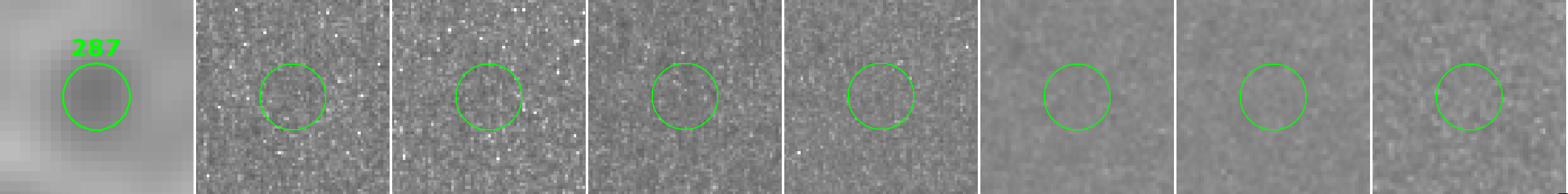}\\
\includegraphics[width=\linewidth]{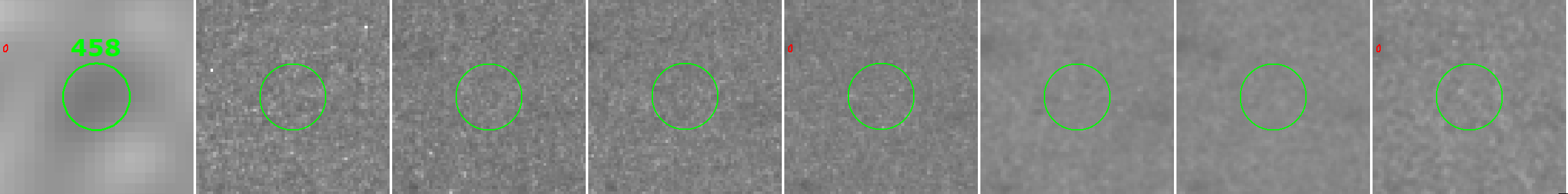}\\
\includegraphics[width=\linewidth]{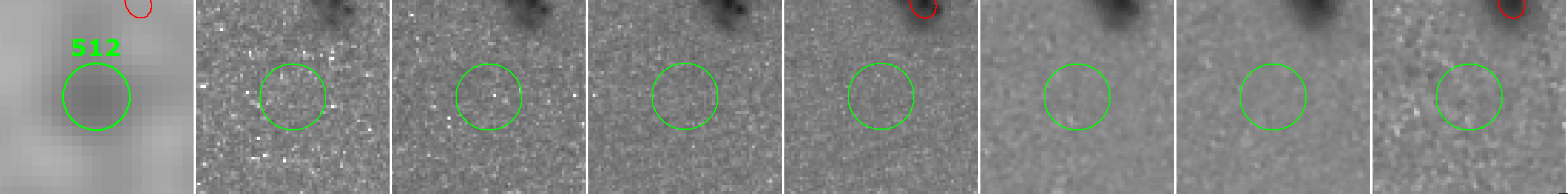}\\
\caption{Negative images of sources with no visible JWST counterpart. (ID~193 was shown in \Pa.)  Image details are as in Figure~1.}

\label{f:invis}
\end{figure*}

\subsection{NIRISS Spectroscopy}
\label{s:niriss}
NIRISS observations of the TDF included not only F200W imaging but also slitless spectroscopy (Estrada-Carpenter \etal, in prep.). The spectral range covered, limited by the F200W blocking filter, was 1.75--2.24~\micron, and the spectral resolution was $R\approx220$. Both perpendicular grism orientations, GR150C and GR150R, were observed.    The NIRISS spectra did not cover the full NIRCam area, but 115 of the 211 radio sources in the NIRCam sample defined in Section~\ref{s:process} have NIRISS spectra at the source position.

Many of the NIRISS spectra show only a featureless continuum, are too noisy to be useful, or have artifacts from nearby contaminating sources. Only 35 show a definite emission line, but a single line cannot be identified without further information.  Photometric redshifts are useful but not definitive for this purpose, and extra redshift precision (as opposed to accuracy) is not useful for this paper.  Therefore single-line redshifts are not used here.  Sixteen sources showed more than one spectral feature, confirming the feature identifications and giving $Q=3$ redshifts that are used here. (In all cases, the secondary features were too weak to merit $Q=4$.) Fifteen of these redshifts (flagged ``j'' in Table~\ref{t:ID}) were new, and one (for ID~348) raised a $Q=3$ Binospec $z=0.8861$ to $Q=4$.   The other $Q=3$  NIRISS redshift (ID 429) is consistent with the Binospec redshift. Altogether 70 sources (69 within the NIRCam sample) have $Q\ge3$ spectroscopic redshifts.

\section{Source matching} 
\label{s:match}
\subsection{Match process}
\label{s:process}

An initial NIRCam object catalog was created with SExtractor
\citep{Bertin1996} on the F444W image. In contrast to
\citetalias{Willner2023}, the SExtractor parameter {\sc
  deblend\_mincont} was set to 0.0002 to give more aggressive source
deblending. Then for each radio source, the nearest source in the
F444W catalog was identified. All potential matches were examined
visually in all NIRCam images to verify or correct the potential
counterparts, and radio sources lacking NIRCam coverage were excluded
from the sample. Six radio sources that are near the edge of the SW coverage and outside the LW coverage (\citealt{Hyun2023} IDs 270/295/364/447/500/544)
are included in the NIRCam sample, but their photometry in the SW filters is
less reliable than for most sources because part of their light may fall
outside the image.  Two sources, IDs~81/99, lack NIRCam
coverage but are covered by NIRISS F200W.  These both have obvious
counterparts and are included in Table~\ref{t:ID} and in the source counts, but they are not members of the NIRCam sample and are
excluded from further analysis because they have no color information. The NIRCam sample thus comprises 211 radio
sources.  This is fewer than four times the Spoke~1 number, probably
because Spokes 2--4 are on average farther from the radio beam
center and therefore have less sensitive radio observations.

Automated {(nearest-neighbor)} matching found an F444W source within
0\farcs3 of the radio position for 198 of the 211 radio sources in the initial sample.  For all of these,
visual inspection showed an obvious identification, usually a galaxy
well resolved on the NIRCam image. The more aggressive deblending in the new F444W catalog identified
the correct counterpart for ID~213, which as shown in Figure~\ref{f:match} is overlapped by a large
spiral galaxy that the \Pa\ automated matching
identified as the counterpart.  ID~60 (Figure~\ref{f:match}) is about half off the LW
frames, but the bright nucleus is on the frame, and SExtractor found it.

Of the 13 sources for which automated matching found no counterpart
within 0\farcs3, five have no recognizable counterpart at all.
Section~\ref{s:missing} discusses those.  Three others (IDs~147/157/318; Figure~\ref{f:match}) have
counterparts within 0\farcs4. ID~147 is 0\farcs33 from its assigned counterpart, which is a red galaxy with an elongated disk and a bright nucleus.  ID~157 is between two galaxies
0\farcs8 apart.  The radio flux is probably the blended flux from both galaxies, but the
counterpart assigned here is the southern galaxy, which is redder and
brighter in F444W than the northern galaxy.  ID~318 is 0\farcs32 from the counterpart assigned here, which is a red galaxy of early-type appearance.  In these three cases, the radio source may be a AGN lobe, or the radio position may simply have been affected by random uncertainties. For the five galaxies outside the LW area, the automated F444W search found unrelated sources up to 1\farcs3 away.  A similar search on the F200W image would have found the correct counterparts though only by luck for ID~270 (Figure~\ref{f:match}).  That source, in addition to being near the SW image edge, is near some diffraction spikes from a bright star outside the image.  SExtractor found the spikes, not the actual source, but the assigned position for the spikes happened to match the source position.  The actual counterpart of ID~270 is bright and extended and separated only 0\farcs16 from the radio position. Identifications look reliable for all five sources.  In particular, there is no reason to believe an F444W image would show different counterparts.  ID~544 is also outside the LW image but only barely, and SExtractor found the outskirts of its correct counterpart only 0\farcs18 from the radio position.

The six sources with radio to F444W offsets between 0\farcs24 and 0\farcs3 are the most likely of the automated matches to be false ones. They are a mixed group as shown in Figure~\ref{f:bigsep}.  Some of them are discussed individually in Appendix~\ref{s:special}, but the final counterpart identifications look reliable as discussed in Section~\ref{s:reliability}.  The only dubious identification is for ID~248, where the radio source is closer to the bright spot in the galaxy's spiral arm, which was not cataloged as a separate source even with the aggressive deblending than to the galaxy's photocenter. In this paper, the counterpart is considered to be the whole galaxy, which is centered 0\farcs26 from the radio position, rather than the bright spot at essentially 0\arcsec\ separation.

\begin{figure}
\includegraphics[width=\linewidth, clip=true, trim=0 0 0 0]{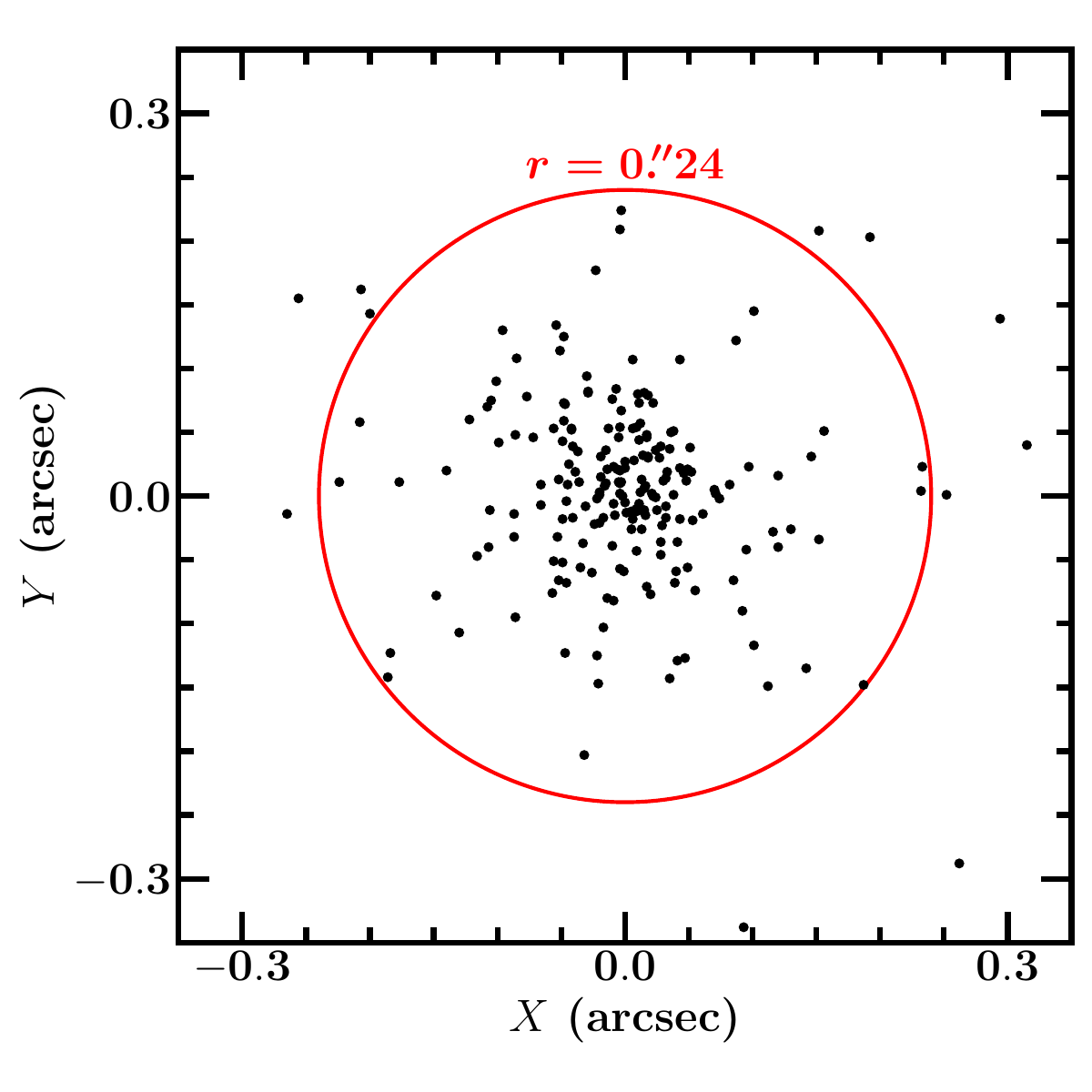}
\caption{Offsets between radio catalog position and identified IR counterpart position. $X$ represents the offset in right ascension, and $Y$ represents the declination offset. IR positions are from F444W if available but from F200W for eight sources outside the F444W image. The red circle has a radius of 0\farcs24, slightly less than the ID~248 (see text) separation. }
\label{f:xy}
\end{figure}

\begin{figure}
\includegraphics[width=\linewidth, clip=true, trim=0 0 0 0]{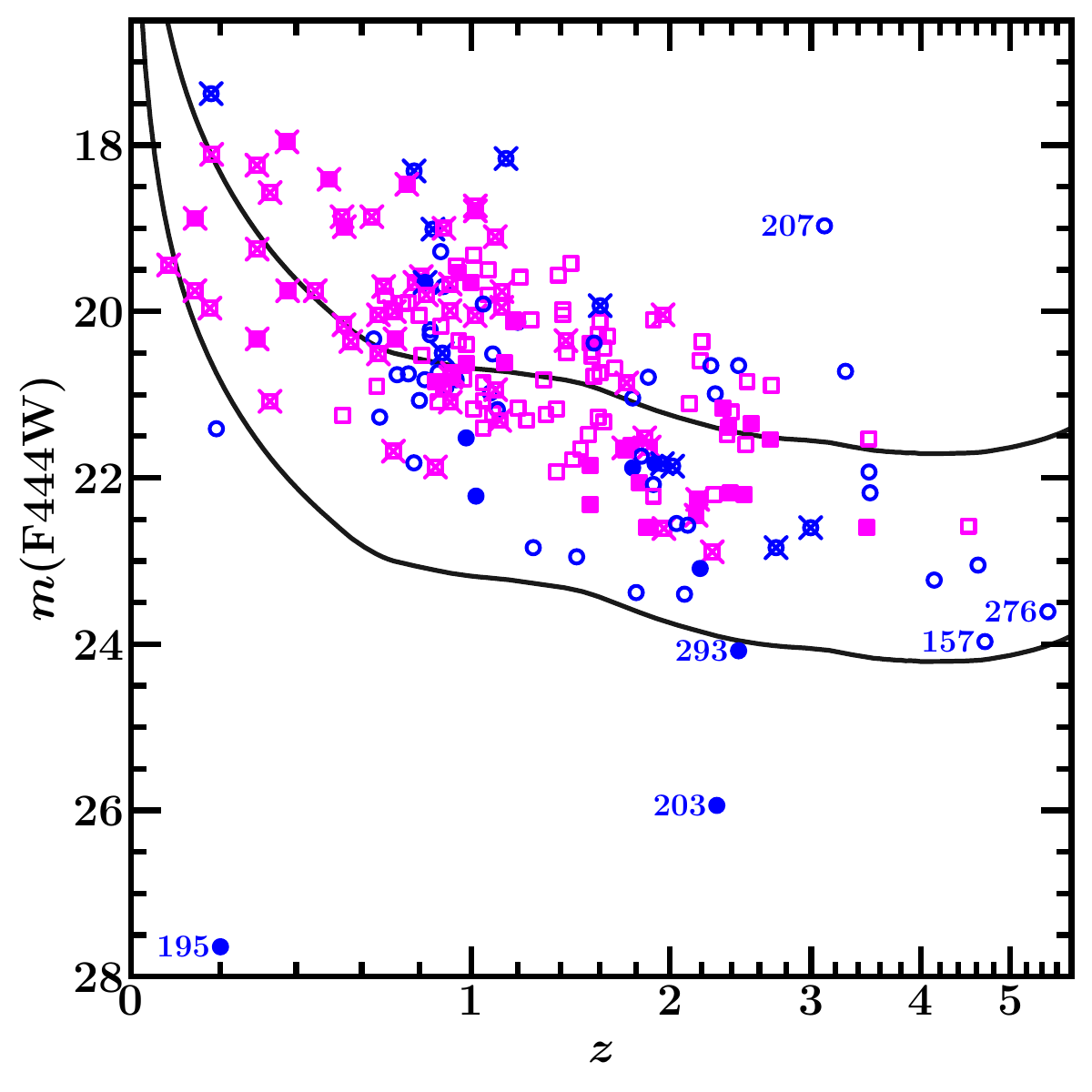}
\caption{F444W apparent AB magnitude versus redshift scaled as $\log(1+z)$ for the host galaxies. 
Point shapes and colors indicate SED classes defined in Section~\ref{s:properties}: blue circles represent type QSO, and magenta squares represent type Gal.  Filled symbols indicate galaxies with excess radio emission as defined in Section~\ref{s:properties}, and $\times$ symbols indicate galaxies with spectroscopic redshifts. 
The solid lines show the F444W magnitudes of stellar populations (\citealt{Bruzual2003} stellar models) with stellar masses $10^{10}$ and $10^{11}$\,\Msol\ that formed (with ${\rm SFR}\propto e^{-t/0.5\,{\rm Gyr}}$) at $z=7$ and evolved passively from then.
A few points far from the main distribution are labeled with their source IDs.
}
\label{f:fz}
\end{figure}

\begin{figure*}
{\bf ~~~~3~GHz\hfill ~~F090W \hfill F115W \hfill F150W \hfill F200W \hfill F277W \hfill F356W \hfill F444W~~~~}\\
\includegraphics[width=\linewidth]{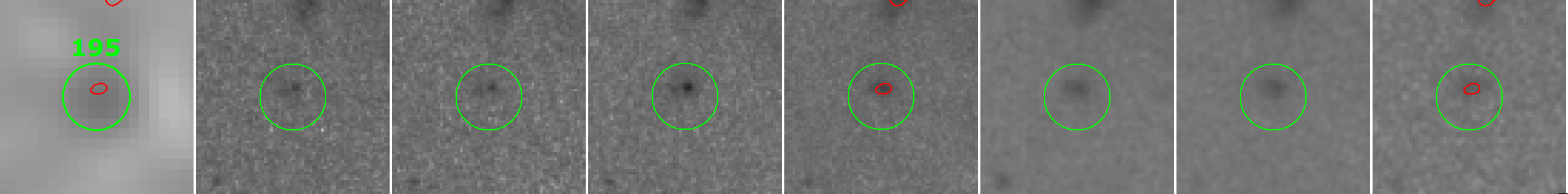}\\
\includegraphics[width=\linewidth]{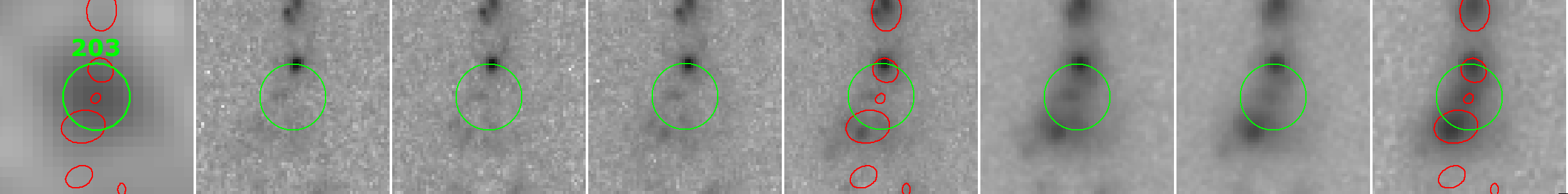}\\
\includegraphics[width=\linewidth]{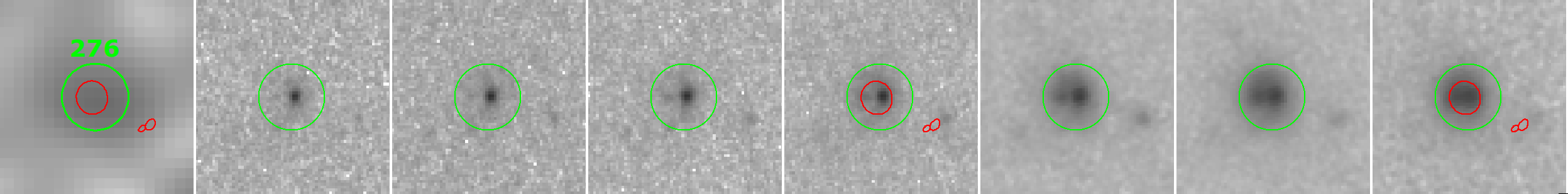}\\
\includegraphics[width=\linewidth]{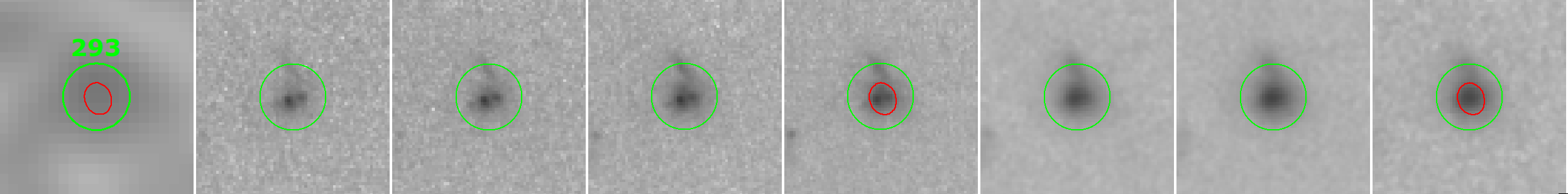}\\
\caption{Negative images of four sources with $\rm [F444W]>  23.5$. ID~157, which has $\rm [F444W]=23.97$ is shown in Figure~\ref{f:match}.  Image details are as in Figure~1.
}
\label{f:faint}
\end{figure*}

\subsection{Sources with no JWST counterpart}
\label{s:missing}

Five radio sources, IDs~56/193/287/458/512, have no visible
counterpart in the NIRCam images.  Two of them, IDs~193/287, are in
Spoke1 and were discussed in \Pa. ID~193 was considered likely to be a radio lobe of the $z=0.17$ Seyfert
galaxy ID~194 rather than an infrared-faint radio source \citep[IFRS;][]{Zinn2011}.
Principal arguments were the radio source being spatially extended,
its proximity to the Seyfert galaxy,
and the low likelihood of finding even one IFRS in the TDF area.

The three no-counterpart sources outside Spoke1 are similar to ID~287, which 
\Pa\ considered most likely to be a spurious radio source.  All four
sources, shown in Figure~\ref{f:invis}, are near the sensitivity limit of the radio survey with
None of these radio sources was found by an independent source search \Pap\ with the {\tt PyBDSF}
\citep{Mohan2015} algorithm.
In radio imaging, problems such as unflagged radio-frequency interference and calibration and deconvolution errors all tend to raise the wings of the noise distribution above Gaussian.  This makes false positives more likely, and the presence of four spurious sources in the NIRCam area is reasonable.

If any of the four radio sources without a counterpart is real, it would be an IFRS \citep{Zinn2011}).  These are thought to be
radio-loud AGN at high redshifts, perhaps $z>5$
\citep{Zinn2011,Herzog2016}.  IFRSs have median radio spectral indices
around $-0.88$ \citep{Herzog2016} and therefore
would have 1.4~GHz flux densities $\sim$10~\mmJy\ (or 20~\mmJy\ for ID~56).
At this radio flux density, the IFRS surface density is unknown but perhaps
$\sim$30\,deg$^{-2}$ \citep{Zinn2011}, and the expected number of such sources in the
4-spoke TDF area is $\sim$0.5, making it unlikely to find more than one IFRS in this area and reasonable to suspect that all four radio sources are spurious.

ID~329 could also be an IFRS, but its 0\farcs28 offset from the  nucleus of an $m(\rm F444W)\approx21.1$~mag galaxy argues for association with the galaxy despite there being no other indication that this galaxy hosts an AGN.

\subsection{Identification Reliability}
\label{s:reliability}

Figure~\ref{f:xy} shows the coordinate offsets between the radio
positions and the positions of the identified NIRCam counterparts. The median offset for the 206 identified sources (including eight identified in the F200W band) is 0\farcs07, consistent with the 0\farcs08 found in \Pa\ and with the VLA position uncertainties \citep{Hyun2023}, confirming the reliability of most of the identifications. 

Figure~\ref{f:fz} shows the F444W magnitudes of the radio-host galaxies as a function of redshift (Section~\ref{s:fits}). Nearly all of the counterparts are brighter than 23.8~mag and would be detectable in less-sensitive 4.4 or 4.5~\micron\ surveys. Finding radio-source counterparts far above the detection limit is strong evidence that few if any detections are spurious, as argued in \Pa. Five sources (labeled in  Figure~\ref{f:fz}) are fainter than 23.8~mag. An image of ID~157 is in Figure~\ref{f:match}, and Figure~\ref{f:faint} shows images of the other four. Some additional comments on IDs~203/276 are in Appendix~\ref{s:special}.

\section{Analysis} \label{s:disc}
\subsection{SED fitting method}
\label{s:fits}

\begin{figure}
\centering
\includegraphics[width=\linewidth]{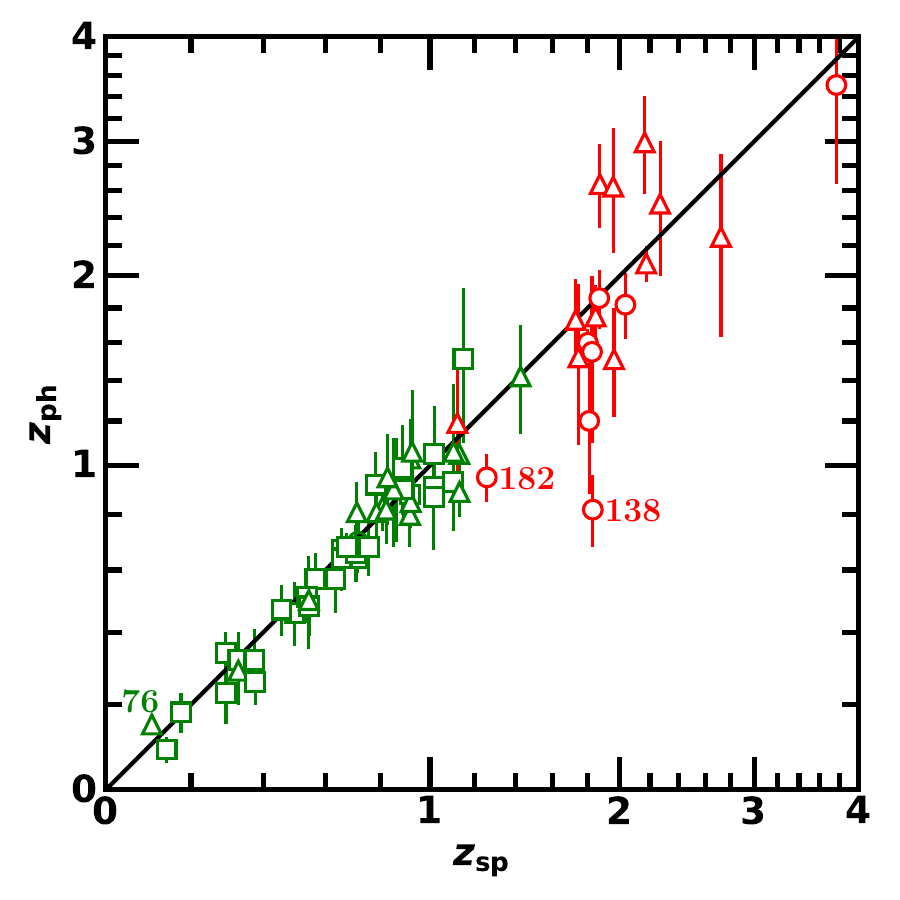}
\vspace{-5.5ex}
\caption{Photometric versus spectroscopic redshifts. Point shapes indicate the \zsp\ quality: squares $Q=4$, triangles $Q=3$, circles $Q=2$. Green indicates \zsp\ from Binospec or Hectospec, red from NIRISS.  Only points with \zph\ uncertainty $\delta \zph<0.32\zph$ are shown.  The solid line shows equality. \red{Three} ${>}2.5\sigma$ outliers are labeled.
 }
\label{f:zz}
\end{figure}

To derive host-galaxy properties, including photometric redshifts, we used photometry from JWST/NIRCam
{(Section~\ref{s:jwst})} and from HST 
{(Section~\ref{s:ground})} in addition to the \citet{Hyun2023} radio flux densities. The NIRCam images were convolved  to
match the PSF of the F444W images separately for each spoke because of the differing position angles and following the method described by \citet[][their Sec.~2.3]{Summers2023}.  Source colors were derived
from SExtractor {\sc iso} magnitudes by running in dual-image mode
with F444W as the detection image.  The colors were applied to the
F444W {\sc auto} magnitudes to derive magnitudes at the seven other
NIRCam wavelengths.  HST magnitudes were derived the same way except
that no image convolution was used.  
For red sources, in many cases the HST and
in a few cases the short-wavelength NIRCam magnitudes were for nearby, unrelated sources rather than the radio-source counterpart.
These magnitudes were replaced by upper limits of half the observed flux densities, i.e., attributing at least half the observed flux to the contaminating source. 
All other upper limits were derived from the pipeline uncertainty (``ERR'') images as the square root of the summed variance within the photometry aperture.  Source ID~194 was saturated in the three longest-wavelength bands, which were treated as unmeasured.   Sources that were near but not off the image edge used the flux density as corrected by SExtractor.

Spectral energy distributions (SEDs) were analyzed with CIGALE \citep{Boquien2019, Yang2020, Yang2022} version 2025.0.\footnote{This work was carried out using cloud computing resources provided by Amazon Web Services.} Table~\ref{tab:CIGALE_parameters} summarizes the key parameters and the values or ranges used. The stellar component was modeled using \citet{Bruzual2003} synthesis models with a \citet{Chabrier2001} initial mass function (IMF) and solar ($Z=0.02$) metallicity. The star-formation history was exponential (\pkg{sfhdelayed}) with $e$-folding timescales given by $\tau_{\rm main}$   and main-population ages given by $\rm age_{\rm main}$.  Dust attenuation followed a modified \citet{Calzetti2000} law (\pkg{dust\_modified\_starburst}) with nebular-line color excess {$E(B-V)_{\rm lines}$} and power-law slope modifications {$\delta$}. IR dust emission was modeled using \citet{Dale2014} templates (\pkg{dale2014}) with emission slopes {$\alpha$}, while potential AGN contributions were incorporated via the SKIRTOR model \citep[\pkg{skirtor2016};][]{Stalevski2012, Stalevski2016}. The AGN-component parameters included viewing angle ${i}$, fractional AGN contribution to the bolometric luminosity ($f_{\rm AGN}$),
polar dust extinction $E(B-V)_{\rm AGN}$, and a power-law index change $\delta_{\rm AGN}$ modifying the rest visible-wavelengths emission of the accretion disk to represent the transitional ADAF-to-disk SED \citep{Lopez2024}.  A value $\delta_{\rm AGN}=0$ indicates pure advection-dominated accretion flow (ADAF), while $\delta_{\rm AGN}=1$ indicates a pure thin disk. 
Radio continuum emission was modeled through synchrotron processes (\pkg{radio}) with radio--IR correlation coefficient $q_{\rm IR}$ for star formation and radio-loudness parameter \textsl{$R_{\rm AGN}$} for AGN activity.

In comparison to \Pa\ (which used CIGALE version 2022.1), the current parameters tend to restrict the options for fitting SEDs.  This is more appropriate for the small number of observation wavelengths available.  Fewer options limit the fitting to common SED types rather than rare types with extreme parameters.  Other changes from \Pa\ include using HST data rather than Subaru/HSC and MMT/MMIRS data and including the radio flux density in the fits.  These changes should give more robust parameter estimates in most cases. Another change, which amounts to a simple scaling of the output values \citep{Madau2014}, was using a Chabrier IMF instead of a \citet{Salpeter1955} IMF.
One other difference is that the results in Table~\ref{t:SEDfits} are CIGALE's ``Bayes'' values and uncertainties (as described in the table note) whereas \Pa\ used the single best-fit model for the parameter values although the Bayes range was also given.

Best-fit parameters for each galaxy were determined through $\chi^{2}$ minimization with uncertainties estimated via CIGALE's Bayes probability distributions, yielding physical parameters including photometric redshift, stellar mass, star formation rate, visual extinction, and AGN contribution.
The ``SED sample'' comprises the 200 radio hosts that have LW photometry available.

\setcounter{table}{1}
\begin{table*}
\hspace*{-5em}
\begin{threeparttable}
\caption{CIGALE Parameters for SED Modeling}
\label{tab:CIGALE_parameters}
\small
    \begin{tabular}{lllll}
    \hline\hline
     Model Properties & Modules & Parameters  & \Pa \textsuperscript{\dag} & This work\textsuperscript{\ddag} \\
    \hline
       \multirow{2}{\widthof{Star-formation}}{Star-formation history} & \pkg{sfhdelayed} & $\tau_{\rm main}$$^{a}$ & 0.1, 0.5, 1, 2.5, 5~Gyr & 0.1, 0.5, 1, 5~Gyr\\
           &                & age$_{\rm main}$$^{b}$ & 0.5, 1, 3, 5, 7, 9~Gyr & 0.5, 1, 3, 5, 7~Gyr\\
           &                & $f_{\rm burst}$$^{c}$ &  0.0, 0.001 & 0.0 \\
           \cline{2-5}
           & \pkg{sfh2exp} & $\tau_{\rm main}$$^{a}$ & 0.1, 0.5, 1, 2.5, 5~Gyr & \multirow{2}{2pt}{N/A}\\
           &                 & age$_{\rm main}$$^{b}$ & 0.5, 1, 3, 5, 7, 9~Gyr & \\
           \cline{2-5}
           & \pkg{sfhdelayedbq} & $\tau_{\rm main}$$^{a}$ & 0.1, 0.5, 1, 2.5, 5~Gyr & \multirow{2}{2pt}{N/A}\\
           &                 & age$_{\rm main}$$^{b}$ & 0.5, 1, 3, 5, 7, 9~Gyr & \\
    \hline
       \multirow{2}{\widthof{Stellar}}{Stellar populations} & \pkg{bc03} & Initial mass function  & Salpeter & Chabrier \\
                          &            & metallicity   & 0.0001, 0.004, 0.02 & 0.02\\
    \hline
       Dust attenuation &  \pkg{\multirow{2}{\widthof{starburst}}{dust\_\linebreak[0]modified\_\linebreak[0]starburst}} & $E(B-V)_{\rm lines}$$^{d}$ & 0.01, 0.1, 0.25, 0.5, 1.0, 1.5 & 0.1, 0.3, 0.6, 1.0, 2.0 \\
                        &                                  & E\_BV\_factor$^{e}$ & 0.5, 1.0 & 0.44 \\
                         &                                 & uv\_bump\_amplitude$^{f}$ & 0, 1, 2, 3 & 3 \\
                         &                                 & powerlaw\_slope$^{g}$ & $-0.5$, $-0.25$, $0.0$ & $-0.5$, $-0.25$, $0.0$ \\
                         &                                 & Ext\_law\_emission\_lines$^{h}$ & Milky Way, SMC & Milky Way \\
    \hline
       Dust emission &  \pkg{dl2014} & $\alpha$$^{i}$ & 1.5,~2.5 & \multirow{4}{2pt}{N/A} \\
                     &               & $u_{\rm min}$$^{j}$ & 0.1, 1, 10, 50 & \\
                     &               & $q_{\rm pah}$$^{k}$ & 2.5, 5.26 & \\
                     &               & $\gamma^{l}$ & 0.1 & \\
    \cline{2-5}
                     &  \pkg{themis} & $\alpha$$^{i}$ & 1.5,~ 2.5 & \multirow{4}{2pt}{N/A}\\
                     &               & u$_{\rm min}$$^{j}$ & 0.1, 1, 10, 50 & \\
                     &               & $q_{\rm pah}$$^{k}$ & 2.5, 5.26 & \\
                     &               & $\gamma^{l}$ & 0.1 & \\         
    \cline{2-5}

                     & \pkg{dale2014}  & $\alpha$$^{i}$ & N/A & 1.5, 2.0, 2.5 \\
    \hline
       AGN & \pkg{skirtor2016} & $i$$^{m}$ & 30\degree, 70\degree & 30\degree, 70\degree \\
       &    & $E(B-V)_{\rm AGN}$$^{n}$ & 0.03 & 0.0, 0.2, 0.4\\
       &     & $\delta_{\rm AGN}$$^{o}$ & $-0.36$ & 0, 1 \\
       &     & $t^{p}$                  & 3, 7, 11 & 7 \\
       &     & lambda\_fracAGN$^{q}$          & 0/0$^r$ & 0.01--100000$^s$ \\
       &     & $f_{\rm AGN}$$^{t}$ & 0.0, 0.3, 0.6, 0.9 & 0.0, 0.1, 0.3, 0.5, 0.7, 0.9 \\
       &     & disk\_type$^{u}$         & 1 & 2 \\
    \hline
       Radio & \pkg{radio} & $q_{\rm IR}$$^{v}$ & \multirow{2}{2pt}{N/A}  & 2.3, 2.5, 2.7 \\
             & & $R_{\rm AGN}$$^{w}$ & & 0.01, 0.1, 1, 10, 100, 1000, 10000 \\
    \hline
    \end{tabular}
    \end{threeparttable}
Note---The SED fitting was performed across a logarithmically spaced redshift grid from $z = 0.01$ to $8$ with 80 discrete values.
\raggedright
  \tablenotemark{\dag}{CIGALE Version 2022.1}
  \tablenotemark{\ddag}{CIGALE Version 2025.0}
  \tablenotemark{a}{$e$-folding time of the main stellar population},
  \tablenotemark{b}{age of the main stellar population}, 
  \tablenotemark{c}{mass fraction of the late burst population}, 
  \tablenotemark{d}{color excess of the nebular lines}, 
  \tablenotemark{e}{factor of $E(B-V)$ for continuum to $E(B-V)$ for lines}, 
  \tablenotemark{f}{amplitude of UV bump},
  \tablenotemark{g}{slope of the power law modifying the attenuation curve}, 
  \tablenotemark{h}{extinction law for the emission lines}, 
  \tablenotemark{i}{power-law slope of $dM/dU \propto U^{-\alpha}$},
  \tablenotemark{j}{minimum radiation field}, 
  \tablenotemark{k}{mass fraction of PAH}, 
  \tablenotemark{l}{fraction illuminated from $U_{\rm min}$ to $U_{\rm max}$},
  \tablenotemark{m}{viewing angle with respect to the AGN axis}, 
  \tablenotemark{n}{$E(B-V)$ for the extinction in the polar direction},
  \tablenotemark{o}{power-law index of modifying the optical slope of the disk for disk type}, 
  \tablenotemark{p}{average edge-on optical depth at 9.7~\micron},
  \tablenotemark{q}{range for computing $f_{\rm AGN}$},
  \tablenotemark{r}{CIGALE default:  dust emission},
  \tablenotemark{s}{wavelength range in \micron},
  \tablenotemark{t}{fraction of luminosity coming from an AGN in the range defined by lambda\_fracAGN}, 
  \tablenotemark{u}{AGN disk spectrum: 1 = \citet{Schartmann2005}, and 2 = \citet{Lopez2024}}, 
  \tablenotemark{v}{IR-to-radio correlation coefficient}, 
  \tablenotemark{w}{radio-loudness parameter}. 
\end{table*}

In most cases, the new photometric redshifts for the Spoke1 sources are close to the ones reported in \Pa. The median difference from \Pa\  \red{$\langle\Delta \zph\rangle=-0.01$}. However, results for seven galaxies out of 59 in common changed enough that $|\Delta\zph|/(1+\zph) > 0.25$, and four of those changes were ${>}2\sigma_{\rm ph}$, where $\sigma_{\rm ph}$ is the uncertainty in the new \zph. Three of the four big changes (for IDs 197/226/363) decreased \zph\ from 2.9--4.2 in \Pa\ to 0.8--1.9 here, but \zph\ for ID~276 increased from 1.1 to 5.5. One possible reason is the addition of the HST data: no HST detection of ID~276 forces the redshift to be higher.  However in most cases, the HST photometry lowers \zph.  The sole Spoke1 source with a catastrophic \zph\ failure in \Pa, ID~363 at $\zsp=1.6019$, now has $\zph=0.84\pm0.56$ instead of \Pa's $\zph=3.01$. (ID~363's \zph\ makes no difference in other parameters because those were based on \zsp\ all along.)
Overall, the new \zph\ estimates give a lower median photometric redshift $\langle\zph\rangle=1.14$ compared to $\langle\zph\rangle=1.37$ in \Pa. (This differs from the overall $\langle z\rangle=1.33$ in \Pa\ because the overall $\langle z\rangle$ includes \zsp.)
Figure~\ref{f:zz} shows that the new photometric redshifts are in good agreement with \zsp, now with many more sources (Table~\ref{t:z}) than in \Pa. 
The only ${>}2.5\sigma_{\rm ph}$ failures \red{are ID~76,} with small $\sigma_{\rm ph}=0.015$, and two sources with $Q=2$ NIRISS \zsp.
The good agreement may be a bit too optimistic because most sources without \zsp\  are about a magnitude fainter at 4.44~\micron\ (Figure~\ref{f:fz}) than sources with \zsp, and the typical \zph\ accuracy is likely worse than Figure~\ref{f:zz} indicates. Nevertheless, nearly all the radio hosts have $m(\rm F444W)<23$~mag, a factor of 100 above the JWST survey limit, and therefore the S/N should be high enough to give good photometric redshifts for most of the galaxies.

In comparing other parameters, the change from Salpeter to Chabrier IMF scales all the new $M_\star$ values lower than \Pa's by a factor of 0.61 \citep{Madau2014}. After taking that into account, and ignoring the six sources
for which large changes in \zph\ produced large changes in $M_\star$, only two sources (ID~268/315) changed $M_\star$ by more than 0.5~dex ($\Delta M_\star=-1.22,-0.51$, respectively). The most likely reason for the ID~268 change is that the stronger SExtractor deblending, in separating the two overlapping galaxies (Figure~\ref{f:tricky}), made the photometry aperture too small. 
For ID~315, \Pa's ``best'' $M_\star$ was above the 84th percentile of the probability density function, and  $\Delta M_\star$ is within the statistical uncertainties.
Overall, however, stellar masses are well determined provided the redshift and absolute photometry are accurate.
  
In contrast to $M_\star$, galaxy parameters SFR, $A_V$, and $f_{\rm AGN}$ are sensitive to the exact SED as well as the derived redshift.  The inclusion of radio data may also have influenced the fits, and $f_{\rm AGN}$, and $A_V$ were limited to discrete values in \Pa\ but allowed to take a continuous range here.
The definition of $f_{\rm AGN}$ also changed.  In this paper, it is the fraction of the bolometric luminosity coming from an AGN, whereas in \Pa\ it was the fraction of the dust-reprocessed luminosity.\footnote{Based on preliminary fitting before changing the $f_{\rm AGN}$ definition, the definition change itself had little effect. This may have been because CIGALE had to deduce dust emission from its energy-balance requirement, not from nonexistent observations of the dust emission.} SFRs for 17 galaxies (again excluding those with large $\Delta z$ and scaling by a factor of 0.63 for Salpeter to Chabrier IMF---\citealt{Madau2014}) changed by more than 0.5~dex, $A_V$ for six galaxies changed by more than 0.7~mag, and $f_{\rm AGN}$ for 23 galaxies changed by more than 0.3. SFR changes were 11 up, 6 down, median $\langle\Delta \log(\rm SFR)\rangle=+0.04$. Changes in $A_V$ were 4 down, 2 up, $\langle\Delta A_V\rangle=-0.15$, and in $f_{\rm AGN}$ were 18 down, 5 up, $\langle \Delta f_{\rm AGN}\rangle=-0.06$. Some of these changes reflect differences in the fitting method, but some reflect real uncertainties in the photometry and limitations in the existing data.  These changes are not trivial, but they do not affect a majority of the galaxies.

\subsection{Source properties}
\label{s:properties}

\begin{figure}
\includegraphics[width=\linewidth]{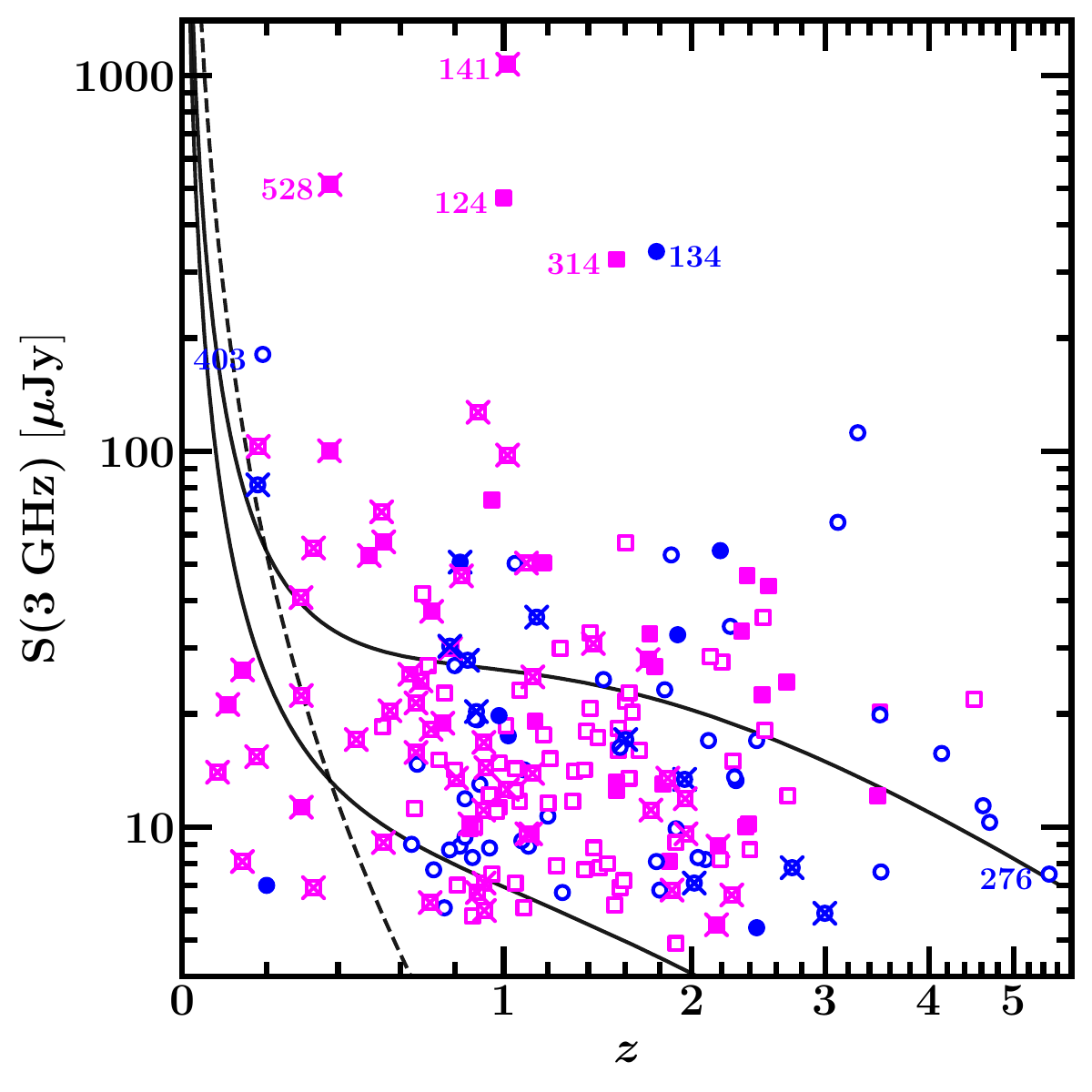}
\vspace{-5ex}
\caption{Observed 3~GHz flux density versus redshift scaled as $\log(1+z)$. Point shapes and colors indicate SED classes defined in Section~\ref{s:nature}: blue circles represent type AGN, and magenta squares represent type Gal.  Filled symbols indicate galaxies with excess radio emission as defined in Section~\ref{s:properties}, and $\times$ symbols mark galaxies with spectroscopic redshifts. 
The six points with highest flux density are labeled.  The solid lines show the flux density from star formation (for a \citealt{Chabrier2001} IMF and including both thermal and non-thermal radio emission per Equation~\ref{e:L3}) for a galaxy on the SFMS \citep{Popesso2023} with stellar masses $10^{11}$\,\Msol\ (upper curve) and $10^{10}$\,\Msol\ (lower curve).  The dashed line shows the flux density for a fixed 1.4~GHz
luminosity density (all non-thermal with spectral index $-0.7$) of $10^{22}$~W~Hz$^{-1}$, which corresponds to a star-formation rate of about 3.3~\Msol~yr$^{-1}$ \citep[][their Eq.~3 converted to a Chabrier IMF]{Mahajan2019}. (Note: the solid lines in Paper~1's Fig.~8 were mis-plotted and should be ignored.)}
\label{f:sz}
\end{figure}

The TDF radio sample spans a wide redshift range, $z\simeq0.08$ to 5.5 with little dependence of 3~GHz flux density on redshift (Figure~\ref{f:sz}). The median redshift \red{$\langle z\rangle=1.10$} is lower than the $\langle z\rangle=1.33$ reported in \Pa\ mainly because of the new CIGALE calculations reported in Section~\ref{s:fits}. An additional factor is that Spoke1 is the closest to the center of the primary beam, and the resulting higher primary-beam response allows on-average fainter and more distant objects to enter the radio sample in this spoke. Another possibility is that the primary beam was centered on a $z=1.4429$ flat-spectrum radio quasar \citep{Hyun2023, Willner2023}, and it is possible that the quasar is part of a cosmic overdensity. If so, Spoke1 would have an excess of objects at $z\simeq1.4$ and therefore higher $\langle z\rangle$,
but no such excess is apparent. The SED sample has 11 radio hosts in the range $1.333<z<1.533$, and three of them are in Spoke1.  This doesn't rule out an overdensity, but significant contribution to the smaller median redshift seems unlikely.

\begin{figure}
\includegraphics[width=\linewidth]{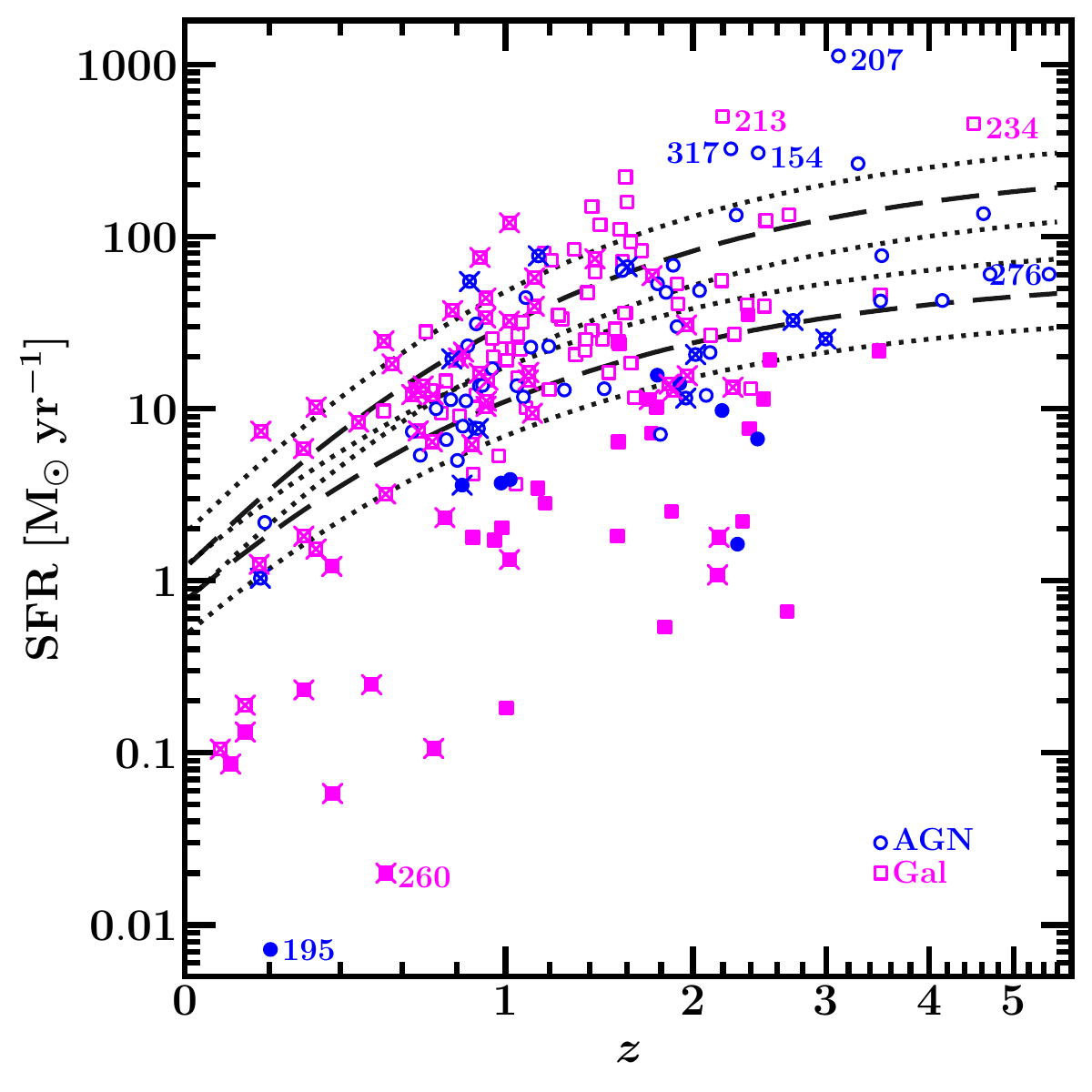}
\vspace{-5ex}
\caption{Photometric star-formation rate versus redshift  scaled as $\log(1+z)$. Point shapes and colors indicate SED classes defined in Section~\ref{s:nature}: blue circles represent type AGN, and magenta squares represent type Gal.  Filled symbols indicate galaxies with excess radio emission as defined in Section~\ref{s:properties}, and $\times$ symbols mark galaxies with spectroscopic redshifts. Long-dashed lines show the galaxy SFMS \citep{Popesso2023} for stellar masses  $10^{11}$\,\Msol\ (upper curve) and $10^{10}$\,\Msol\ (lower curve), and the dotted lines show intervals of 0.2~dex on either side. Seven points with high or low SFR and three points with $z>5$ are labeled.}
\label{f:z_sfr}
\end{figure}

\begin{figure*}
\includegraphics[width=\linewidth]{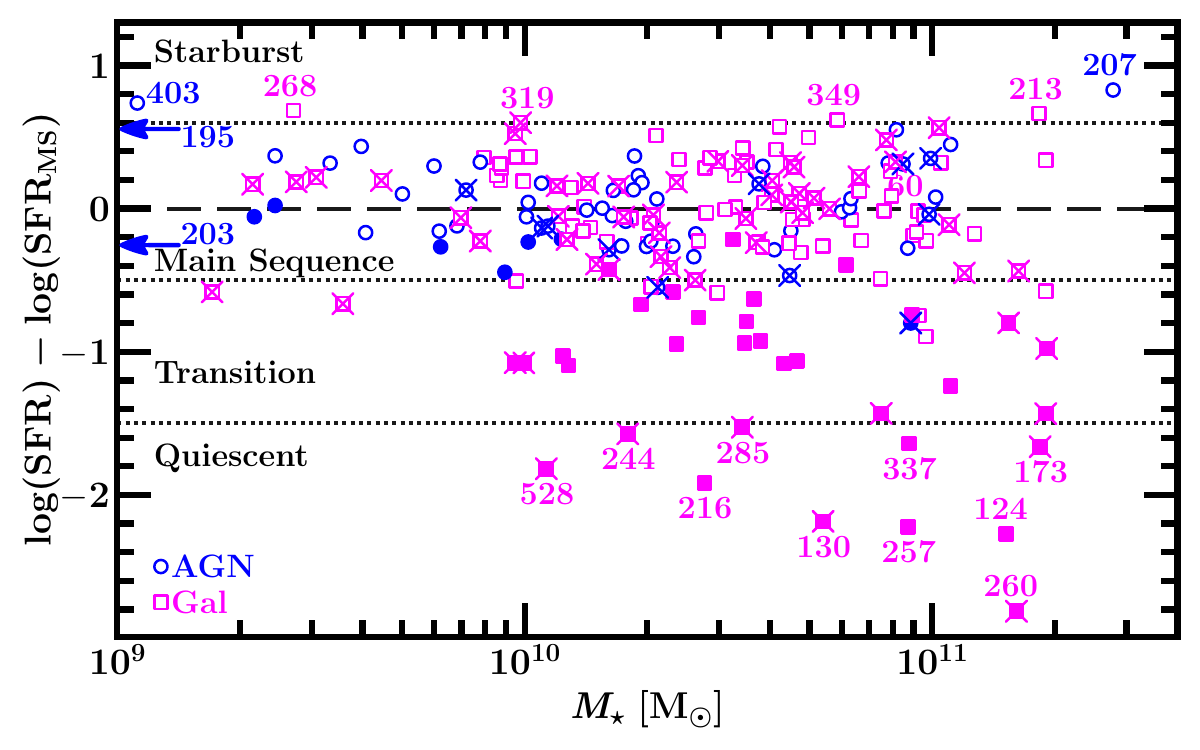}
\vspace{-5ex}
\caption{Star-formation rates from the CIGALE fits relative to the star-formation main sequence. 
The horizontal dashed line marks the SFMS.  The upper dotted line marks 0.6\,dex above the SFMS, commonly taken as the boundary for a starburst \citep[e.g.,][]{Rodighiero2011}, and the other two dotted lines mark the lower boundary of the SFMS and the upper boundary of the quiescent region \citep[$\Delta_{\rm SFR}=-0.5$ and $-1.5$, respectively:][]{Renzini2015}.
Point shapes and colors indicate SED classes defined in Section~\ref{s:nature}: blue circles represent type QSO, and magenta squares represent type Gal.  Filled symbols indicate galaxies with excess radio emission as defined in Section~\ref{s:properties}, and $\times$ symbols indicate galaxies with spectroscopic redshifts. Arrows mark three galaxies that have masses smaller the plot limit. ID numbers are indicated for those three low-mass galaxies and for starburst and quiescent galaxies. }
\label{f:delsfr}
\end{figure*}

The star formation rates (SFRs) from the SED fitting (Section~\ref{s:fits}) range from $\sim$0.007 to $\sim$1100\,\Msol\,yr$^{-1}$ as shown in Figure~\ref{f:z_sfr}, although the highest SFRs could be much lower and still fit the observed photometry. 
(Relative uncertainties on four of the five SFRs that are $>$300\,\Msol\,yr$^{-1}$ are $>$100\%.)
Figure~\ref{f:delsfr} compares the SFRs to the star-formation main sequence (SFMS) of \citet[][]{Popesso2023}. Six host galaxies (3\% of the sample) are starbursts, 154 (77\%) are normal star-forming galaxies with $\rm -0.5 \le \Delta_{SFR} < 0.6 $,  24 (12\%) are galaxies in transition with $\rm -1.5 \le \Delta_{SFR} < -0.5$, and 10 (5\%) are quiescent with $\rm \Delta_{SFR} < -1.5$. (The range definitions  were taken from \citealt{Rodighiero2011} and \citealt{Renzini2015}.) These compare with fractions 5\%, 58\%, 29\%, and 8\% found for the 58 galaxies available in \Pa.  The major reason for the change was the revised CIGALE calculations discussed in Section~\ref{s:fits}, which overall moved some galaxies from the Transition to the MS category.  The main effect of the change from the \citet{Speagle2014} SFMS used in \Pa\ to the  \citet{Popesso2023} SFMS is to lower the SFMS SFR for $M_\star \ga5\times10^{10}$~\Msol. Using the \citeauthor{Speagle2014}\ SFMS would move 
the three starbursts in this mass range into the MS category and seven other galaxies into lower categories but would not have a drastic effect overall.

SFRs can also be derived directly from radio flux densities \citep[e.g.,][]{Condon1992,Murphy2011,Mahajan2019,Delvecchio2021} provided no AGN contributes to the radio flux.  Because the radio luminosity comes from both thermal ($L_T$) and non-thermal ($L_{\rm NT}$) emission, and different calibrations exist, the conversion from observed 3~GHz luminosities to SFRs is complex and uncertain.  
In radio surveys, the redshift distribution of sources and the fractions that are star-forming versus AGN depend on the survey's flux-density limit. One consequence of that is that the spectral index $\alpha$ of the radio-source sample also depends on flux density.  Examples include Fig.~1a of \citet{Tompkins2023} or Figs.~4--5 of \citet{Windhorst1990}. Those authors' samples had no uniform SED or redshift information.  Rather than using  empirical relations between $\alpha$ and flux density, this paper uses the redshift and SFR measurements for each source to derive the physically motivated rest-frame $L(\rm1.4~GHz)$, treating the thermal and non-thermal components separately. The differing redshift and radio luminosity distributions of the star-forming and AGN populations then account for the median spectral index changing with flux density \citep[][]{Tompkins2023}.

The thermal luminosity $L_T{\rm(1.4~GHz)}$ is a straightforward count of ultraviolet photons \citep{Rubin1968} and ought to be fairly independent of the exact physical conditions, whereas $L_{\rm NT}$ depends on the complex interplay of supernovae and their remnants \citep{Condon1992}.
Nevertheless, the sum of these two components correlates well with other SFR tracers for a wide variety of local galaxies \citep{Mahajan2019}. Those authors' most-favored relation, converted to Chabrier IMF, is:
\begin{multline}
L{\rm(1.4~GHz)/(10^{22}~W~Hz^{-1})} \equiv L_{\rm NT} + L_T\\ = \rm (SFR/3.31~\Msol~yr^{-1})^{(1/0.75)}\quad.
\label{e:L}
\end{multline}
Equation~\ref{e:L} is for $L\rm(1.4~GHz)$ in the rest frame. To apply it at nonzero redshifts and observing frequencies other than 1.4~GHz, K corrections are required, and $L_{\rm NT}$ and $L_T$ have different ones.
For the thermal component, \citet[][their Eq.~11]{Murphy2011} give, after converting from a Kroupa to a Chabrier IMF \citep{Madau2014}:
\begin{equation}
L_T(\nu/{\rm GHz})= \rm (SFR/44.73) (\nu/{\rm 1.4~GHz})^{-0.1}\quad.
\label{e:Lt}
\end{equation}
From the observed 3~GHz flux density, the K correction for $L_T$ is $[(1+z)*(3/1.4)]^{-0.1}$, and the correction for $L_{\rm NT}$ has the same form except with an exponent of $-0.7$ (assuming that spectral index for non-thermal emission). The total luminosity emitted at any frequency is $L_T+L_{\rm NT}$, and for the frequency $\nu$ at which the observed 3~GHz flux was emitted, 
\begin{multline}
L(3[1+z]~{\rm GHz})/{(10^{22}~{\rm W~Hz^{-1}})} = \\ {\rm(SFR/44.73)}*[(3/1.4)*(1+z)]^{-0.1}  \\ +[({\rm SFR/3.31)^{1/0.75}-(SFR/44.73)}]\\ *[(3/1.4)*(1+z)]^{-0.7}\quad.
\label{e:L3}
\end{multline}
For a $z=0$ galaxy with $M_*=10^{11}$~\Msol, the SFMS $\rm SFR = 1.2$~\Msol~yr$^{-1}$, and 16\% of  the observed 3~GHz flux is thermal.  At higher redshifts, the SFMS SFR is larger, and the non-thermal radio emission rises faster than linearly with SFR.  This outweighs the steep spectral index ($-0.7$ assumed) of the non-thermal emission, and at all $z\ga1$, only $\sim$8\% of the observed 3~GHz emission is thermal. This differs from the conclusion of \citet[][their Fig.~20 and related discussion]{Delhaize2017} because those authors didn't consider a non-linear dependence of $L_{\rm NT}$ on SFR.  At the relevant luminosities, ignoring the thermal emission and simply using Equation~\ref{e:L} with a spectral index of $-0.7$  would have made $\la$10\% difference.  

The whole approach above assumes that high-redshift galaxies have the same ratio of radio emission to SFR as local galaxies. That seems to be true. \citet{Delhaize2017} measured SFRs in the COSMOS field from total infrared luminosity $L_{\rm TIR}$ derived from Herschel data. From $z=0$ to $z=3$, the ratio of far-IR to radio luminosity, expressed as $q_{\rm TIR}\equiv \log(L_{\rm TIR}/3.75\times10^{12}\rm~Hz) - \log(L(1.4~\rm GHz)/W~Hz^{-1})$, changed by about ${-}0.7$~dex. While Equation~\ref{e:L} has no explicit redshift dependence, for a fixed galaxy mass, the SFMS gives higher SFR at higher redshift.  For a galaxy with the median mass ($\langle M_\star\rangle=2.7\times10^{10}$~\Msol) of the SED sample, the SFR and therefore $L_{\rm TIR}$ are $\sim$70 times higher at $z=3$ than at $z=0$. However, the non-linearity in Equation~\ref{e:L} means that $L(1.4~\rm GHz)$ increases by about 300 times for a net $\Delta q_{\rm TIR}=-0.61$, about the same as the \citet{Delhaize2017} empirical result.  The \citet{Delvecchio2021} COSMOS results are not directly comparable because they attributed the observed $z$-dependence of $q_{\rm TIR}$ to  higher-redshift samples having generally higher masses.   Both COSMOS studies give absolute values of $q_{\rm TIR}$ close to the value used here. For a galaxy with the median $M_\star$ and redshift ($\langle z\rangle=1.08$) of the SED sample and on the SFMS, $q_{\rm TIR}=2.55$ from the relation used here (based on Eq.~6 of \citealt{Mahajan2019} after multiplying their $L_{\rm FIR}$ by a factor of 2.26---the median $L_{\rm TIR}/L_{\rm FIR}$ for their SFRS sample). The COSMOS studies give  $q_{\rm TIR}=2.51$ \citep{Delhaize2017} and $q_{\rm TIR}=2.58$ \citep{Delvecchio2021}. Neither COSMOS study took account of the non-linear relation between SFR and $L(1.4\rm~GHz)$ despite its being well established in local galaxies \citep[][and references therein]{Mahajan2019}.  This work uses Equation~\ref{e:L3} because it is wholly empirical, doesn't depend on fitting for $M_\star$, and is consistent with local galaxies of a wide variety of types.

\begin{figure*}
\includegraphics[width=\linewidth]{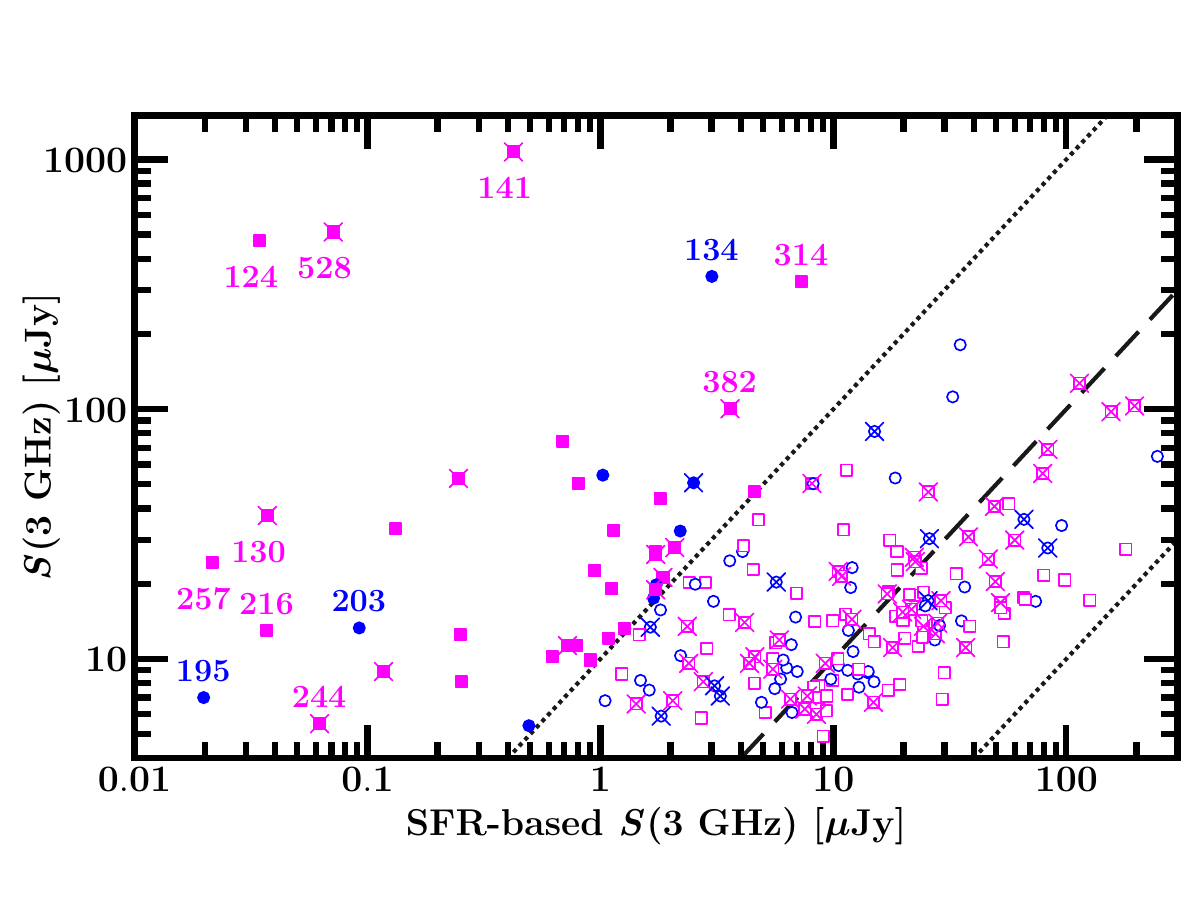}
\vspace{-5ex}
\caption{Comparison of observed radio flux density to that expected from star formation.  The ordinate shows the observed 3~GHz flux density, and the abscissa shows the flux density expected from the SFR derived from the CIGALE SED fitting.  The predicted radio emission for a given SFR was taken from Equation~\ref{e:L3}.  Dashed and dotted lines show  equality and $\pm$1\,dex scatter. Point shapes and colors indicate SED classes defined in Section~\ref{s:nature}: blue circles represent type AGN, and magenta squares represent type Gal.  Filled symbols indicate galaxies with excess radio emission as defined in Section~\ref{s:properties}, and $\times$ symbols mark galaxies with spectroscopic redshifts. Some of the more extreme radio-loud points are labeled with their source IDs.}
\label{f:S3_sfr}
\end{figure*}

For each source analyzed by CIGALE, inserting the derived SFR and redshift into Equation~\ref{e:L3} predicts the radio luminosity at the rest frequency corresponding to observed 3~GHz. Figure~\ref{f:S3_sfr} compares these predictions to the observed flux densities.  Out of the 200 radio sources in the SED sample, 42 (21\%) exhibit radio emission $>$10$\times$ the value predicted from the SFR. These sources are designated radio-loud (RL). The typical factor-of-two uncertainties in the CIGALE-derived SFRs mean that the factor-of-10 threshold corresponds to $\sim$3.8$\sigma$, ensuring that RL sources are unlikely to be explained by uncertainties in the derived SFR. Their excess emission could arise from an AGN---AGN are included in the 42 RL sources---or from star formation hidden behind dust and therefore not contributing to the observed visible--NIR SED. The remaining 158 (79\%) sources, which have radio emission within a factor of 10 of the prediction from their SFR, are designated radio-quiet (RQ), and there is no strong need to invoke any process other than star formation to produce their radio emission.

The 21\% RL fraction found here is more than double the 8.8\% found by \citet{Algera2020}.
That is despite those authors having used a seemingly less strict criterion for radio loudness ($S(3~{\rm GHz})>6\times$ the value implied by SFR) than used here ($>$10$\times$). 
Part of the difference may be the slightly greater depth of the \citeauthor{Algera2020}\ radio data (0.5~\mmJy\ RMS at beam center, 2.5~\mmJy\ at beam edge versus 1.0--2.9~\mmJy\ here). A larger effect probably lies in different procedures for deriving SFR. \citeauthor{Algera2020}\ based their SFRs on Herschel far-IR data, which can find star formation hidden by dust. For galaxies with the stellar masses of the radio hosts, obscured star formation is far larger than unobscured star formation \citep{Whitaker2017}. The obscuring dust  hides much of the UV emission from young stars, and the TDF's lack of Herschel data gives CIGALE no direct way to include that emission. In such circumstances, the RL fraction based on CIGALE fits would be an overestimate, and the radio flux would be a better SFR indicator than the CIGALE fits. 
Another study \citep{Delhaize2017} is not directly comparable to ours or that of \citeauthor{Algera2020}\ because \citeauthor{Delhaize2017}\ used a radio survey $\sim$5$\times$ shallower than ours, and the RL fraction is higher at shallower limits \citep{Algera2020}.
\citeauthor{Delhaize2017}\ didn't give a specific RL fraction, but they found 510 RL galaxies in their star-forming sample. In addition, they found 791 galaxies that were radio-detected but show no evidence of star formation, making them RL. Those two sets are already 15.9\% of their 8184 radio-detected galaxies, and some fraction of the 1636 radio-detected HLAGN they found have to be added.  In that same radio survey, \citet[][their Fig.~10]{Smolcic2017id} found a 24\% radio-excess fraction though without the benefit of far-IR observations. Despite disagreement on the exact RL fractions, studies agree that some sources are RL because of AGN radio emission.

\subsection{Nature of the sources}
\label{s:nature}

\setcounter{table}{4}
\begin{table}
\caption{Source SED classifications\label{t:class}}
\begin{tabular}{l r@{\quad\quad} r r}
\hline
Class   & $N$    &  \multicolumn{2}{c}{Nature of} \\
&&\multicolumn{2}{c}{radio emission}\\
        &        &    RL      &    RQ    \\
\hline
AGN     & 61  &  9      &    52     \\
Gal     & 139 &  33      &    106     \\
\hline
\end{tabular}
\end{table}

\begin{figure*}
\centering
\includegraphics[width=.72\linewidth]{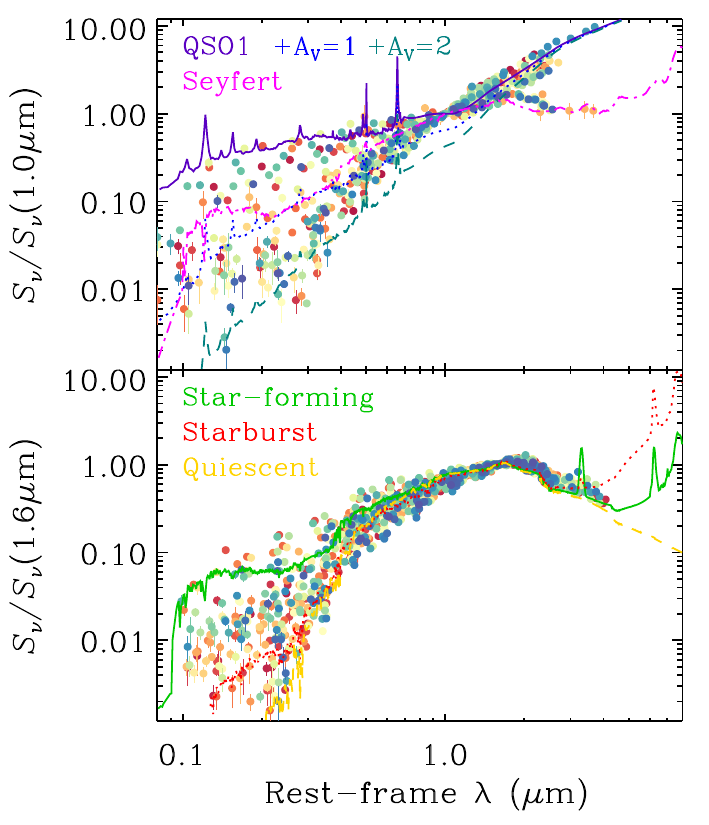}
\caption{Rest-frame SEDs of the radio-source host galaxies. Filled-circle colors identify individual host galaxies.  (The color choices are arbitrary.)  Panels show galaxies in each SED class: QSO top, Gal bottom. Flux densities are normalized at rest-frame 1.6\,$\mu$m.  Lines in each panel show templates from the SWIRE library \citep{Polletta2007}. The top panel has a type~1 QSO template,  TQSO1, in purple (solid line). The same template reddened with $A_{V}=1$ is in blue (dotted line) and with $A_{V}=2$ in teal (dashed line). The bottom panel shows templates for a star-forming galaxy in green (solid line, top in UV, middle in IR), a starburst galaxy in red (dotted line, middle in UV, top in IR), and an old stellar population in yellow (dashed line, bottom in both UV and IR).}
\label{f:rfseds}
\end{figure*}

The source SED shapes fall into two broad categories:  SEDs that rise beyond rest-frame 1.6\,\micron, consistent with being AGN-dominated (labeled ``AGN''), 
and  stellar-dominated SEDs (labeled ``Gal'') with falling SEDs beyond rest 1.6\,\micron. 
This classification is based on the rest visible--NIR SED, where AGN-heated dust emission produces an SED  redder than that from a stellar population \citep{Sanders1989} and a dip at $\sim$1\,\micron. Such hot dust ($T_d\ga1000$\,K) is usually associated with an AGN. For power-law SEDs, we assumed that the AGN dominates also at rest-frame visible wavelengths and classified those objects as AGN. In some cases, the SED steepens at shorter wavelengths, consistent with a reddened AGN \citep{Gregg2002}. 
Figure~\ref{f:rfseds} shows the SEDs of  sources in the two categories, and
Table~\ref{t:class} shows the number of sources in each SED category.
Overall, 31\% of radio-host galaxies have AGN-type SEDs.

\begin{figure}
\includegraphics[width=\linewidth]{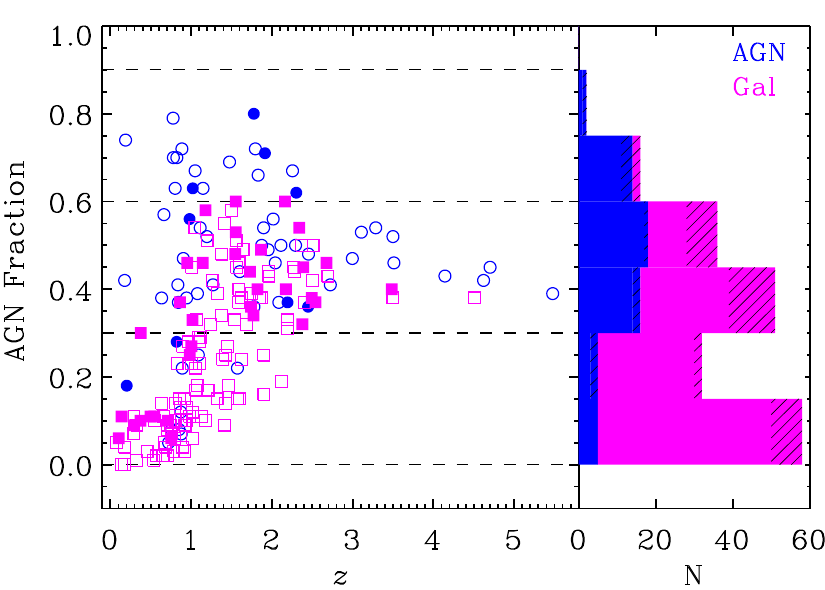}
\caption{CIGALE AGN fraction versus redshift. Sources classified AGN are shown as blue circles and Gal as magenta squares. Filled symbols represent sources with a radio excess. The right panel shows the number of sources of each SED class, coded with the preceding colors, for each level of AGN fraction. Hatched regions indicate RL sources. }
\label{f:agn_fraction}
\end{figure}

The SED-shape classification agrees with CIGALE's $f_{\rm AGN}$ for \red{76}\% of the sample, better than the 64\% found in \Pa.  This is using a more stringent definition of AGN according to CIGALE, $f_{\rm AGN}\ge0.45$, than \Pa's $f_{\rm AGN}\ge0.3$. (\Pa\ used discrete values of $f_{\rm AGN}$ while this work uses a continuous range.  The limit 0.45 is halfway between \Pa's discrete values of 0.3 and 0.6.) Figure~\ref{f:agn_fraction} compares the two classifications. Of the \red{61} galaxies classified AGN by SED templates, \red{84}\% have $f_{\rm AGN}\ge0.3$, but only \red{57}\% have $f_{\rm AGN}\ge0.45$. (For the 138 galaxies classified ``Gal,'' the same fractions are 42\% and 17\%.)  Of the 58 galaxies with $f_{\rm AGN}\ge0.45$, 60\% fit AGN templates.  While the overall agreement is fair, the two classification methods are looking at different characteristics of the SED, and Section~\ref{s:agn} considers both methods.

Absent mid-IR data (or low redshift), no SED classification will identify an AGN whose emission does not contribute to visible--NIR wavelengths. This might be the case for massive host galaxies dominated by stellar emission (\eg, some powerful radio galaxies---\citealt{Haas2008}) or when dust obscures the AGN at rest-visible wavelengths, \ie, at $z\gtrsim2$, the dust would emit at $\lambda>5$\,\micron, where there are no available data for this field.

\subsection{AGN Identification}
\label{s:agn}

\begin{deluxetable*}{lrcccccc}
\tablecolumns{8}
\tablewidth{\textwidth}
\setlength{\tabcolsep}{2em}
\tablecaption{AGN detections}
\label{t:agn}
\tablehead{\colhead{}&
\colhead{ SED }&\colhead{ $f_{\rm AGN}$}&\colhead{X-rays}&\colhead{Pt. Src.}&\colhead{ RL }&\colhead{ VLBA}&\colhead{ Spec.}
}
\startdata
SED &\bf 61& 35& 22& 8& 9& 1& 3\\
$f_{\rm AGN}$ & 84&\bf 58& 15& 8& 16& 3& 2\\
X-ray & 98& 102&\bf 59& 12& 13& 1& 6\\
Pt.\ src.\ & 71& 68& 65&\bf 18& 4& 1& 2\\
RL & 94& 84& 88& 56&\bf 42& 7& 0\\
VLBA & 68& 63& 66& 25& 43&\bf 8& 0\\
Spec.\ & 64& 62& 59& 22& 48& 14&\bf 6\\
\enddata
\tablecomments{All numbers are for the SED sample, i.e.,  the 200 galaxies in Table~\ref{t:SEDfits}.  Numbers along the table diagonal (in bold) show the number of radio hosts identified as AGN by the method in the corresponding row and column labels.  Numbers above or right of the diagonal show the number identified by both of the methods that intersect at that number.  Numbers below or left of the diagonal show the number identified by either of the intersecting methods.  Among the six radio hosts in the NIRCam sample but not the SED sample, two (IDs~295/364) are Chandra sources, two (IDs~447/500) show pointlike nuclei confirmed by \galfit, and ID~500 also shows an AGN spectrum.}
\end{deluxetable*}

\begin{deluxetable}{lhccr}
\tablecolumns{5}
\tablewidth{\linewidth}
\setlength{\tabcolsep}{1.7em}
\tablecaption{AGN detection probabilities}
\label{t:prob}
\tablehead{&&\colhead{$p_1$}&\colhead{$p_2$}&\colhead{$N$}}
\startdata
SED--$f_{\rm AGN}$	&	101	&	0.60	&	0.57	&	101	\\
SED--X-ray	&	164	&	0.37	&	0.36	&	164	\\
SED--Pt. Src.	&	137	&	0.44	&	0.13	&	137	\\
SED--RL	&	285	&	0.21	&	0.15	&	285	\\
$f_{\rm AGN}$--X-ray	&	228	&	0.25	&	0.26	&	228	\\
$f_{\rm AGN}$--Pt. Src	&	131	&	0.44	&	0.14	&	131	\\
$f_{\rm AGN}$--RL	&	152	&	0.38	&	0.28	&	152	\\
X-ray--Pt.\ Src.	&	89	&	0.67	&	0.20	&	89	\\
X-ray--RL	&	191	&	0.31	&	0.22	&	191	\\
Pt. Src.--RL	&	189	&	0.10	&	0.22	&	189	\\        
\enddata
\tablecomments{$p_1$ is the probability of detecting an AGN by the first method in each row given that it was detected by the second method, and $p_2$ is the converse.  For example, an $f_{\rm AGN}$ detection implies a 60\% probability of an SED detection, and an SED detection implies a 56\% chance of an $f_{\rm AGN}$ detection. The number $N$ is the number of AGN in a notional comprehensive AGN population if the two detection probabilities are independent.}
\end{deluxetable}

An AGN can make its presence known in multiple ways other than the SED. One of the most reliable AGN identification methods is X-ray
detection. The TDF has been observed by XMM-Newton and NuSTAR
\citep{Zhao2024,Silver2025} and by Chandra (W.~P. Maksym \etal, in prep.). Matching the three catalogs to the entire \citet{Hyun2023} 3~GHz
catalog gave 42 matches within 2\arcsec\ for XMM, 5 within 2\arcsec\
for NuSTAR, and 128 within 1\arcsec\ for Chandra.  (The matching radii
were chosen based on positional accuracies of each X-ray catalog.) Within the SED sample, there are 58 Chandra matches to radio-host galaxies, of which 10 are also XMM sources, and XMM detected one additional source (ID~201) that was not detected by Chandra.  Chandra detected two additional sources (IDs~295/364) within the NIRCam area but not in the SED sample because of missing LW photometry. Of the 61 sources detected by Chandra or XMM, which observe relatively soft X-rays, one (ID~194) is also a NuSTAR source.  A few radio hosts have 
Chandra sources just outside the 1\arcsec\ radius, but all of these X-ray sources have obvious counterparts other than the radio host and are not plausible matches
to the radio host.

Some X-rays could come from X-ray binaries (XRBs, i.e., related to star formation) rather than an AGN, but this is not significant for any of the radio hosts.  X-ray luminosities  $L_X(\hbox{0.5--7~keV}) > 10^{42}$~ergs~s$^{-1}$ require an AGN \citep[e.g.,][]{Persic2004,Colbert2004},  and all but \red{10} of the X-ray-detected radio hosts are above that threshold. For all \red{10} hosts below that threshold, the observed $L_X>L_X(\rm XRB)$ where $L_X(\rm XRB)$ is the X-ray luminosity expected \citep{Mineo2014} based on the galaxy's SFR (Table~\ref{t:SEDfits}).

Another AGN detection method is looking for a pointlike nuclear source in the NIRCam images \citep{Ortiz2024}.  The \citeauthor{Ortiz2024}\  list includes 24 objects in the SED sample. An independent
examination of the radio-host galaxies (by a different examiner using the eight individual NIRCam images instead of a color composite) found 47 more (plus two not in the SED sample) as well as 17 of the
24 that \citeauthor{Ortiz2024}\  had already found.  These searches were subjective and therefore uncertain---as indicated by the two searches finding different lists.  Nevertheless, finding a pointlike IR source is a strong indication of an AGN, and there are 71 such sources in the SED sample.
\galfit\ decomposition of these 71 hosts by the  method of \citet{Ortiz2024}\footnote{The bright Seyfert galaxy ID~194   (which is Source~1 in the \citeauthor{Ortiz2024}\ list) cannot be decomposed in F356W, F410M, or F444W because the images are saturated near the galaxy's center.  F277W is worse than F444W for detecting an AGN signature because AGN are redder than starlight, but even so, \galfit\ shows an unresolved nucleus in this galaxy.} found 18 hosts best fit by an unresolved point source, while the remaining 53 are best fit by two S\'ersic profiles or are indeterminate. While many of these 53 galaxies may have an AGN contribution to the central light emission, only the 18 sources unresolved by \galfit\ are counted as AGN here.

Pointlike nuclear sources can also be found by very long baseline radio interferometry (VLBI).  \citet{Saikia2025} reported 4.8~GHz Very Long Baseline Array (VLBA) observations of 106 radio sources from the \citet{Hyun2023} list.  Of those, 61 are in the NIRCam sample, and eight (all in the SED sample) were detected.  The non-detections may simply reflect lack of 4.8 GHz sensitivity---23 of the VLBA non-detections have likely XMM, NuSTAR, or Chandra counterparts---but the high brightness temperatures of the detections require a radio AGN \citep{Saikia2025}.

Spectra can also indicate an AGN via high-excitation or broad emission lines or by  line ratios \citep{Baldwin1981}.  Among the 54 objects having $Q\ge3$ ground-based spectra, seven (flag `z' in Table~\ref{t:ID}) have line ratios indicative of ``Seyfert'' classification according to the \citet[][their Figs.~2 and~4]{Lamareille2010} criteria or are broad-line QSOs (the latter being IDs~142/363).  Six are in the SED sample, and all of those are X-ray sources. 

Radio loudness (Figure~\ref{f:S3_sfr}) categorizes 42 hosts (21\%, identified in Table~\ref{t:SEDfits}) in the SED sample as AGN.  This includes eight not identified by other methods and raises the fraction with an AGN signature from 62\% to 66\%. The 21\% RL fraction is close to the 24\% radio-excess fraction in the VLA-COSMOS survey \citep[][their Fig.~10]{Smolcic2017id}. The slightly lower RL fraction is probably a combination of our stricter criterion for RL and the higher flux-density limit of VLA-COSMOS. Our overall $\sim$66\% fraction with AGN signatures is higher than the 42\% found by \citet{Smolcic2017id}. Much of the difference is in the X-ray detections, 30\% in this sample compared to 11\% by \citeauthor{Smolcic2017id}, because the TDF Chandra observations are deeper than the COSMOS ones. Another difference is that \citeauthor{Smolcic2017id}\ found AGN indications in the SED, including mid-IR, for only 21\% of their sample compared to $\sim$42\% here. This probably reflects the better sensitivity and wavelength coverage of the NIRCam observations compared to the IRAC observations available to \citeauthor{Smolcic2017id}. Entirely omitting our $f_{\rm AGN}$ and SED criteria would still find at least one AGN signature in 46\% of our sample, and that has to be too low. Regardless of AGN evidence, radio emission is consistent with star formation in 79\% of the SED sample galaxies (Figure~\ref{f:S3_sfr}, Section~\ref{s:properties}).

The AGN-detection methods used here have differing reliabilities.  X-ray detection should be reliable with at most one or two non-AGN included.  The weakest X-ray AGN case is ID~283 with $\log L_X=41.12$ and $\log L_X(\rm SFR)=40.87$. VLBA detection is also reliable \citep{Saikia2025}. Central point-source detection is relatively new and hard to assess, but the presence of X-ray or VLBA sources in 13 of the 18 pointlike sources suggests the reliability is high.  More likely the number of AGN visible as pointlike sources is underestimated because 26 of the 53 sources with a pointlike appearance but resolved by \galfit\ have X-ray (24) or VLBA (3) detections. (ID~141 has both.)  Perhaps these radio hosts contain an AGN embedded in a compact stellar bulge.  The SED-template and CIGALE-fit reliability is also hard to assess.  The templates show overlap between the respective SEDs, probably because both reddening and an AGN can produce red colors.  CIGALE uses a different process but is based on the same data, which can be equally ambiguous. The reliability of both methods probably drops at $z\ga1.75$ because the F444W band samples the stellar peak at 1.6~\micron\ rather than longer rest wavelengths where AGN dust emission is more likely to dominate. Of the \red{84} hosts with SED or $f_{\rm AGN}$ signatures, \red{29} have confirmation of an AGN from X-rays (\red{26}) or VLBI (3). This is a lower limit on reliability because the SED selections are best at finding highly obscured AGN, but the obscuration makes X-ray detection more difficult \citep[e.g.,][]{Hickox2009}.
A genuine radio excess should be a reliable AGN indicator, but the expected radio flux density depends on the SFR  from the CIGALE fit.  Requiring the excess to be a factor of 10 should mitigate that uncertainty.  AGN spectral features should be reliable but exist for only a few galaxies, all of which are also X-ray sources.

Table~\ref{t:agn} summarizes the AGN detections including how many counterparts are detected by each pair of methods.  The ``flags'' column in Table~\ref{t:ID} marks most of the AGN detections, and Table~\ref{t:SEDfits} shows the $f_{\rm AGN}$, RL, and SED-template detections.  The most productive single method is using the SED templates, but that is only $\sim$\red{46}\% complete and not entirely reliable. X-ray detection is reliable and almost as productive.

The combined detection methods can be used to estimate the AGN population and the detection completeness of each method. If one imagines a population of $N$ AGN in the NIRCam area,\footnote{The imagined population size $N$ can be larger than the sample size if one or both of the detection methods are incomplete because an incomplete detection method $i$ requires large $N$ to produce the observed $N_i$.} the number $N_i$ detected by any method will be $N p_i$, where $p_i$ is the probability that method $i$ will detect an AGN in the population. If two detection methods $i,j$ are {\em independent}, the number $N_{ij}$ detected by both methods (shown above or to the right of the diagonal in Table~\ref{t:agn}) will be $N p_i p_j$.  Therefore
\begin{align}
\label{e:p}
p_i = N_{ij}/N_j~,\quad {\rm and}\\
N = N_i/p_i = N_j/p_j\quad.
\label{e:n}
\end{align}
Table~\ref{t:prob} gives the results for the pairs of methods that have significant numbers of objects in the sample.
The notional AGN population $N$ can be inflated if either detection method is unreliable, classifying non-AGN as AGN.  If one method in a comparison is reliable and the other isn't, the completeness of the reliable method will be underestimated.
 
The hypothesis of independence won't be true in general. In fact, $N$ is a crude measure of independence.  If two methods $i,j$ produced exactly the same list of AGN, $p_i$ and $p_j$ would both be one (Equation~\ref{e:p}), and $N$ would be the minimum number $N_{ij}$ (Equation~\ref{e:n}). The lowest $N$ in Table~\ref{t:prob} is for X-ray and nuclear point sources: 20\% of X-ray sources show point-source nuclei, and 67\% of radio counterparts with point-source nuclei are X-ray sources. The reason these two selection methods are correlated is easy to understand. These sources should be mostly AGN with luminous nuclei and little obscuration (Type~1). Another low-$N$ pair is SED--$f_{\rm AGN}$.  These are highly correlated because they are based on the same SED, though classified in different ways.  In contrast, X-ray and radio loudness might be somewhat independent \citep[\eg,][]{Pennock2025} with X-rays finding luminous, unobscured AGN while radio loudness finds AGN that have developed jets and/or radio lobes. Indeed Table~\ref{t:prob} shows that this combination has one of the highest values of $N$. The value $p_r=22\%$ is reasonably consistent with 20\% of optically selected QSOs being RL \citep{Kellermann2016}, though the RL fraction found in any sample depends on the exact definition of the term and on the stellar masses of the sample galaxies \citep{Best2005}. The largest $N$ is SED--RL, which also makes sense because RL can find heavily obscured AGN, while SED requires the AGN to dominate the NIR emission. 

In all, the various detection methods identify 132 host galaxies out of the 200 in the SED sample as having at least one AGN signature.  This 66\% compares with 23\% from \citet{Algera2020}, who however did not use pointlike nuclei, VLBA detection, or spectra as identification criteria.  Leaving those out would still give 130 AGN (66\%) in the SED sample. Restricting the SED sample to the 160 sources with $S(\rm3~GHz)<30$~\mmJy\ (Table~\ref{t:ID}) gives 95 AGN by the \citeauthor{Algera2020}\ selection methods and 97 by the methods used here.  In more detail, the shallower 160~ks Chandra data\footnote{The COSMOS-Legacy observations \citep{Civano2016} comprise 4.6~Ms of Chandra exposure time but spread over multiple fields of view.} used by \citeauthor{Algera2020}\ identified an AGN fraction of 7.4\%.  The corresponding fraction in the SED sample, excluding the XMM source not detected by Chandra, is 29\% (or 24\% if all \red{10} $L_X<10^{42}$~erg~s$^{-1}$ were arbitrarily excluded).  For SED fitting, \citet{Algera2020} used {\tt AGNfitter} \citep{CalistroRivera2016}, which found an AGN fraction of only 10.4\% and only 6.8\% based on only the rest visible--NIR SED, i.e., not considering mid-IR dust emission.  The corresponding fractions found here are $\sim$30\% for either SED templates or CIGALE $f_{\rm AGN}$ alone and $\sim$42\% for at least one of the two.   Overall it appears that the \citet{Algera2020} AGN fractions refer to an AGN dominating the emission, whereas the fractions reported here refer to indications that an AGN is present, even when not dominant. 

Even when an AGN is present, it is not necessarily the source of the radio emission.  For example, ID~242 is an X-ray source, yet its 3~GHz flux density is only $\sim$26\% of what its $\rm SFR\approx220$~\Msol~yr$^{-1}$ predicts, leaving little room for AGN radio emission. Another example is ID~312, likewise an X-ray source, which has 3~GHz flux density $\sim$32\% of what its $\rm SFR\approx44$~\Msol~yr$^{-1}$ predicts.  At the other extreme, ID~141 (Figure~\ref{f:tricky}) is a double-lobe radio source as well as a Chandra X-ray source with a pointlike nucleus; there is no doubt  where its radio flux is coming from. 
Although AGN SEDs  dominate at $z>2.8$ (Figure~\ref{f:agn_fraction}), only one of 12 sources in that range is RL. However, for three others, the radio flux is $>$6$\times$ larger expected from SFR, and a majority of the rest have uncertain SFRs. Furthermore, the radio emission from a given SFR is little tested at these redshifts. AGN radio emission may therefore be important for high-redshift sources.

\section{Summary and Conclusions}
\label{s:summ}

Even {$\sim$50~minute} exposure times with
JWST/\linebreak[0]NIRCam can detect counterparts for nearly all radio sources with $S(3\rm~GHz)>
5$~\mmJy. Out of 213 TDF 3~GHz sources with any NIRCam or NIRISS coverage, 208 have likely counterparts. One of the sources with no counterpart is likely to be a radio lobe of a bright Seyfert galaxy, and the other four radio sources are faint and might be spurious.  Therefore, the detection rate is at least 97.7\% and plausibly 100\%.

A simple position search in F444W with radius 0\farcs3 would have produced correct counterparts for 198 out of 205 sources within the F444W image area with no definite false matches but one ambiguous match (ID~248 at 0\farcs26 separation). This result was achieved by using an F444W object catalog that aggressively deblended sources with multiple peaks. As a caution, one source (ID~270, outside the F444W image) matched an F200W catalog object at a distance of 0\farcs16, but the catalog object is a diffraction spike, not the real counterpart.  Sources at separations up to 0\farcs4 from the radio position may in some cases be correct counterparts but need to be individually checked.

CIGALE fitting of the 200 radio-host galaxies that have photometric coverage extending to 4.4~\micron\ provided photometric redshifts and galaxy properties.  Comparing \zph\ to \zsp\ for galaxies with \zsp\ $Q\ge3$ and \zph\ available\footnote{ID~500 has \zsp\ but no \zph\ because it lacks LW data.} found two galaxies (3\%) with discrepancies greater than 2.5 times the \zph\ uncertainty (with $\sim$0.8 expected from Gaussian statistics).  Galaxies that lack spectroscopic redshifts tend to be fainter than galaxies that have \zsp\ but by $\la$1~mag in the median, and nearly all the radio hosts have $m(\rm F444W)$ a factor of 100 above the JWST survey limit. Therefore the S/N for most galaxies should be high enough to give good photometric redshifts.

For the galaxies with CIGALE fits, $\sim$79\% have radio flux densities no more than an order of magnitude greater than that expected from star formation. For the remaining 21\%, the excess radio flux could come either from an AGN or from star formation hidden behind dust. Here the expected radio flux was based on the non-linear SFR--$L$(1.4~GHz) relation established for local galaxies.  The non-linearity may provide a natural explanation for previously reported mass and redshift dependence, but testing that will require examination of samples that have far-IR data.  

Fully 66\% of the sample members have at least one indicator of an AGN's presence, but this does not mean that the AGN dominates the radio or visible--NIR luminosity. The higher AGN fraction found in this study compared to previous ones is partly due to the deep X-ray observations available for the TDF and partly to counting any AGN signature rather than requiring the AGN to dominate either the radio or NIR emission.

While the bulk of the radio-host galaxies are the same population as seen in previous surveys, searching at 4.44~\micron\ finds even the reddest galaxies that may be missed in shorter-wavelength searches.  More study is needed, in particular spectra of these host galaxies, but they likely constitute the highest-redshift and dustiest fraction of the population.  Monitoring for variability would also be valuable for identifying AGN, and the TDF is by design the most suitable field for such studies.

Another avenue for future research is understanding the radio population at $S(3\rm~GHz)\ll5$~\mmJy.  As shown in Figure~\ref{f:fz}, nearly all of \mmJy\ radio sources have NIR counterparts more than a factor of 100 brighter than the JWST NIRCam detection limit $AB\lesssim28$--29~mag. Given the steep slope of the NIR galaxy counts, a large population of faint galaxies is yet to be detected at radio wavelengths. Presumably these are a mix of fainter star-forming galaxies and weaker AGN \citep[\eg,][]{Tompkins2023} than studied here.  Detecting this population, and thereby delineating the early co-evolution of galaxy assembly and supermassive black hole growth, will be one of the main jobs for the future facilities such as the Next Generation VLA and the Square Kilometer Array, and JWST should have ample sensitivity to identify and study their counterparts.

\bigskip

This paper is dedicated to the memory of Dr.\ Giovanni G.\ Fazio, who died suddenly on 2026 February 12.  Giovanni was a full participant in the PEARLS collaboration, a wonderful colleague, and a great inspiration to the team.

This work is based on observations made with the NASA/ESA/CSA James Webb Space Telescope. The data were obtained from the Mikulski Archive for Space Telescopes at the Space Telescope Science Institute, which is operated by the Association of Universities for Research in Astronomy, Inc. (AURA), under NASA contract NAS 5-03127 for JWST. These observations are associated with JWST GTO program 2738. The National Radio Astronomy Observatory is a facility of the National Science Foundation operated under cooperative agreement by Associated Universities, Inc. 
Observations reported here were obtained at the MMT Observatory, a joint facility of the Smithsonian Institution and the University of Arizona.
SHC, RAJ, RAW, and HBH acknowledge support from NASA JWST Interdisciplinary Scientist grants NNX14AN10G, 80NSSC18K0200,  NAG5-12460, and 21-SMDSS21-0013, respectively, from  NASA Goddard Space Flight Center (GSFC).
RO, RAJ, and AMK acknowledge support from grants HST-GO-15278.* and HST-GO-16252.* from STScI, which is operated by AURA under contract NAS\,5-26555 from NASA.
CNAW acknowledges funding from the JWST/NIRCam contract NASS-0215 to the University of Arizona.
We also acknowledge the indigenous peoples of Arizona, including the Akimel O'odham (Pima) and Pee Posh (Maricopa) Indian Communities, whose care and keeping of the land has enabled us to be at ASU's Tempe campus in the Salt River Valley, where much of our work was conducted.


Data presented in this article were obtained from the Mikulski Archive for Space Telescopes (MAST) at the Space Telescope Science Institute. 
The NIRCam observations used here can be accessed via \dataset[10.17909/jtd6-af15]{http://dx.doi.org/10.17909/jtd6-af15}, the NIRISS observations via \dataset[10.17909/7xzs-bb33]{https://doi.org/10.17909/7xzs-bb33}, and the HST observations via 
\dataset[10.17909/wv13-qc14]{https://doi.org/10.17909/wv13-qc14}.

\software{
CIGALE: \citep{Boquien2019,Yang2020,Yang2022}}
\software{
SourceExtractor: \citep{Bertin1996} 
\url{https://www.astromatic.net/software/sextractor/} or
\url{https://sextractor.readthedocs.io/en/latest/}}
\software{
PyBDSF \citep{Mohan2015}}
\software{
JWST calibration pipeline version 1.7.2 \citep{pipeline} \url{https://zenodo.org/badge/DOI/10.5281/zenodo.7071140.svg}
}
\software{\galfit\ \citep{Peng2002,Peng2010} \url{https://users.obs.carnegiescience.edu/peng/work/galfit/galfit.html}}
\software{{\sc xcsao} \citep{Kurtz1998}}

\facilities{Hubble Space Telescope, James Webb Space Telescope, Mikulski Archive
\url{https://archive.stsci.edu}, MMT/Binospec, MMT/Hectospec, VLA}

\FloatBarrier

\appendix
\restartappendixnumbering

\section{Individual sources of interest}
\label{s:special}

\begin{figure*}
{\bf ~~~~3~GHz\hfill ~~F090W \hfill F115W \hfill F150W \hfill F200W \hfill F277W \hfill F356W \hfill F444W~~~~}\\
\includegraphics[width=\linewidth]{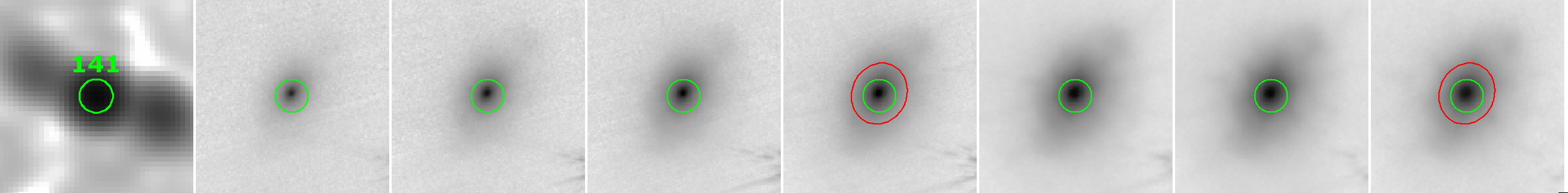}\\
\includegraphics[width=\linewidth]{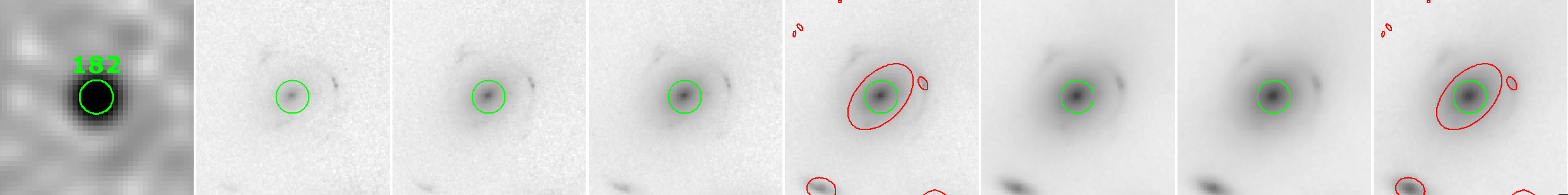}\\
\includegraphics[width=\linewidth]{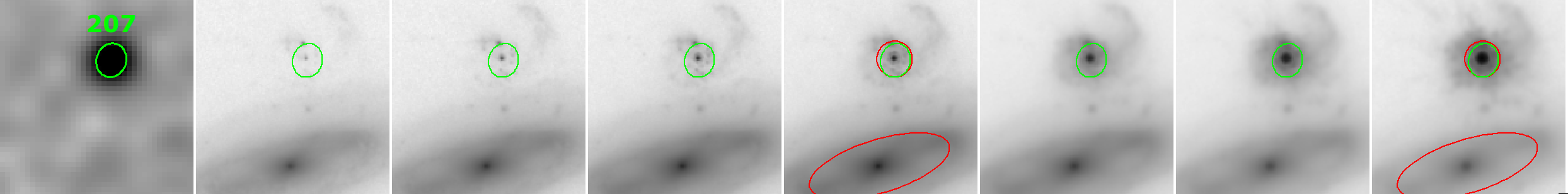}\\
\includegraphics[width=\linewidth]{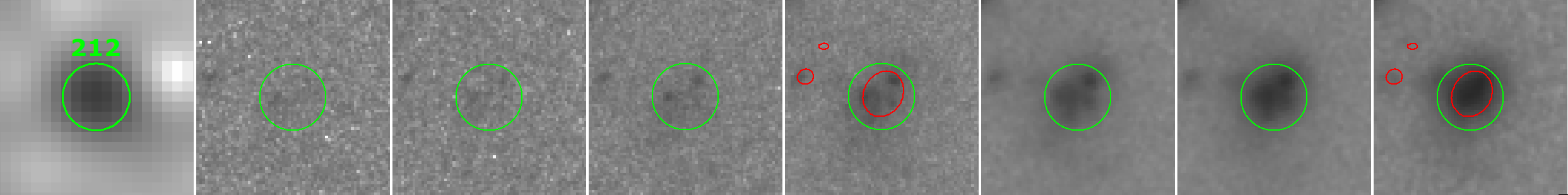}\\
\includegraphics[width=\linewidth]{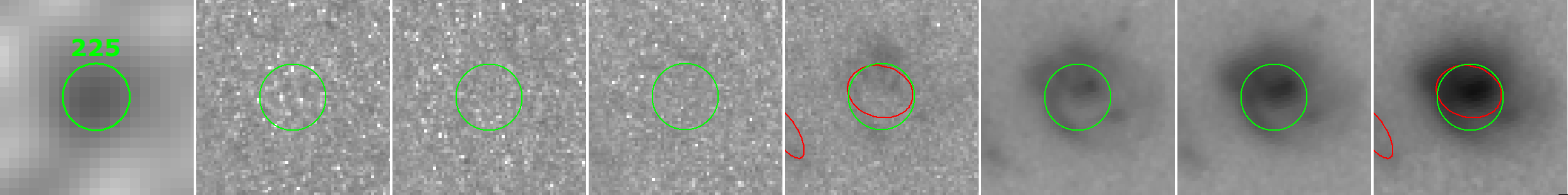}\\
\includegraphics[width=\linewidth]{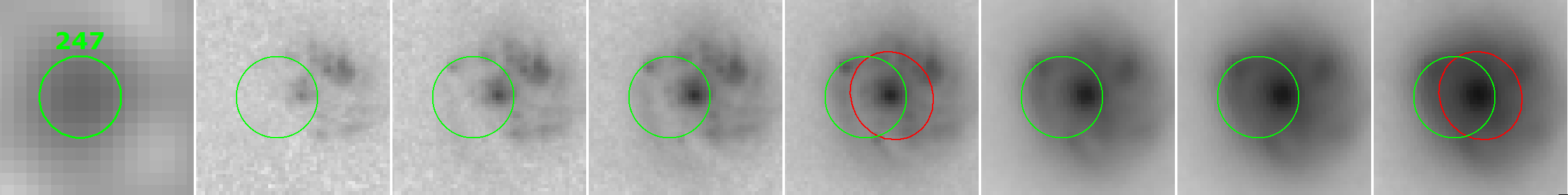}\\
\includegraphics[width=\linewidth]{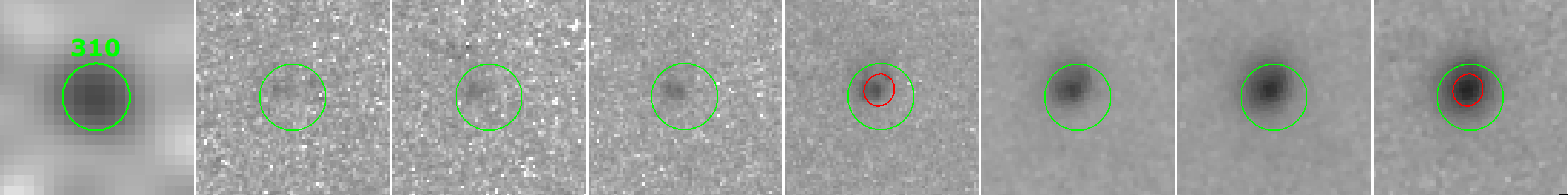}\\
\includegraphics[width=\linewidth]{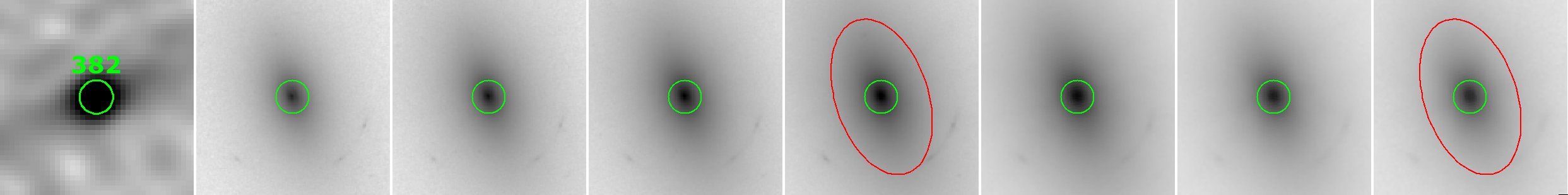}\\
\vspace{-4ex}
\caption{Negative images of sources discussed in Appendix~\ref{s:special} and not shown in other figures. Thumbnails for IDs~212/225/247/310 are 2\arcsec\ on a side; others are 4\arcsec. Other image details are as in Figure~1.}
\label{f:tricky}
\end{figure*}

Sources of special interest, either for their identification or otherwise, are described below by ID. They are  shown in Figure~\ref{f:tricky} when not otherwise indicated.

\begin{itemize}
\item[60:] (Figure~\ref{f:match}) the counterpart is a ring galaxy at
  \red{$\zph=1.39$.} The ring shows bright
  spots that may be \hii\ regions, but the radio flux comes from the
  nucleus (or at least well interior to the ring) and is consistent
  with the expected flux from star formation. \red{The source has a
  pointlike nucleus and is a Chandra X-ray source, but the SED shows
  little indication of an AGN.}

\item[141:] with $S(3~\rm GHz)=1.07$~mJy, this is the brightest radio source in the NIRCam area and the ninth-brightest in the whole \citet{Hyun2023} sample. The source is a classic double-lobed radio galaxy at $\zsp=1.0175$ and is an X-ray source with a  VLBA-detected nucleus. Both \citet{Ortiz2024} and the independent search for this work noticed a pointlike signature in the infrared images, but \galfit\ prefers a bulge over a point source. Despite all the evidence for an AGN, the galaxy's SED matches an early-type galaxy template, and CIGALE's $f_{\rm AGN}=0.33\pm0.12$.

\item[182:] this is the gravitational lens source described by \citet{Adams2025} and in the \citet{Ferrami2025} list. Based on independent photometry and SED fitting, those authors derived $\zph=1.25\pm0.14$, a little higher than $\zph=0.95\pm0.10$ found here.   

\item[203:] (Figure~\ref{f:faint}) the \Pa\ automated search did not find the counterpart, but the improved search used here found it 0\farcs19 from the radio position.

\item[207:] this object has $\rm SFR = 1120 \pm 560$~\Msol~yr$^{-1}$, the highest in the sample albeit with large uncertainty and dependent on $\zph =3.11\pm1.08$, also uncertain. The pointlike nucleus was found by \citet{Ortiz2024} and independently in this work.  The spiral structure is clumpy, and the asymmetric arm to the northwest evinces past interaction.  The large spiral galaxy $\sim$2\farcs2 to the south is at $\zsp=0.1742$ and therefore unrelated.

\item[212:] at $\zph=4.14\pm0.30$, this is the fifth-highest-redshift source in the sample. The source is clumpy, but all clumps are red and have similar colors.

\item[213:] (Figure~\ref{f:match}) the automated search in \Pa\ found the large galaxy rather than the actual radio counterpart, but the new search found the correct counterpart. The new photometry gives $\zph=2.20\pm1.83$, but this may be affected by contamination from the spiral ($\zsp=0.3608$).  The radio source appears extended or double, especially in JWST F356W, and therefore is probably a background galaxy rather than a spatially offset AGN \citep{Barrows2025}. The red color is also more indicative of a background galaxy, but spectroscopy is needed to tell for sure.

\item[225:] the counterpart is very red, and only searches at $\lambda\ga2.5$~\micron\ would have a chance of finding  it. With $\zph=3.50\pm0.29$, $A_V\approx2.7$, and $\rm SFR\approx 46$~\Msol~yr$^{-1}$, both redshift and dust content likely contribute to the red color. The dust extinction is the fourth-highest in the sample, and the SFR is at the 80th percentile. The northern outskirts of the galaxy are visible in the JWST F200W image (rest $\lambda\approx420$~nm), but the nucleus is not, suggesting higher dust content there.

\item[247:] this is a Chandra source with a pointlike nucleus. The bright spot corresponding to the nucleus would have been identified as the radio counterpart even in HST F606W (but not F435W, where the nucleus is invisible), but only JWST F277W and longer give good evidence that the source is a face-on spiral galaxy.

\item[268:]  (Figure~\ref{f:match}) the automated search found the counterpart identified in \Pa, but the aperture size is too small to include all of the galaxy's light.

\item[276:] (Figure~\ref{f:faint}) at $\zph=5.48\pm0.70$, this is the highest-redshift source in the sample. The NIRCam images show two pointlike sources separated by 0\farcs17 with the eastern one being very red.  The automated search found only a single source because the two are barely resolved at 4.44~\micron, and the photometry is a blend of both sources.  Because they have such different colors, \zph\ and other parameters from the SED fitting are unreliable. The radio position is closer to the blue source, which is visible in HST F606W but not HST F435W. If that is because the Lyman break falls between those filters, the redshift for the blue source would be between 2.6 and 4.

\item[310:] at $\zph=4.62\pm1.10$, this is the third-highest-redshift source in the sample.  The source is very red, but there is a faint, bluer companion just to its east, best seen in the JWST F150W image. The companion dominates the JWST F115W image, and a source, probably the companion, is faintly visible in HST F606W but not HST F435W. The comments for ID~276 apply to this source but much less strongly.  In particular, the source-blending for ID~310 is minor for JWST F150W and longer wavelengths.

\item[382:]  the 3~GHz source is unresolved  \citep{Hyun2023}, but the radio image shows a faint hint of a two-sided jet at position angle $\sim$115\degr. The counterpart galaxy is one of the lenses described by \citet{Ferrami2025}. The redshift measurement $\zsp=0.3741$ used here is the same one quoted by \citeauthor{Ferrami2025}, not an independent measurement.

\end{itemize}

\bibliography{TDFref}{}

@ARTICLE{Planck2018,
       author = {{Planck Collaboration} and {Aghanim}, N. and {Akrami}, Y. and {Ashdown}, M. and {Aumont}, J. and {Baccigalupi}, C. and {Ballardini}, M. and {Banday}, A.~J. and {Barreiro}, R.~B. and {Bartolo}, N. and {Basak}, S. and {Battye}, R. and {Benabed}, K. and {Bernard}, J. -P. and {Bersanelli}, M. and {Bielewicz}, P. and {Bock}, J.~J. and {Bond}, J.~R. and {Borrill}, J. and {Bouchet}, F.~R. and {Boulanger}, F. and {Bucher}, M. and {Burigana}, C. and {Butler}, R.~C. and {Calabrese}, E. and {Cardoso}, J. -F. and {Carron}, J. and {Challinor}, A. and {Chiang}, H.~C. and {Chluba}, J. and {Colombo}, L.~P.~L. and {Combet}, C. and {Contreras}, D. and {Crill}, B.~P. and {Cuttaia}, F. and {de Bernardis}, P. and {de Zotti}, G. and {Delabrouille}, J. and {Delouis}, J. -M. and {Di Valentino}, E. and {Diego}, J.~M. and {Dor{\'e}}, O. and {Douspis}, M. and {Ducout}, A. and {Dupac}, X. and {Dusini}, S. and {Efstathiou}, G. and {Elsner}, F. and {En{\ss}lin}, T.~A. and {Eriksen}, H.~K. and {Fantaye}, Y. and {Farhang}, M. and {Fergusson}, J. and {Fernandez-Cobos}, R. and {Finelli}, F. and {Forastieri}, F. and {Frailis}, M. and {Fraisse}, A.~A. and {Franceschi}, E. and {Frolov}, A. and {Galeotta}, S. and {Galli}, S. and {Ganga}, K. and {G{\'e}nova-Santos}, R.~T. and {Gerbino}, M. and {Ghosh}, T. and {Gonz{\'a}lez-Nuevo}, J. and {G{\'o}rski}, K.~M. and {Gratton}, S. and {Gruppuso}, A. and {Gudmundsson}, J.~E. and {Hamann}, J. and {Handley}, W. and {Hansen}, F.~K. and {Herranz}, D. and {Hildebrandt}, S.~R. and {Hivon}, E. and {Huang}, Z. and {Jaffe}, A.~H. and {Jones}, W.~C. and {Karakci}, A. and {Keih{\"a}nen}, E. and {Keskitalo}, R. and {Kiiveri}, K. and {Kim}, J. and {Kisner}, T.~S. and {Knox}, L. and {Krachmalnicoff}, N. and {Kunz}, M. and {Kurki-Suonio}, H. and {Lagache}, G. and {Lamarre}, J. -M. and {Lasenby}, A. and {Lattanzi}, M. and {Lawrence}, C.~R. and {Le Jeune}, M. and {Lemos}, P. and {Lesgourgues}, J. and {Levrier}, F. and {Lewis}, A. and {Liguori}, M. and {Lilje}, P.~B. and {Lilley}, M. and {Lindholm}, V. and {L{\'o}pez-Caniego}, M. and {Lubin}, P.~M. and {Ma}, Y. -Z. and {Mac{\'\i}as-P{\'e}rez}, J.~F. and {Maggio}, G. and {Maino}, D. and {Mandolesi}, N. and {Mangilli}, A. and {Marcos-Caballero}, A. and {Maris}, M. and {Martin}, P.~G. and {Martinelli}, M. and {Mart{\'\i}nez-Gonz{\'a}lez}, E. and {Matarrese}, S. and {Mauri}, N. and {McEwen}, J.~D. and {Meinhold}, P.~R. and {Melchiorri}, A. and {Mennella}, A. and {Migliaccio}, M. and {Millea}, M. and {Mitra}, S. and {Miville-Desch{\^e}nes}, M. -A. and {Molinari}, D. and {Montier}, L. and {Morgante}, G. and {Moss}, A. and {Natoli}, P. and {N{\o}rgaard-Nielsen}, H.~U. and {Pagano}, L. and {Paoletti}, D. and {Partridge}, B. and {Patanchon}, G. and {Peiris}, H.~V. and {Perrotta}, F. and {Pettorino}, V. and {Piacentini}, F. and {Polastri}, L. and {Polenta}, G. and {Puget}, J. -L. and {Rachen}, J.~P. and {Reinecke}, M. and {Remazeilles}, M. and {Renzi}, A. and {Rocha}, G. and {Rosset}, C. and {Roudier}, G. and {Rubi{\~n}o-Mart{\'\i}n}, J.~A. and {Ruiz-Granados}, B. and {Salvati}, L. and {Sandri}, M. and {Savelainen}, M. and {Scott}, D. and {Shellard}, E.~P.~S. and {Sirignano}, C. and {Sirri}, G. and {Spencer}, L.~D. and {Sunyaev}, R. and {Suur-Uski}, A. -S. and {Tauber}, J.~A. and {Tavagnacco}, D. and {Tenti}, M. and {Toffolatti}, L. and {Tomasi}, M. and {Trombetti}, T. and {Valenziano}, L. and {Valiviita}, J. and {Van Tent}, B. and {Vibert}, L. and {Vielva}, P. and {Villa}, F. and {Vittorio}, N. and {Wandelt}, B.~D. and {Wehus}, I.~K. and {White}, M. and {White}, S.~D.~M. and {Zacchei}, A. and {Zonca}, A.},
        title = "{Planck 2018 results. VI. Cosmological parameters}",
      journal = {\aap},
         year = 2020,
        month = sep,
       volume = {641},
          eid = {A6},
        pages = {A6},
          doi = {10.1051/0004-6361/201833910},
archivePrefix = {arXiv},
       eprint = {1807.06209},
 primaryClass = {astro-ph.CO},
       adsurl = {https://ui.adsabs.harvard.edu/abs/2020A&A...641A...6P}
}

@ARTICLE{Chabrier2001, 
       author = {{Chabrier}, Gilles},
        title = "{The Galactic Disk Mass Budget. I. Stellar Mass Function and Density}",
      journal = {\apj},
         year = 2001,
        month = jun,
       volume = {554},
       number = {2},
        pages = {1274-1281},
          doi = {10.1086/321401},
archivePrefix = {arXiv},
       eprint = {astro-ph/0107018},
 primaryClass = {astro-ph},
       adsurl = {https://ui.adsabs.harvard.edu/abs/2001ApJ...554.1274C}
}

@ARTICLE{Lopez2024,
       author = {{L{\'o}pez}, I.~E. and {Yang}, G. and {Mountrichas}, G. and {Brusa}, M. and {Alexander}, D.~M. and {Baldi}, R.~D. and {Bertola}, E. and {Bonoli}, S. and {Comastri}, A. and {Shankar}, F. and {Acharya}, N. and {Alonso Tetilla}, A.~V. and {Lapi}, A. and {Laloux}, B. and {L{\'o}pez L{\'o}pez}, X. and {Mu{\~n}oz Rodr{\'\i}guez}, I. and {Musiimenta}, B. and {Osorio Clavijo}, N. and {Sala}, L. and {Sengupta}, D.},
        title = "{A CIGALE module tailored (not only) for low-luminosity active galactic nuclei}",
      journal = {\aap},
         year = 2024,
        month = dec,
       volume = {692},
          eid = {A209},
        pages = {A209},
          doi = {10.1051/0004-6361/202450510},
archivePrefix = {arXiv},
       eprint = {2404.16938},
 primaryClass = {astro-ph.GA},
       adsurl = {https://ui.adsabs.harvard.edu/abs/2024A&A...692A.209L}
}

@MISC{Mohan2015,
       author = {{Mohan}, Niruj and {Rafferty}, David},
        title = "{PyBDSF: Python Blob Detection and Source Finder}",
 howpublished = {Astrophysics Source Code Library, record ascl:1502.007},
         year = 2015,
        month = feb,
          eid = {ascl:1502.007},
        pages = {ascl:1502.007},
archivePrefix = {ascl},
       eprint = {1502.007},
       adsurl = {https://ui.adsabs.harvard.edu/abs/2015ascl.soft02007M}
}

@ARTICLE{Madau2014,
       author = {{Madau}, Piero and {Dickinson}, Mark},
        title = "{Cosmic Star-Formation History}",
      journal = {\araa},
         year = 2014,
        month = aug,
       volume = {52},
        pages = {415-486},
          doi = {10.1146/annurev-astro-081811-125615},
archivePrefix = {arXiv},
       eprint = {1403.0007},
 primaryClass = {astro-ph.CO},
       adsurl = {https://ui.adsabs.harvard.edu/abs/2014ARA&A..52..415M}
}

@ARTICLE{Windhorst1985,
       author = {{Windhorst}, R.~A. and {Miley}, G.~K. and {Owen}, F.~N. and {Kron}, R.~G. and {Koo}, D.~C.},
        title = "{Sub-millijansky 1.4 GHz source counts and multicolor studies of weak rado galaxy populations.}",
      journal = {\apj},
         year = 1985,
        month = feb,
       volume = {289},
        pages = {494-513},
          doi = {10.1086/162911},
       adsurl = {https://ui.adsabs.harvard.edu/abs/1985ApJ...289..494W}
}

@ARTICLE{Fomalont2006,
       author = {{Fomalont}, E.~B. and {Kellermann}, K.~I. and {Cowie}, L.~L. and {Capak}, P. and {Barger}, A.~J. and {Partridge}, R.~B. and {Windhorst}, R.~A. and {Richards}, E.~A.},
        title = "{The Radio/Optical Catalog of the SSA 13 Field}",
      journal = {\apjs},
         year = 2006,
        month = dec,
       volume = {167},
       number = {2},
        pages = {103-160},
          doi = {10.1086/508169},
       adsurl = {https://ui.adsabs.harvard.edu/abs/2006ApJS..167..103F}
}

@ARTICLE{Russell2008,
       author = {{Russell}, J. and {Ryan}, R.~E., Jr. and {Cohen}, S.~H. and {Windhorst}, R.~A. and {Waddington}, I.},
        title = "{Optical Morphologies of Millijansky Radio Galaxies Observed by the Hubble Space Telescope and in the Very Large Array FIRST Survey}",
      journal = {\apjs},
         year = 2008,
        month = dec,
       volume = {179},
       number = {2},
        pages = {306-318},
          doi = {10.1086/592045},
       adsurl = {https://ui.adsabs.harvard.edu/abs/2008ApJS..179..306R}
}

@ARTICLE{Windhorst1995,
       author = {{Windhorst}, R.~A. and {Fomalont}, E.~B. and {Kellermann}, K.~I. and {Partridge}, R.~B. and {Richards}, E. and {Franklin}, B.~E. and {Pascarelle}, S.~M. and {Griffiths}, R.~E.},
        title = "{Identification of faint radio sources with optically luminous interacting disk galaxies}",
      journal = {\nat},
         year = 1995,
        month = jun,
       volume = {375},
       number = {6531},
        pages = {471-474},
          doi = {10.1038/375471a0},
       adsurl = {https://ui.adsabs.harvard.edu/abs/1995Natur.375..471W}
}

@ARTICLE{Tompkins2023,
       author = {{Tompkins}, Scott A. and {Driver}, Simon P. and {Robotham}, Aaron S.~G. and {Windhorst}, Rogier A. and {Lagos}, Claudia del P. and {Vernstrom}, T. and {Hopkins}, Andrew M.},
        title = "{The cosmic radio background from 150 MHz to 8.4 GHz and its division into AGN and star-forming galaxy flux}",
      journal = {\mnras},
         year = 2023,
        month = may,
       volume = {521},
       number = {1},
        pages = {332-353},
          doi = {10.1093/mnras/stad116},
       adsurl = {https://ui.adsabs.harvard.edu/abs/2023MNRAS.521..332T}
}

@ARTICLE{Fabricant2019,
       author = {{Fabricant}, Daniel and {Fata}, Robert and {Epps}, Harland and {Gauron}, Thomas and {Mueller}, Mark and {Zajac}, Joseph and {Amato}, Stephen and {Barberis}, Jack and {Bergner}, Henry and {Brennan}, Patricia and {Brown}, Warren and {Chilingarian}, Igor and {Geary}, John and {Kradinov}, Vladimir and {McLeod}, Brian and {Smith}, Matthew and {Woods}, Deborah},
        title = "{Binospec: A Wide-field Imaging Spectrograph for the MMT}",
      journal = {\pasp},
         year = 2019,
        month = jul,
       volume = {131},
       number = {1001},
        pages = {075004},
          doi = {10.1088/1538-3873/ab1d78},
archivePrefix = {arXiv},
       eprint = {1905.03320},
 primaryClass = {astro-ph.IM},
       adsurl = {https://ui.adsabs.harvard.edu/abs/2019PASP..131g5004F}
}

@ARTICLE{Newman2013,
       author = {{Newman}, Jeffrey A. and {Cooper}, Michael C. and {Davis}, Marc and {Faber}, S.~M. and {Coil}, Alison L. and {Guhathakurta}, Puragra and {Koo}, David C. and {Phillips}, Andrew C. and {Conroy}, Charlie and {Dutton}, Aaron A. and {Finkbeiner}, Douglas P. and {Gerke}, Brian F. and {Rosario}, David J. and {Weiner}, Benjamin J. and {Willmer}, C.~N.~A. and {Yan}, Renbin and {Harker}, Justin J. and {Kassin}, Susan A. and {Konidaris}, N.~P. and {Lai}, Kamson and {Madgwick}, Darren S. and {Noeske}, K.~G. and {Wirth}, Gregory D. and {Connolly}, A.~J. and {Kaiser}, N. and {Kirby}, Evan N. and {Lemaux}, Brian C. and {Lin}, Lihwai and {Lotz}, Jennifer M. and {Luppino}, G.~A. and {Marinoni}, C. and {Matthews}, Daniel J. and {Metevier}, Anne and {Schiavon}, Ricardo P.},
        title = "{The DEEP2 Galaxy Redshift Survey: Design, Observations, Data Reduction, and Redshifts}",
      journal = {\apjs},
         year = 2013,
        month = sep,
       volume = {208},
       number = {1},
          eid = {5},
        pages = {5},
          doi = {10.1088/0067-0049/208/1/5},
archivePrefix = {arXiv},
       eprint = {1203.3192},
 primaryClass = {astro-ph.CO},
       adsurl = {https://ui.adsabs.harvard.edu/abs/2013ApJS..208....5N}
}

@ARTICLE{Smolcic2017id,
       author = {{Smol{\v{c}}i{\'c}}, V. and {Delvecchio}, I. and {Zamorani}, G. and {Baran}, N. and {Novak}, M. and {Delhaize}, J. and {Schinnerer}, E. and {Berta}, S. and {Bondi}, M. and {Ciliegi}, P. and {Capak}, P. and {Civano}, F. and {Karim}, A. and {Le Fevre}, O. and {Ilbert}, O. and {Laigle}, C. and {Marchesi}, S. and {McCracken}, H.~J. and {Tasca}, L. and {Salvato}, M. and {Vardoulaki}, E.},
        title = "{The VLA-COSMOS 3 GHz Large Project: Multiwavelength counterparts and the composition of the faint radio population}",
      journal = {\aap},
         year = 2017,
        month = jun,
       volume = {602},
          eid = {A2},
        pages = {A2},
          doi = {10.1051/0004-6361/201630223},
archivePrefix = {arXiv},
       eprint = {1703.09719},
 primaryClass = {astro-ph.GA},
       adsurl = {https://ui.adsabs.harvard.edu/abs/2017A&A...602A...2S}
}

@ARTICLE{Algera2020,
       author = {{Algera}, H.~S.~B. and {van der Vlugt}, D. and {Hodge}, J.~A. and {Smail}, I.~R. and {Novak}, M. and {Radcliffe}, J.~F. and {Riechers}, D.~A. and {R{\"o}ttgering}, H. and {Smol{\v{c}}i{\'c}}, V. and {Walter}, F.},
        title = "{A Multiwavelength Analysis of the Faint Radio Sky (COSMOS-XS): the Nature of the Ultra-faint Radio Population}",
      journal = {\apj},
         year = 2020,
        month = nov,
       volume = {903},
       number = {2},
          eid = {139},
        pages = {139},
          doi = {10.3847/1538-4357/abb77a},
archivePrefix = {arXiv},
       eprint = {2009.13531},
 primaryClass = {astro-ph.GA},
       adsurl = {https://ui.adsabs.harvard.edu/abs/2020ApJ...903..139A}
}

@ARTICLE{Jansen2018,
       author = {{Jansen}, Rolf A. and {Windhorst}, Rogier A.},
        title = "{The James Webb Space Telescope North Ecliptic Pole Time-domain Field. I. Field Selection of a JWST Community Field for Time-domain Studies}",
      journal = {\pasp},
         year = 2018,
        month = dec,
       volume = {130},
       number = {994},
        pages = {124001},
          doi = {10.1088/1538-3873/aae476},
archivePrefix = {arXiv},
       eprint = {1807.05278},
 primaryClass = {astro-ph.GA},
       adsurl = {https://ui.adsabs.harvard.edu/abs/2018PASP..130l4001J}
}

@ARTICLE{windhorst2023,
       author = {{Windhorst}, Rogier A. and {Cohen}, Seth H. and {Jansen}, Rolf A. and {Summers}, Jake and {Tompkins}, Scott and {Conselice}, Christopher J. and {Driver}, Simon P. and {Yan}, Haojing and {Coe}, Dan and {Frye}, Brenda and {Grogin}, Norman and {Koekemoer}, Anton and {Marshall}, Madeline A. and {O'Brien}, Rosalia and {Pirzkal}, Nor and {Robotham}, Aaron and {Ryan}, Russell E. and {Willmer}, Christopher N.~A. and {Carleton}, Timothy and {Diego}, Jose M. and {Keel}, William C. and {Porto}, Paolo and {Redshaw}, Caleb and {Scheller}, Sydney and {Wilkins}, Stephen M. and {Willner}, S.~P. and {Zitrin}, Adi and {Adams}, Nathan J. and {Austin}, Duncan and {Arendt}, Richard G. and {Beacom}, John F. and {Bhatawdekar}, Rachana A. and {Bradley}, Larry D. and {Broadhurst}, Tom and {Cheng}, Cheng and {Civano}, Francesca and {Dai}, Liang and {Dole}, Herv{\'e} and {D'Silva}, Jordan C.~J. and {Duncan}, Kenneth J. and {Fazio}, Giovanni G. and {Ferrami}, Giovanni and {Ferreira}, Leonardo and {Finkelstein}, Steven L. and {Furtak}, Lukas J. and {Gim}, Hansung B. and {Griffiths}, Alex and {Hammel}, Heidi B. and {Harrington}, Kevin C. and {Hathi}, Nimish P. and {Holwerda}, Benne W. and {Honor}, Rachel and {Huang}, Jia-Sheng and {Hyun}, Minhee and {Im}, Myungshin and {Joshi}, Bhavin A. and {Kamieneski}, Patrick S. and {Kelly}, Patrick and {Larson}, Rebecca L. and {Li}, Juno and {Lim}, Jeremy and {Ma}, Zhiyuan and {Maksym}, Peter and {Manzoni}, Giorgio and {Meena}, Ashish Kumar and {Milam}, Stefanie N. and {Nonino}, Mario and {Pascale}, Massimo and {Petric}, Andreea and {Pierel}, Justin D.~R. and {del Carmen Polletta}, Maria and {R{\"o}ttgering}, Huub J.~A. and {Rutkowski}, Michael J. and {Smail}, Ian and {Straughn}, Amber N. and {Strolger}, Louis-Gregory and {Swirbul}, Andi and {Trussler}, James A.~A. and {Wang}, Lifan and {Welch}, Brian and {B. Wyithe}, J. Stuart and {Yun}, Min and {Zackrisson}, Erik and {Zhang}, Jiashuo and {Zhao}, Xiurui},
        title = "{JWST PEARLS. Prime Extragalactic Areas for Reionization and Lensing Science: Project Overview and First Results}",
      journal = {\aj},
         year = 2023,
        month = jan,
       volume = {165},
       number = {1},
          eid = {13},
        pages = {13},
          doi = {10.3847/1538-3881/aca163},
archivePrefix = {arXiv},
       eprint = {2209.04119},
 primaryClass = {astro-ph.CO},
       adsurl = {https://ui.adsabs.harvard.edu/abs/2023AJ....165...13W}
}

@ARTICLE{Hyun2023,
       author = {{Hyun}, Minhee and {Im}, Myungshin and {Smail}, Ian R. and {Cotton}, William D. and {Birkin}, Jack E. and {Kikuta}, Satoshi and {Shim}, Hyunjin and {Willmer}, Christopher N.~A. and {Condon}, James J. and {Windhorst}, Rogier A. and {Cohen}, Seth H. and {Jansen}, Rolf A. and {Ly}, Chun and {Matsuda}, Yuichi and {Fazio}, Giovanni G. and {Swinbank}, A.~M. and {Yan}, Haojing},
        title = "{The JCMT SCUBA-2 Survey of the James Webb Space Telescope North Ecliptic Pole Time-Domain Field}",
      journal = {\apjs},
         year = 2023,
        month = jan,
       volume = {264},
       number = {1},
          eid = {19},
        pages = {19},
          doi = {10.3847/1538-4365/ac9bf4},
archivePrefix = {arXiv},
       eprint = {2301.02786},
 primaryClass = {astro-ph.GA},
       adsurl = {https://ui.adsabs.harvard.edu/abs/2023ApJS..264...19H}
}

@ARTICLE{Bertin1996,
       author = {{Bertin}, E. and {Arnouts}, S.},
        title = "{SExtractor: Software for source extraction.}",
      journal = {\aaps},
         year = 1996,
        month = jun,
       volume = {117},
        pages = {393-404},
          doi = {10.1051/aas:1996164},
       adsurl = {https://ui.adsabs.harvard.edu/abs/1996A&AS..117..393B}
}

@ARTICLE{Seymour2007,
       author = {{Seymour}, Nick and {Stern}, Daniel and {De Breuck}, Carlos and {Vernet}, Joel and {Rettura}, Alessandro and {Dickinson}, Mark and {Dey}, Arjun and {Eisenhardt}, Peter and {Fosbury}, Robert and {Lacy}, Mark and {McCarthy}, Pat and {Miley}, George and {Rocca-Volmerange}, Brigitte and {R{\"o}ttgering}, Huub and {Stanford}, S. Adam and {Teplitz}, Harry and {van Breugel}, Wil and {Zirm}, Andrew},
        title = "{The Massive Hosts of Radio Galaxies across Cosmic Time}",
      journal = {\apjs},
         year = 2007,
        month = aug,
       volume = {171},
       number = {2},
        pages = {353-375},
          doi = {10.1086/517887},
archivePrefix = {arXiv},
       eprint = {astro-ph/0703224},
 primaryClass = {astro-ph},
       adsurl = {https://ui.adsabs.harvard.edu/abs/2007ApJS..171..353S}
}

@ARTICLE{Willner2012,
       author = {{Willner}, S.~P. and {Ashby}, M.~L.~N. and {Barmby}, P. and {Chapman}, S.~C. and {Coil}, A.~L. and {Cooper}, M.~C. and {Huang}, J. -S. and {Ivison}, R. and {Koo}, D.~C.},
        title = "{A Fully Identified Sample of AEGIS20 Microjansky Radio Sources}",
      journal = {\apj},
         year = 2012,
        month = sep,
       volume = {756},
       number = {1},
          eid = {72},
        pages = {72},
          doi = {10.1088/0004-637X/756/1/72},
archivePrefix = {arXiv},
       eprint = {1206.4571},
 primaryClass = {astro-ph.CO},
       adsurl = {https://ui.adsabs.harvard.edu/abs/2012ApJ...756...72W}
}

@ARTICLE{Spinrad1985,
       author = {{Spinrad}, H. and {Djorgovski}, S. and {Marr}, J. and {Aguilar}, L.},
        title = "{A third update of the status of the 3 CR sources : further new redshifts and new identifications of distant galaxies.}",
      journal = {\pasp},
         year = 1985,
        month = oct,
       volume = {97},
        pages = {932-961},
          doi = {10.1086/131647},
       adsurl = {https://ui.adsabs.harvard.edu/abs/1985PASP...97..932S}
}

@ARTICLE{Mahajan2019,
       author = {{Mahajan}, Smriti and {Ashby}, M.~L.~N. and {Willner}, S.~P. and {Barmby}, P. and {Fazio}, G.~G. and {Maragkoudakis}, A. and {Raychaudhury}, S. and {Zezas}, A.},
        title = "{The Star Formation Reference Survey - III. A multiwavelength view of star formation in nearby galaxies}",
      journal = {\mnras},
         year = 2019,
        month = jan,
       volume = {482},
       number = {1},
        pages = {560-577},
          doi = {10.1093/mnras/sty2699},
archivePrefix = {arXiv},
       eprint = {1810.01336},
 primaryClass = {astro-ph.GA},
       adsurl = {https://ui.adsabs.harvard.edu/abs/2019MNRAS.482..560M}
}

@ARTICLE{Speagle2014,
       author = {{Speagle}, J.~S. and {Steinhardt}, C.~L. and {Capak}, P.~L. and {Silverman}, J.~D.},
        title = "{A Highly Consistent Framework for the Evolution of the Star-Forming ``Main Sequence'' from z \raisebox{-0.5ex}\textasciitilde 0-6}",
      journal = {\apjs},
         year = 2014,
        month = oct,
       volume = {214},
       number = {2},
          eid = {15},
        pages = {15},
          doi = {10.1088/0067-0049/214/2/15},
archivePrefix = {arXiv},
       eprint = {1405.2041},
 primaryClass = {astro-ph.GA},
       adsurl = {https://ui.adsabs.harvard.edu/abs/2014ApJS..214...15S}
}

@ARTICLE{Bruzual2003,
       author = {{Bruzual}, G. and {Charlot}, S.},
        title = "{Stellar population synthesis at the resolution of 2003}",
      journal = {\mnras},
         year = 2003,
        month = oct,
       volume = {344},
       number = {4},
        pages = {1000-1028},
          doi = {10.1046/j.1365-8711.2003.06897.x},
archivePrefix = {arXiv},
       eprint = {astro-ph/0309134},
 primaryClass = {astro-ph},
       adsurl = {https://ui.adsabs.harvard.edu/abs/2003MNRAS.344.1000B}
}

@ARTICLE{Dale2014,
       author = {{Dale}, Daniel A. and {Helou}, George and {Magdis}, Georgios E. and {Armus}, Lee and {D{\'\i}az-Santos}, Tanio and {Shi}, Yong},
        title = "{A Two-parameter Model for the Infrared/Submillimeter/Radio Spectral Energy Distributions of Galaxies and Active Galactic Nuclei}",
      journal = {\apj},
         year = 2014,
        month = mar,
       volume = {784},
       number = {1},
          eid = {83},
        pages = {83},
          doi = {10.1088/0004-637X/784/1/83},
archivePrefix = {arXiv},
       eprint = {1402.1495},
 primaryClass = {astro-ph.GA},
       adsurl = {https://ui.adsabs.harvard.edu/abs/2014ApJ...784...83D}
}

@ARTICLE{Calzetti2000,
       author = {{Calzetti}, Daniela and {Armus}, Lee and {Bohlin}, Ralph C. and {Kinney}, Anne L. and {Koornneef}, Jan and {Storchi-Bergmann}, Thaisa},
        title = "{The Dust Content and Opacity of Actively Star-forming Galaxies}",
      journal = {\apj},
         year = 2000,
        month = apr,
       volume = {533},
       number = {2},
        pages = {682-695},
          doi = {10.1086/308692},
archivePrefix = {arXiv},
       eprint = {astro-ph/9911459},
 primaryClass = {astro-ph},
       adsurl = {https://ui.adsabs.harvard.edu/abs/2000ApJ...533..682C}
}

@ARTICLE{Stalevski2012,
       author = {{Stalevski}, Marko and {Fritz}, Jacopo and {Baes}, Maarten and {Nakos}, Theodoros and {Popovi{\'c}}, Luka {\v{C}}.},
        title = "{3D radiative transfer modelling of the dusty tori around active galactic nuclei as a clumpy two-phase medium}",
      journal = {\mnras},
         year = 2012,
        month = mar,
       volume = {420},
       number = {4},
        pages = {2756-2772},
          doi = {10.1111/j.1365-2966.2011.19775.x},
archivePrefix = {arXiv},
       eprint = {1109.1286},
 primaryClass = {astro-ph.CO},
       adsurl = {https://ui.adsabs.harvard.edu/abs/2012MNRAS.420.2756S}
}

@ARTICLE{Stalevski2016,
       author = {{Stalevski}, Marko and {Ricci}, Claudio and {Ueda}, Yoshihiro and {Lira}, Paulina and {Fritz}, Jacopo and {Baes}, Maarten},
        title = "{The dust covering factor in active galactic nuclei}",
      journal = {\mnras},
         year = 2016,
        month = may,
       volume = {458},
       number = {3},
        pages = {2288-2302},
          doi = {10.1093/mnras/stw444},
archivePrefix = {arXiv},
       eprint = {1602.06954},
 primaryClass = {astro-ph.GA},
       adsurl = {https://ui.adsabs.harvard.edu/abs/2016MNRAS.458.2288S}
}

@ARTICLE{Boquien2019,
       author = {{Boquien}, M. and {Burgarella}, D. and {Roehlly}, Y. and {Buat}, V. and {Ciesla}, L. and {Corre}, D. and {Inoue}, A.~K. and {Salas}, H.},
        title = "{CIGALE: a python Code Investigating GALaxy Emission}",
      journal = {\aap},
         year = 2019,
        month = feb,
       volume = {622},
          eid = {A103},
        pages = {A103},
          doi = {10.1051/0004-6361/201834156},
archivePrefix = {arXiv},
       eprint = {1811.03094},
 primaryClass = {astro-ph.GA},
       adsurl = {https://ui.adsabs.harvard.edu/abs/2019A&A...622A.103B}
}

@ARTICLE{Yang2020,
       author = {{Yang}, G. and {Boquien}, M. and {Buat}, V. and {Burgarella}, D. and {Ciesla}, L. and {Duras}, F. and {Stalevski}, M. and {Brandt}, W.~N. and {Papovich}, C.},
        title = "{X-CIGALE: Fitting AGN/galaxy SEDs from X-ray to infrared}",
      journal = {\mnras},
         year = 2020,
        month = jan,
       volume = {491},
       number = {1},
        pages = {740-757},
          doi = {10.1093/mnras/stz3001},
archivePrefix = {arXiv},
       eprint = {2001.08263},
 primaryClass = {astro-ph.GA},
       adsurl = {https://ui.adsabs.harvard.edu/abs/2020MNRAS.491..740Y}
}

@ARTICLE{Yang2022,
       author = {{Yang}, Guang and {Boquien}, M{\'e}d{\'e}ric and {Brandt}, W.~N. and {Buat}, V{\'e}ronique and {Burgarella}, Denis and {Ciesla}, Laure and {Lehmer}, Bret D. and {Ma{\l}ek}, Katarzyna and {Mountrichas}, George and {Papovich}, Casey and {Pons}, Estelle and {Stalevski}, Marko and {Theul{\'e}}, Patrice and {Zhu}, Shifu},
        title = "{Fitting AGN/Galaxy X-Ray-to-radio SEDs with CIGALE and Improvement of the Code}",
      journal = {\apj},
         year = 2022,
        month = mar,
       volume = {927},
       number = {2},
          eid = {192},
        pages = {192},
          doi = {10.3847/1538-4357/ac4971},
archivePrefix = {arXiv},
       eprint = {2201.03718},
 primaryClass = {astro-ph.GA},
       adsurl = {https://ui.adsabs.harvard.edu/abs/2022ApJ...927..192Y}
}

@ARTICLE{Salpeter1955,
       author = {{Salpeter}, Edwin E.},
        title = "{The Luminosity Function and Stellar Evolution.}",
      journal = {\apj},
         year = 1955,
        month = jan,
       volume = {121},
        pages = {161},
          doi = {10.1086/145971},
       adsurl = {https://ui.adsabs.harvard.edu/abs/1955ApJ...121..161S}
}

@ARTICLE{Condon98,
       author = {{Condon}, J.~J. and {Cotton}, W.~D. and {Greisen}, E.~W. and {Yin}, Q.~F. and {Perley}, R.~A. and {Taylor}, G.~B. and {Broderick}, J.~J.},
        title = "{The NRAO VLA Sky Survey}",
      journal = {\aj},
         year = 1998,
        month = may,
       volume = {115},
       number = {5},
        pages = {1693-1716},
          doi = {10.1086/300337},
       adsurl = {https://ui.adsabs.harvard.edu/abs/1998AJ....115.1693C}
}

@INPROCEEDINGS{Becker94,
       author = {{Becker}, Robert H. and {White}, Richard L. and {Helfand}, David J.},
        title = "{The VLA's FIRST Survey}",
    booktitle = {Astronomical Data Analysis Software and Systems III},
         year = 1994,
       editor = {{Crabtree}, D.~R. and {Hanisch}, R.~J. and {Barnes}, J.},
       series = {Astronomical Society of the Pacific Conference Series},
       volume = {61},
        month = jan,
        pages = {165},
       adsurl = {https://ui.adsabs.harvard.edu/abs/1994ASPC...61..165B}
}

@ARTICLE{Sawicki02,
       author = {{Sawicki}, Marcin},
        title = "{The 1.6 Micron Bump as a Photometric Redshift Indicator}",
      journal = {\aj},
         year = 2002,
        month = dec,
       volume = {124},
       number = {6},
        pages = {3050-3060},
          doi = {10.1086/344682},
archivePrefix = {arXiv},
       eprint = {astro-ph/0209437},
 primaryClass = {astro-ph},
       adsurl = {https://ui.adsabs.harvard.edu/abs/2002AJ....124.3050S}
}

@ARTICLE{Fabricant2005,
       author = {{Fabricant}, Daniel and {Fata}, Robert and {Roll}, John and {Hertz}, Edward and {Caldwell}, Nelson and {Gauron}, Thomas and {Geary}, John and {McLeod}, Brian and {Szentgyorgyi}, Andrew and {Zajac}, Joseph and {Kurtz}, Michael and {Barberis}, Jack and {Bergner}, Henry and {Brown}, Warren and {Conroy}, Maureen and {Eng}, Roger and {Geller}, Margaret and {Goddard}, Richard and {Honsa}, Michael and {Mueller}, Mark and {Mink}, Douglas and {Ordway}, Mark and {Tokarz}, Susan and {Woods}, Deborah and {Wyatt}, William and {Epps}, Harland and {Dell'Antonio}, Ian},
        title = "{Hectospec, the MMT's 300 Optical Fiber-Fed Spectrograph}",
      journal = {\pasp},
         year = 2005,
        month = dec,
       volume = {117},
       number = {838},
        pages = {1411-1434},
          doi = {10.1086/497385},
archivePrefix = {arXiv},
       eprint = {astro-ph/0508554},
 primaryClass = {astro-ph},
       adsurl = {https://ui.adsabs.harvard.edu/abs/2005PASP..117.1411F}
}

@ARTICLE{Murphy2011,
       author = {{Murphy}, E.~J. and {Condon}, J.~J. and {Schinnerer}, E. and {Kennicutt}, R.~C. and {Calzetti}, D. and {Armus}, L. and {Helou}, G. and {Turner}, J.~L. and {Aniano}, G. and {Beir{\~a}o}, P. and {Bolatto}, A.~D. and {Brandl}, B.~R. and {Croxall}, K.~V. and {Dale}, D.~A. and {Donovan Meyer}, J.~L. and {Draine}, B.~T. and {Engelbracht}, C. and {Hunt}, L.~K. and {Hao}, C. -N. and {Koda}, J. and {Roussel}, H. and {Skibba}, R. and {Smith}, J. -D.~T.},
        title = "{Calibrating Extinction-free Star Formation Rate Diagnostics with 33 GHz Free-free Emission in NGC 6946}",
      journal = {\apj},
         year = 2011,
        month = aug,
       volume = {737},
       number = {2},
          eid = {67},
        pages = {67},
          doi = {10.1088/0004-637X/737/2/67},
archivePrefix = {arXiv},
       eprint = {1105.4877},
 primaryClass = {astro-ph.CO},
       adsurl = {https://ui.adsabs.harvard.edu/abs/2011ApJ...737...67M}
}

@ARTICLE{Polletta2007,
       author = {{Polletta}, M. and {Tajer}, M. and {Maraschi}, L. and {Trinchieri}, G. and {Lonsdale}, C.~J. and {Chiappetti}, L. and {Andreon}, S. and {Pierre}, M. and {Le F{\`e}vre}, O. and {Zamorani}, G. and {Maccagni}, D. and {Garcet}, O. and {Surdej}, J. and {Franceschini}, A. and {Alloin}, D. and {Shupe}, D.~L. and {Surace}, J.~A. and {Fang}, F. and {Rowan-Robinson}, M. and {Smith}, H.~E. and {Tresse}, L.},
        title = "{Spectral Energy Distributions of Hard X-Ray Selected Active Galactic Nuclei in the XMM-Newton Medium Deep Survey}",
      journal = {\apj},
         year = 2007,
        month = jul,
       volume = {663},
       number = {1},
        pages = {81-102},
          doi = {10.1086/518113},
archivePrefix = {arXiv},
       eprint = {astro-ph/0703255},
 primaryClass = {astro-ph},
       adsurl = {https://ui.adsabs.harvard.edu/abs/2007ApJ...663...81P}
}

@ARTICLE{Zinn2011,
       author = {{Zinn}, P. -C. and {Middelberg}, E. and {Ibar}, E.},
        title = "{Infrared-faint radio sources: a cosmological view. AGN number counts, the cosmic X-ray background and SMBH formation}",
      journal = {\aap},
         year = 2011,
        month = jul,
       volume = {531},
          eid = {A14},
        pages = {A14},
          doi = {10.1051/0004-6361/201016264},
archivePrefix = {arXiv},
       eprint = {1104.0564},
 primaryClass = {astro-ph.CO},
       adsurl = {https://ui.adsabs.harvard.edu/abs/2011A&A...531A..14Z}
}

@ARTICLE{Herzog2016,
       author = {{Herzog}, A. and {Norris}, R.~P. and {Middelberg}, E. and {Seymour}, N. and {Spitler}, L.~R. and {Emonts}, B.~H.~C. and {Franzen}, T.~M.~O. and {Hunstead}, R. and {Intema}, H.~T. and {Marvil}, J. and {Parker}, Q.~A. and {Sirothia}, S.~K. and {Hurley-Walker}, N. and {Bell}, M. and {Bernardi}, G. and {Bowman}, J.~D. and {Briggs}, F. and {Cappallo}, R.~J. and {Callingham}, J.~R. and {Deshpande}, A.~A. and {Dwarakanath}, K.~S. and {For}, B. -Q. and {Greenhill}, L.~J. and {Hancock}, P. and {Hazelton}, B.~J. and {Hindson}, L. and {Johnston-Hollitt}, M. and {Kapi{\'n}ska}, A.~D. and {Kaplan}, D.~L. and {Lenc}, E. and {Lonsdale}, C.~J. and {McKinley}, B. and {McWhirter}, S.~R. and {Mitchell}, D.~A. and {Morales}, M.~F. and {Morgan}, E. and {Morgan}, J. and {Oberoi}, D. and {Offringa}, A. and {Ord}, S.~M. and {Prabu}, T. and {Procopio}, P. and {Udaya Shankar}, N. and {Srivani}, K.~S. and {Staveley-Smith}, L. and {Subrahmanyan}, R. and {Tingay}, S.~J. and {Wayth}, R.~B. and {Webster}, R.~L. and {Williams}, A. and {Williams}, C.~L. and {Wu}, C. and {Zheng}, Q. and {Bannister}, K.~W. and {Chippendale}, A.~P. and {Harvey-Smith}, L. and {Heywood}, I. and {Indermuehle}, B. and {Popping}, A. and {Sault}, R.~J. and {Whiting}, M.~T.},
        title = "{The radio spectral energy distribution of infrared-faint radio sources}",
      journal = {\aap},
         year = 2016,
        month = oct,
       volume = {593},
          eid = {A130},
        pages = {A130},
          doi = {10.1051/0004-6361/201527000},
archivePrefix = {arXiv},
       eprint = {1607.02707},
 primaryClass = {astro-ph.GA},
       adsurl = {https://ui.adsabs.harvard.edu/abs/2016A&A...593A.130H}
}

@ARTICLE{Strazzullo2010,
       author = {{Strazzullo}, Veronica and {Pannella}, Maurilio and {Owen}, Frazer N. and {Bender}, Ralf and {Morrison}, Glenn E. and {Wang}, Wei-Hao and {Shupe}, David L.},
        title = "{The Deep Swire Field. IV. First Properties of the sub-mJy Galaxy Population: Redshift Distribution, AGN Activity, and Star Formation}",
      journal = {\apj},
         year = 2010,
        month = may,
       volume = {714},
       number = {2},
        pages = {1305-1323},
          doi = {10.1088/0004-637X/714/2/1305},
archivePrefix = {arXiv},
       eprint = {1003.4734},
 primaryClass = {astro-ph.CO},
       adsurl = {https://ui.adsabs.harvard.edu/abs/2010ApJ...714.1305S}
}

@ARTICLE{Owen2018,
       author = {{Owen}, Frazer N.},
        title = "{Deep JVLA Imaging of GOODS-N at 20 cm}",
      journal = {\apjs},
         year = 2018,
        month = apr,
       volume = {235},
       number = {2},
          eid = {34},
        pages = {34},
          doi = {10.3847/1538-4365/aab4a1},
archivePrefix = {arXiv},
       eprint = {1803.05455},
 primaryClass = {astro-ph.GA},
       adsurl = {https://ui.adsabs.harvard.edu/abs/2018ApJS..235...34O}
}

@ARTICLE{Sanders1989,
       author = {{Sanders}, D.~B. and {Phinney}, E.~S. and {Neugebauer}, G. and {Soifer}, B.~T. and {Matthews}, K.},
        title = "{Continuum Energy Distributions of Quasars: Shapes and Origins}",
      journal = {\apj},
         year = 1989,
        month = dec,
       volume = {347},
        pages = {29},
          doi = {10.1086/168094},
       adsurl = {https://ui.adsabs.harvard.edu/abs/1989ApJ...347...29S}
}

@ARTICLE{Rodighiero2011,
       author = {{Rodighiero}, G. and {Daddi}, E. and {Baronchelli}, I. and {Cimatti}, A. and {Renzini}, A. and {Aussel}, H. and {Popesso}, P. and {Lutz}, D. and {Andreani}, P. and {Berta}, S. and {Cava}, A. and {Elbaz}, D. and {Feltre}, A. and {Fontana}, A. and {F{\"o}rster Schreiber}, N.~M. and {Franceschini}, A. and {Genzel}, R. and {Grazian}, A. and {Gruppioni}, C. and {Ilbert}, O. and {Le Floch}, E. and {Magdis}, G. and {Magliocchetti}, M. and {Magnelli}, B. and {Maiolino}, R. and {McCracken}, H. and {Nordon}, R. and {Poglitsch}, A. and {Santini}, P. and {Pozzi}, F. and {Riguccini}, L. and {Tacconi}, L.~J. and {Wuyts}, S. and {Zamorani}, G.},
        title = "{The Lesser Role of Starbursts in Star Formation at z = 2}",
      journal = {\apjl},
         year = 2011,
        month = oct,
       volume = {739},
       number = {2},
          eid = {L40},
        pages = {L40},
          doi = {10.1088/2041-8205/739/2/L40},
archivePrefix = {arXiv},
       eprint = {1108.0933},
 primaryClass = {astro-ph.CO},
       adsurl = {https://ui.adsabs.harvard.edu/abs/2011ApJ...739L..40R}
}

@ARTICLE{Renzini2015,
       author = {{Renzini}, Alvio and {Peng}, Ying-jie},
        title = "{An Objective Definition for the Main Sequence of Star-forming Galaxies}",
      journal = {\apjl},
         year = 2015,
        month = mar,
       volume = {801},
       number = {2},
          eid = {L29},
        pages = {L29},
          doi = {10.1088/2041-8205/801/2/L29},
archivePrefix = {arXiv},
       eprint = {1502.01027},
 primaryClass = {astro-ph.GA},
       adsurl = {https://ui.adsabs.harvard.edu/abs/2015ApJ...801L..29R}
}

@ARTICLE{Haas2008,
       author = {{Haas}, Martin and {Willner}, S.~P. and {Heymann}, Frank and {Ashby}, M.~L.~N. and {Fazio}, G.~G. and {Wilkes}, Belinda J. and {Chini}, Rolf and {Siebenmorgen}, Ralf},
        title = "{Near- and Mid-Infrared Photometry of High-Redshift 3CR Sources}",
      journal = {\apj},
         year = 2008,
        month = nov,
       volume = {688},
       number = {1},
        pages = {122-127},
          doi = {10.1086/592085},
archivePrefix = {arXiv},
       eprint = {0807.3966},
 primaryClass = {astro-ph},
       adsurl = {https://ui.adsabs.harvard.edu/abs/2008ApJ...688..122H}
}

@ARTICLE{Gregg2002,
       author = {{Gregg}, Michael D. and {Lacy}, Mark and {White}, Richard L. and {Glikman}, Eilat and {Helfand}, David and {Becker}, Robert H. and {Brotherton}, Michael S.},
        title = "{The Reddest Quasars}",
      journal = {\apj},
         year = 2002,
        month = jan,
       volume = {564},
       number = {1},
        pages = {133-142},
          doi = {10.1086/324145},
archivePrefix = {arXiv},
       eprint = {astro-ph/0107441},
 primaryClass = {astro-ph},
       adsurl = {https://ui.adsabs.harvard.edu/abs/2002ApJ...564..133G}
}

@ARTICLE{Cotton2018,
       author = {{Cotton}, W.~D. and {Condon}, J.~J. and {Kellermann}, K.~I. and {Lacy}, M. and {Perley}, R.~A. and {Matthews}, A.~M. and {Vernstrom}, T. and {Scott}, Douglas and {Wall}, J.~V.},
        title = "{The Angular Size Distribution of {\ensuremath{\mu}}Jy Radio Sources}",
      journal = {\apj},
         year = 2018,
        month = mar,
       volume = {856},
       number = {1},
          eid = {67},
        pages = {67},
          doi = {10.3847/1538-4357/aaaec4},
archivePrefix = {arXiv},
       eprint = {1802.04209},
 primaryClass = {astro-ph.GA},
       adsurl = {https://ui.adsabs.harvard.edu/abs/2018ApJ...856...67C}
}

@ARTICLE{Ivison2007,
       author = {{Ivison}, R.~J. and {Chapman}, S.~C. and {Faber}, S.~M. and {Smail}, Ian and {Biggs}, A.~D. and {Conselice}, C.~J. and {Wilson}, G. and {Salim}, S. and {Huang}, J. -S. and {Willner}, S.~P.},
        title = "{AEGIS20: A Radio Survey of the Extended Groth Strip}",
      journal = {\apjl},
         year = 2007,
        month = may,
       volume = {660},
       number = {1},
        pages = {L77-L80},
          doi = {10.1086/517917},
archivePrefix = {arXiv},
       eprint = {astro-ph/0607271},
 primaryClass = {astro-ph},
       adsurl = {https://ui.adsabs.harvard.edu/abs/2007ApJ...660L..77I}
}

@ARTICLE{Rigby2023,
       author = {{Rigby}, Jane and {Perrin}, Marshall and {McElwain}, Michael and {Kimble}, Randy and {Friedman}, Scott and {Lallo}, Matt and {Doyon}, Ren{\'e} and {Feinberg}, Lee and {Ferruit}, Pierre and {Glasse}, Alistair and et al.},
        title = "{The Science Performance of JWST as Characterized in Commissioning}",
      journal = {\pasp},
         year = 2023,
        month = apr,
       volume = {135},
       number = {1046},
          eid = {048001},
        pages = {048001},
          doi = {10.1088/1538-3873/acb293},
archivePrefix = {arXiv},
       eprint = {2207.05632},
 primaryClass = {astro-ph.IM},
       adsurl = {https://ui.adsabs.harvard.edu/abs/2023PASP..135d8001R}
}

@ARTICLE{Saxena2019,
       author = {{Saxena}, A. and {R{\"o}ttgering}, H.~J.~A. and {Duncan}, K.~J. and {Hill}, G.~J. and {Best}, P.~N. and {Indahl}, B.~L. and {Marinello}, M. and {Overzier}, R.~A. and {Pentericci}, L. and {Prandoni}, I. and {Dannerbauer}, H. and {Barrena}, R.},
        title = "{The nature of faint radio galaxies at high redshifts}",
      journal = {\mnras},
         year = 2019,
        month = sep,
       volume = {489},
       number = {4},
        pages = {5053-5075},
          doi = {10.1093/mnras/stz2516},
archivePrefix = {arXiv},
       eprint = {1906.00746},
 primaryClass = {astro-ph.GA},
       adsurl = {https://ui.adsabs.harvard.edu/abs/2019MNRAS.489.5053S}
}

@ARTICLE{Kondapally2021,
       author = {{Kondapally}, R. and {Best}, P.~N. and {Hardcastle}, M.~J. and {Nisbet}, D. and {Bonato}, M. and {Sabater}, J. and {Duncan}, K.~J. and {McCheyne}, I. and {Cochrane}, R.~K. and {Bowler}, R.~A.~A. and {Williams}, W.~L. and {Shimwell}, T.~W. and {Tasse}, C. and {Croston}, J.~H. and {Goyal}, A. and {Jamrozy}, M. and {Jarvis}, M.~J. and {Mahatma}, V.~H. and {R{\"o}ttgering}, H.~J.~A. and {Smith}, D.~J.~B. and {Wo{\l}owska}, A. and {Bondi}, M. and {Brienza}, M. and {Brown}, M.~J.~I. and {Br{\"u}ggen}, M. and {Chambers}, K. and {Garrett}, M.~A. and {G{\"u}rkan}, G. and {Huber}, M. and {Kunert-Bajraszewska}, M. and {Magnier}, E. and {Mingo}, B. and {Mostert}, R. and {Nikiel-Wroczy{\'n}ski}, B. and {O'Sullivan}, S.~P. and {Paladino}, R. and {Ploeckinger}, T. and {Prandoni}, I. and {Rosenthal}, M.~J. and {Schwarz}, D.~J. and {Shulevski}, A. and {Wagenveld}, J.~D. and {Wang}, L.},
        title = "{The LOFAR Two-meter Sky Survey: Deep Fields Data Release 1. III. Host-galaxy identifications and value added catalogues}",
      journal = {\aap},
         year = 2021,
        month = apr,
       volume = {648},
          eid = {A3},
        pages = {A3},
          doi = {10.1051/0004-6361/202038813},
archivePrefix = {arXiv},
       eprint = {2011.08201},
 primaryClass = {astro-ph.GA},
       adsurl = {https://ui.adsabs.harvard.edu/abs/2021A&A...648A...3K}
}

@ARTICLE{gaia3,
       author = {{Gaia Collaboration} and {Vallenari}, A. and {Brown}, A.~G.~A. and {Prusti}, T. and {de Bruijne}, J.~H.~J. and {Arenou}, F. and {Babusiaux}, C. and {Biermann}, M. and {Creevey}, O.~L. and {Ducourant}, C. and et al.},
        title = "{Gaia Data Release 3. Summary of the content and survey properties}",
      journal = {\aap},
         year = 2023,
        month = jun,
       volume = {674},
          eid = {A1},
        pages = {A1},
          doi = {10.1051/0004-6361/202243940},
archivePrefix = {arXiv},
       eprint = {2208.00211},
 primaryClass = {astro-ph.GA},
       adsurl = {https://ui.adsabs.harvard.edu/abs/2023A&A...674A...1G}
}

@ARTICLE{Chambers1996,
       author = {{Chambers}, K.~C. and {Miley}, G.~K. and {van Breugel}, W.~J.~M. and {Huang}, J. -S.},
        title = "{Ultra--Steep-Spectrum Radio Sources. I. 4C Objects}",
      journal = {\apjs},
         year = 1996,
        month = oct,
       volume = {106},
        pages = {215},
          doi = {10.1086/192337},
       adsurl = {https://ui.adsabs.harvard.edu/abs/1996ApJS..106..215C}
}

@ARTICLE{Miley2008,
       author = {{Miley}, George and {De Breuck}, Carlos},
        title = "{Distant radio galaxies and their environments}",
      journal = {\aapr},
         year = 2008,
        month = feb,
       volume = {15},
       number = {2},
        pages = {67-144},
          doi = {10.1007/s00159-007-0008-z},
archivePrefix = {arXiv},
       eprint = {0802.2770},
 primaryClass = {astro-ph},
       adsurl = {https://ui.adsabs.harvard.edu/abs/2008A&ARv..15...67M}
}

@ARTICLE{Smolcic2017data,
       author = {{Smol{\v{c}}i{\'c}}, V. and {Novak}, M. and {Bondi}, M. and {Ciliegi}, P. and {Mooley}, K.~P. and {Schinnerer}, E. and {Zamorani}, G. and {Navarrete}, F. and {Bourke}, S. and {Karim}, A. and {Vardoulaki}, E. and {Leslie}, S. and {Delhaize}, J. and {Carilli}, C.~L. and {Myers}, S.~T. and {Baran}, N. and {Delvecchio}, I. and {Miettinen}, O. and {Banfield}, J. and {Balokovi{\'c}}, M. and {Bertoldi}, F. and {Capak}, P. and {Frail}, D.~A. and {Hallinan}, G. and {Hao}, H. and {Herrera Ruiz}, N. and {Horesh}, A. and {Ilbert}, O. and {Intema}, H. and {Jeli{\'c}}, V. and {Kl{\"o}ckner}, H. -R. and {Krpan}, J. and {Kulkarni}, S.~R. and {McCracken}, H. and {Laigle}, C. and {Middleberg}, E. and {Murphy}, E.~J. and {Sargent}, M. and {Scoville}, N.~Z. and {Sheth}, K.},
        title = "{The VLA-COSMOS 3 GHz Large Project: Continuum data and source catalog release}",
      journal = {\aap},
         year = 2017,
        month = jun,
       volume = {602},
          eid = {A1},
        pages = {A1},
          doi = {10.1051/0004-6361/201628704},
archivePrefix = {arXiv},
       eprint = {1703.09713},
 primaryClass = {astro-ph.GA},
       adsurl = {https://ui.adsabs.harvard.edu/abs/2017A&A...602A...1S}
}

@software{pipeline,
  author       = {Bushouse, Howard and
                  Eisenhamer, Jonathan and
                  Dencheva, Nadia and
                  Davies, James and
                  Greenfield, Perry and
                  Morrison, Jane and
                  Hodge, Phil and
                  Simon, Bernie and
                  Grumm, David and
                  Droettboom, Michael and
                  Slavich, Edward and
                  Sosey, Megan and
                  Pauly, Tyler and
                  Miller, Todd and
                  Jedrzejewski, Robert and
                  Hack, Warren and
                  Davis, David and
                  Crawford, Steven and
                  Law, David and
                  Gordon, Karl and
                  Regan, Michael and
                  Cara, Mihai and
                  MacDonald, Ken and
                  Bradley, Larry and
                  Shanahan, Clare and
                  Jamieson, William and
                  Teodoro, Mairan and
                  Williams, Thomas},
  title        = {JWST Calibration Pipeline},
  month        = sep,
  year         = 2022,
  note         = {{If you use this software in your work, please cite 
                   it using the following metadata.}},
  publisher    = {Zenodo},
  version      = {1.7.2},
  doi          = {10.5281/zenodo.7071140},
  url          = {https://doi.org/10.5281/zenodo.7071140}
}

@INPROCEEDINGS{Windhorst1990, 
       author = {{Windhorst}, Rogier and {Mathis}, Doug and {Neuschaefer}, Lyman},
        title = "{The evolution of weak radio galaxies at radio and optical wavelengths.}",
    booktitle = {Evolution of the Universe of Galaxies},
         year = 1990,
       editor = {{Kron}, Richard G.},
       series = {Astronomical Society of the Pacific Conference Series},
       volume = {10},
        month = jan,
        pages = {389-403},
       adsurl = {https://ui.adsabs.harvard.edu/abs/1990ASPC...10..389W}
}

@ARTICLE{Ortiz2024,
       author = {{Ortiz}, Rafael and {Windhorst}, Rogier A. and {Cohen}, Seth H. and {Willner}, Steven P. and {Jansen}, Rolf A. and {Carleton}, Timothy and {Kamieneski}, Patrick S. and {Rutkowski}, Michael J. and {Smith}, Brent M. and {Summers}, Jake and {Cheng}, Cheng and {Coe}, Dan and {Conselice}, Christopher J. and {Diego}, Jose M. and {Driver}, Simon P. and {D'Silva}, Jordan C.~J. and {Frye}, Brenda L. and {Gim}, Hansung B. and {Grogin}, Norman A. and {Hammel}, Heidi B. and {Hathi}, Nimish P. and {Holwerda}, Benne W. and {Hyun}, Minhee and {Im}, Myungshin and {Keel}, William C. and {Koekemoer}, Anton M. and {Li}, Juno and {Marshall}, Madeline A. and {McCabe}, Tyler J. and {McLeod}, Noah J. and {Milam}, Stefanie N. and {O'Brien}, Rosalia and {Pirzkal}, Nor and {Robotham}, Aaron S.~G. and {Ryan}, Russell E. and {Willmer}, Christopher N.~A. and {Yan}, Haojing and {Yun}, Min S. and {Zitrin}, Adi},
        title = "{PEARLS: Discovery of Point-source Features within Galaxies in the North Ecliptic Pole Time Domain Field}",
      journal = {\apj},
         year = 2024,
        month = oct,
       volume = {974},
       number = {2},
          eid = {258},
        pages = {258},
          doi = {10.3847/1538-4357/ad6d5e},
archivePrefix = {arXiv},
       eprint = {2404.10709},
 primaryClass = {astro-ph.GA},
       adsurl = {https://ui.adsabs.harvard.edu/abs/2024ApJ...974..258O}
}

@ARTICLE{Willner2023,
       author = {{Willner}, S.~P. and {Gim}, Hansung B. and {Polletta}, Maria del Carmen and {Cohen}, Seth H. and {Willmer}, Christopher N.~A. and {Zhao}, Xiurui and {D'Silva}, Jordan C.~J. and {Jansen}, Rolf A. and {Koekemoer}, Anton M. and {Summers}, Jake and {Windhorst}, Rogier A. and {Coe}, Dan and {Conselice}, Christopher J. and {Driver}, Simon P. and {Frye}, Brenda and {Grogin}, Norman A. and {Marshall}, Madeline A. and {Nonino}, Mario and {Ortiz}, Rafael and {Pirzkal}, Nor and {Robotham}, Aaron and {Rutkowski}, Michael J. and {Ryan}, Russell E. and {Tompkins}, Scott and {Yan}, Haojing and {Hammel}, Heidi B. and {Milam}, Stefanie N. and {Adams}, Nathan J. and {Beacom}, John F. and {Bhatawdekar}, Rachana and {Cheng}, Cheng and {Civano}, F. and {Cotton}, W. and {Hyun}, Minhee and {Kikuta}, Satoshi and {Nyland}, K.~E. and {Peters}, W.~M. and {Petric}, Andreea and {R{\"o}ttgering}, Huub J.~A. and {Shimwell}, T. and {Yun}, Min S.},
        title = "{PEARLS: JWST Counterparts of Microjansky Radio Sources in the Time Domain Field}",
      journal = {\apj},
         year = 2023,
        month = dec,
       volume = {958},
       number = {2},
          eid = {176},
        pages = {176},
          doi = {10.3847/1538-4357/acfdfb},
archivePrefix = {arXiv},
       eprint = {2309.13008},
 primaryClass = {astro-ph.GA},
       adsurl = {https://ui.adsabs.harvard.edu/abs/2023ApJ...958..176W}
}

@ARTICLE{Lamareille2010,
       author = {{Lamareille}, F.},
        title = "{Spectral classification of emission-line galaxies from the Sloan Digital Sky Survey. I. An improved classification for high-redshift galaxies}",
      journal = {\aap},
         year = 2010,
        month = jan,
       volume = {509},
          eid = {A53},
        pages = {A53},
          doi = {10.1051/0004-6361/200913168},
archivePrefix = {arXiv},
       eprint = {0910.4814},
 primaryClass = {astro-ph.CO},
       adsurl = {https://ui.adsabs.harvard.edu/abs/2010A&A...509A..53L}
}

@ARTICLE{Baldwin1981,
       author = {{Baldwin}, J.~A. and {Phillips}, M.~M. and {Terlevich}, R.},
        title = "{Classification parameters for the emission-line spectra of extragalactic objects.}",
      journal = {\pasp},
         year = 1981,
        month = feb,
       volume = {93},
        pages = {5-19},
          doi = {10.1086/130766},
       adsurl = {https://ui.adsabs.harvard.edu/abs/1981PASP...93....5B}
}

@ARTICLE{Zhao2024,
       author = {{Zhao}, Xiurui and {Civano}, Francesca and {Willmer}, Christopher N.~A. and {Bonoli}, Silvia and {Chen}, Chien-Ting and {Creech}, Samantha and {Dupke}, Renato and {Fornasini}, Francesca M. and {Jansen}, Rolf A. and {Kikuta}, Satoshi and {Koekemoer}, Anton M. and {Laha}, Sibasish and {Marchesi}, Stefano and {O'Brien}, Rosalia and {Silver}, Ross and {Willner}, S.~P. and {Windhorst}, Rogier A. and {Yan}, Haojing and {Alcaniz}, Jailson and {Benitez}, Narciso and {Carneiro}, Saulo and {Cenarro}, Javier and {Crist{\'o}bal-Hornillos}, David and {Ederoclite}, Alessandro and {Hern{\'a}n-Caballero}, Antonio and {L{\'o}pez-Sanjuan}, Carlos and {Mar{\'\i}n-Franch}, Antonio and {de Oliveira}, Claudia Mendes and {Moles}, Mariano and {Sodr{\'e}}, Jr., Laerte and {Taylor}, Keith and {Varela}, Jes{\'u}s and {Rami{\'o}}, H{\'e}ctor V{\'a}zquez},
        title = "{PEARLS: NuSTAR and XMM-Newton Extragalactic Survey of the JWST North Ecliptic Pole Time-domain Field II}",
      journal = {\apj},
         year = 2024,
        month = apr,
       volume = {965},
       number = {2},
          eid = {188},
        pages = {188},
          doi = {10.3847/1538-4357/ad2b61},
archivePrefix = {arXiv},
       eprint = {2402.13508},
 primaryClass = {astro-ph.HE},
       adsurl = {https://ui.adsabs.harvard.edu/abs/2024ApJ...965..188Z}
}

@ARTICLE{Popesso2023,
       author = {{Popesso}, P. and {Concas}, A. and {Cresci}, G. and {Belli}, S. and {Rodighiero}, G. and {Inami}, H. and {Dickinson}, M. and {Ilbert}, O. and {Pannella}, M. and {Elbaz}, D.},
        title = "{The main sequence of star-forming galaxies across cosmic times}",
      journal = {\mnras},
         year = 2023,
        month = feb,
       volume = {519},
       number = {1},
        pages = {1526-1544},
          doi = {10.1093/mnras/stac3214},
archivePrefix = {arXiv},
       eprint = {2203.10487},
 primaryClass = {astro-ph.GA},
       adsurl = {https://ui.adsabs.harvard.edu/abs/2023MNRAS.519.1526P}
}

@ARTICLE{Delvecchio2021,
       author = {{Delvecchio}, I. and {Daddi}, E. and {Sargent}, M.~T. and {Jarvis}, M.~J. and {Elbaz}, D. and {Jin}, S. and {Liu}, D. and {Whittam}, I.~H. and {Algera}, H. and {Carraro}, R. and {D'Eugenio}, C. and {Delhaize}, J. and {Kalita}, B.~S. and {Leslie}, S. and {Moln{\'a}r}, D. Cs. and {Novak}, M. and {Prandoni}, I. and {Smol{\v{c}}i{\'c}}, V. and {Ao}, Y. and {Aravena}, M. and {Bournaud}, F. and {Collier}, J.~D. and {Randriamampandry}, S.~M. and {Randriamanakoto}, Z. and {Rodighiero}, G. and {Schober}, J. and {White}, S.~V. and {Zamorani}, G.},
        title = "{The infrared-radio correlation of star-forming galaxies is strongly M$_{{\ensuremath{\star}}}$-dependent but nearly redshift-invariant since z {\ensuremath{\sim}} 4}",
      journal = {\aap},
         year = 2021,
        month = mar,
       volume = {647},
          eid = {A123},
        pages = {A123},
          doi = {10.1051/0004-6361/202039647},
archivePrefix = {arXiv},
       eprint = {2010.05510},
 primaryClass = {astro-ph.GA},
       adsurl = {https://ui.adsabs.harvard.edu/abs/2021A&A...647A.123D}
}

@ARTICLE{Condon1992,
       author = {{Condon}, J.~J.},
        title = "{Radio emission from normal galaxies.}",
      journal = {\araa},
         year = 1992,
        month = jan,
       volume = {30},
        pages = {575-611},
          doi = {10.1146/annurev.aa.30.090192.003043},
       adsurl = {https://ui.adsabs.harvard.edu/abs/1992ARA&A..30..575C}
}

@ARTICLE{Saikia2025,
       author = {{Saikia}, Payaswini and {Wrzosek}, Ramon and {Gelfand}, Joseph and {Brisken}, Walter and {Cotton}, William and {Willner}, S.~P. and {Gim}, Hansung B. and {Windhorst}, Rogier A. and {Estrada-Carpenter}, Vicente and {Katkov}, Ivan Yu. and {Zaw}, Ingyin and {Nicandro Rosenthal}, Michael J. and {Shafi}, Hanaan and {Kellermann}, Kenneth and {Condon}, James and {Koekemoer}, Anton M. and {Conselice}, Christopher J. and {Ortiz}, III, Rafael and {Willmer}, Christopher N.~A. and {Frye}, Brenda and {Grogin}, Norman A. and {Hammel}, Heidi B. and {Cohen}, Seth H. and {Jansen}, Rolf A. and {Summers}, Jake and {D'Silva}, Jordan C.~J. and {Driver}, Simon P. and {Pirzkal}, Nor and {Yan}, Haojing and {Yun}, Min S.},
        title = "{Peering into the Heart of Darkness with VLBA: Radio-quiet Active Galactic Nucleus in the JWST North Ecliptic Pole Time-domain Field}",
      journal = {\apj},
         year = 2025,
        month = aug,
       volume = {989},
       number = {1},
          eid = {29},
        pages = {29},
          doi = {10.3847/1538-4357/ade709},
archivePrefix = {arXiv},
       eprint = {2506.18112},
 primaryClass = {astro-ph.HE},
       adsurl = {https://ui.adsabs.harvard.edu/abs/2025ApJ...989...29S}
}

@ARTICLE{Rubin1968,
       author = {{Rubin}, Robert H.},
        title = "{A Discussion of the Sizes and Excitation of H II Regions}",
      journal = {\apj},
         year = 1968,
        month = oct,
       volume = {154},
        pages = {391},
          doi = {10.1086/149766},
       adsurl = {https://ui.adsabs.harvard.edu/abs/1968ApJ...154..391R}
}

@ARTICLE{Oke1983,
       author = {{Oke}, J.~B. and {Gunn}, J.~E.},
        title = "{Secondary standard stars for absolute spectrophotometry.}",
      journal = {\apj},
         year = 1983,
        month = mar,
       volume = {266},
        pages = {713-717},
          doi = {10.1086/160817},
       adsurl = {https://ui.adsabs.harvard.edu/abs/1983ApJ...266..713O}
}

@software{Bushouse2023,
  author       = {Bushouse, Howard and
                  Eisenhamer, Jonathan and
                  Dencheva, Nadia and
                  Davies, James and
                  Greenfield, Perry and
                  Morrison, Jane and
                  Hodge, Phil and
                  Simon, Bernie and
                  Grumm, David and
                  Droettboom, Michael and
                  Slavich, Edward and
                  Sosey, Megan and
                  Pauly, Tyler and
                  Miller, Todd and
                  Jedrzejewski, Robert and
                  Hack, Warren and
                  Davis, David and
                  Crawford, Steven and
                  Law, David and
                  Gordon, Karl and
                  Regan, Michael and
                  Cara, Mihai and
                  MacDonald, Ken and
                  Bradley, Larry and
                  Shanahan, Clare and
                  Jamieson, William and
                  Teodoro, Mairan and
                  Williams, Thomas},
  title        = {JWST Calibration Pipeline},
  month        = jul,
  year         = 2023,
  publisher    = {Zenodo},
  version      = {1.11.3},
  doi          = {10.5281/zenodo.8157276},
  url          = {https://doi.org/10.5281/zenodo.8157276},
}

@ARTICLE{Robotham2023,
       author = {{Robotham}, A.~S.~G. and {D'Silva}, J.~C.~J. and {Windhorst}, R.~A. and {Jansen}, R.~A. and {Summers}, J. and {Driver}, S.~P. and {Wilmer}, C.~N.~A. and {Bellstedt}, S.},
        title = "{Dynamic Wisp Removal in JWST NIRCam Images}",
      journal = {\pasp},
         year = 2023,
        month = aug,
       volume = {135},
       number = {1050},
          eid = {085003},
        pages = {085003},
          doi = {10.1088/1538-3873/acea42},
archivePrefix = {arXiv},
       eprint = {2305.01175},
 primaryClass = {astro-ph.IM},
       adsurl = {https://ui.adsabs.harvard.edu/abs/2023PASP..135h5003R}
}

@ARTICLE{Lyke2020,
       author = {{Lyke}, Brad W. and {Higley}, Alexandra N. and {McLane}, J.~N. and {Schurhammer}, Danielle P. and {Myers}, Adam D. and {Ross}, Ashley J. and {Dawson}, Kyle and {Chabanier}, Sol{\`e}ne and {Martini}, Paul and {Busca}, Nicol{\'a}s G. and {Mas des Bourboux}, H{\'e}lion du and {Salvato}, Mara and {Streblyanska}, Alina and {Zarrouk}, Pauline and {Burtin}, Etienne and {Anderson}, Scott F. and {Bautista}, Julian and {Bizyaev}, Dmitry and {Brandt}, W.~N. and {Brinkmann}, Jonathan and {Brownstein}, Joel R. and {Comparat}, Johan and {Green}, Paul and {de la Macorra}, Axel and {Mu{\~n}oz Guti{\'e}rrez}, Andrea and {Hou}, Jiamin and {Newman}, Jeffrey A. and {Palanque-Delabrouille}, Nathalie and {P{\^a}ris}, Isabelle and {Percival}, Will J. and {Petitjean}, Patrick and {Rich}, James and {Rossi}, Graziano and {Schneider}, Donald P. and {Smith}, Alexander and {Vivek}, M. and {Weaver}, Benjamin Alan},
        title = "{The Sloan Digital Sky Survey Quasar Catalog: Sixteenth Data Release}",
      journal = {\apjs},
         year = 2020,
        month = sep,
       volume = {250},
       number = {1},
          eid = {8},
        pages = {8},
          doi = {10.3847/1538-4365/aba623},
archivePrefix = {arXiv},
       eprint = {2007.09001},
 primaryClass = {astro-ph.GA},
       adsurl = {https://ui.adsabs.harvard.edu/abs/2020ApJS..250....8L}
}

@ARTICLE{OBrien2024,
       author = {{O'Brien}, Rosalia and {Jansen}, Rolf A. and {Grogin}, Norman A. and {Cohen}, Seth H. and {Smith}, Brent M. and {Silver}, Ross M. and {Maksym}, W.~P. and {Windhorst}, Rogier A. and {Carleton}, Timothy and {Koekemoer}, Anton M. and {Hathi}, Nimish P. and {Willmer}, Christopher N.~A. and {Frye}, Brenda L. and {Alpaslan}, M. and {Ashby}, M.~L.~N. and {Ashcraft}, T.~A. and {Bonoli}, S. and {Brisken}, W. and {Cappelluti}, N. and {Civano}, F. and {Conselice}, C.~J. and {Dhillon}, V.~S. and {Driver}, S.~P. and {Duncan}, K.~J. and {Dupke}, R. and {Elvis}, M. and {Fazio}, G.~G. and {Finkelstein}, S.~L. and {Gim}, H.~B. and {Griffiths}, A. and {Hammel}, H.~B. and {Hyun}, M. and {Im}, M. and {Jones}, V.~R. and {Kim}, D. and {Ladjelate}, B. and {Larson}, R.~L. and {Malhotra}, S. and {Marshall}, M.~A. and {Milam}, S.~N. and {Pierel}, J.~D.~R. and {Rhoads}, J.~E. and {Rodney}, S.~A. and {R{\"o}ttgering}, H.~J.~A. and {Rutkowski}, M.~J. and {Ryan}, R.~E. and {Ward}, M.~J. and {White}, C.~W. and {van Weeren}, R.~J. and {Zhao}, X. and {Summers}, J. and {D'Silva}, J.~C.~J. and {Ortiz}, R. and {Robotham}, A.~S.~G. and {Coe}, D. and {Nonino}, M. and {Pirzkal}, N. and {Yan}, H. and {Acharya}, T.},
        title = "{TREASUREHUNT: Transients and Variability Discovered with HST in the JWST North Ecliptic Pole Time-domain Field}",
      journal = {\apjs},
         year = 2024,
        month = may,
       volume = {272},
       number = {1},
          eid = {19},
        pages = {19},
          doi = {10.3847/1538-4365/ad3948},
archivePrefix = {arXiv},
       eprint = {2401.04944},
 primaryClass = {astro-ph.GA},
       adsurl = {https://ui.adsabs.harvard.edu/abs/2024ApJS..272...19O}
}

@ARTICLE{Schartmann2005,
       author = {{Schartmann}, M. and {Meisenheimer}, K. and {Camenzind}, M. and {Wolf}, S. and {Henning}, T.},
        title = "{Three-dimensional models of clumpy tori in Seyfert galaxies.}",
      journal = {Astronomische Nachrichten},
         year = 2005,
        month = aug,
       volume = {326},
        pages = {555-555},
       adsurl = {https://ui.adsabs.harvard.edu/abs/2005AN....326..555S}
}

@ARTICLE{Pennock2025,
       author = {{Pennock}, Clara M. and {Aird}, James and {Barlow-Hall}, Cassandra L.},
        title = "{Exploring the X-ray--radio connection for AGN via measurements of the multi-dimensional luminosity function}",
      journal = {\mnras, submitted},
         year = 2025,
        month = jul,
          eid = {arXiv:2507.03085},
        pages = {arXiv:2507.03085},
          doi = {10.48550/arXiv.2507.03085},
archivePrefix = {arXiv},
       eprint = {2507.03085},
 primaryClass = {astro-ph.HE},
       adsurl = {https://ui.adsabs.harvard.edu/abs/2025arXiv250703085P}
}

@ARTICLE{Barrows2025,
       author = {{Barrows}, R. Scott and {Comerford}, Julia M.},
        title = "{Spatially Offset Active Galactic Nuclei in the Very Large Array Sky Survey: Tracers of Galaxy Mergers and Wandering Massive Black Holes}",
      journal = {\apj, in press},
         year = 2025,
        month = sep,
          eid = {arXiv:2509.09768},
        pages = {arXiv:2509.09768},
          doi = {10.48550/arXiv.2509.09768},
archivePrefix = {arXiv},
       eprint = {2509.09768},
 primaryClass = {astro-ph.GA},
       adsurl = {https://ui.adsabs.harvard.edu/abs/2025arXiv250909768B}
}

@ARTICLE{Kellermann2016,
       author = {{Kellermann}, K.~I. and {Condon}, J.~J. and {Kimball}, A.~E. and {Perley}, R.~A. and {Ivezi{\'c}}, {\v{Z}}eljko},
        title = "{Radio-loud and Radio-quiet QSOs}",
      journal = {\apj},
         year = 2016,
        month = nov,
       volume = {831},
       number = {2},
          eid = {168},
        pages = {168},
          doi = {10.3847/0004-637X/831/2/168},
archivePrefix = {arXiv},
       eprint = {1608.04586},
 primaryClass = {astro-ph.GA},
       adsurl = {https://ui.adsabs.harvard.edu/abs/2016ApJ...831..168K}
}

@ARTICLE{Best2005,
       author = {{Best}, P.~N. and {Kauffmann}, G. and {Heckman}, T.~M. and {Brinchmann}, J. and {Charlot}, S. and {Ivezi{\'c}}, {\v{Z}}. and {White}, S.~D.~M.},
        title = "{The host galaxies of radio-loud active galactic nuclei: mass dependences, gas cooling and active galactic nuclei feedback}",
      journal = {\mnras},
         year = 2005,
        month = sep,
       volume = {362},
       number = {1},
        pages = {25-40},
          doi = {10.1111/j.1365-2966.2005.09192.x},
archivePrefix = {arXiv},
       eprint = {astro-ph/0506269},
 primaryClass = {astro-ph},
       adsurl = {https://ui.adsabs.harvard.edu/abs/2005MNRAS.362...25B}
}

@ARTICLE{Persic2004,
       author = {{Persic}, M. and {Rephaeli}, Y. and {Braito}, V. and {Cappi}, M. and {Della Ceca}, R. and {Franceschini}, A. and {Gruber}, D.~E.},
        title = "{2-10 keV luminosity of high-mass binaries as a gauge of ongoing star-formation rate}",
      journal = {\aap},
         year = 2004,
        month = jun,
       volume = {419},
        pages = {849-862},
          doi = {10.1051/0004-6361:20034500},
archivePrefix = {arXiv},
       eprint = {astro-ph/0402568},
 primaryClass = {astro-ph},
       adsurl = {https://ui.adsabs.harvard.edu/abs/2004A&A...419..849P}
}

@ARTICLE{Colbert2004,
       author = {{Colbert}, Edward J.~M. and {Heckman}, Timothy M. and {Ptak}, Andrew F. and {Strickland}, David K. and {Weaver}, Kimberly A.},
        title = "{Old and Young X-Ray Point Source Populations in Nearby Galaxies}",
      journal = {\apj},
         year = 2004,
        month = feb,
       volume = {602},
       number = {1},
        pages = {231-248},
          doi = {10.1086/380899},
archivePrefix = {arXiv},
       eprint = {astro-ph/0305476},
 primaryClass = {astro-ph},
       adsurl = {https://ui.adsabs.harvard.edu/abs/2004ApJ...602..231C}
}

@ARTICLE{Adams2025,
       author = {{Adams}, Nathan J. and {Ferrami}, Giovanni and {Westcott}, Lewi and {Harvey}, Thomas and {Estrada-Carpenter}, Vicente and {Conselice}, Christopher J. and {Austin}, Duncan and {Wyithe}, J. Stuart B. and {Goolsby}, Caio M. and {Li}, Qiong and {Rusakov}, Vadim and {Windhorst}, Rogier A. and {Cohen}, Seth H. and {Jansen}, Rolf A. and {Summers}, Jake and {O'Brein}, Roselia and {Koekemoer}, Anton M. and {Driver}, Simon P. and {Frye}, Brenda and {Hathi}, Nimish P. and {Coe}, Dan and {Grogin}, Norman A. and {Marshall}, Madeline A. and {Pirzkal}, Nor and {Ryan}, Jr., Russell E. and {Willmer}, Christopher N.~A. and {Yan}, Haojing and {Holwerda}, Benne W. and {Kamieneski}, Patrick S. and {Broadhurst}, Tom and {Maksym}, W. Peter and {Saikia}, Payaswini and {Gelfand}, Joseph D.},
        title = "{JWSTs PEARLS: NIRCam imaging and NIRISS spectroscopy of a $z=3.6$ star-forming galaxy lensed into a near-Einstein Ring by a $z=1.258$ massive elliptical galaxy}",
      journal = {arXiv e-prints},
         year = 2025,
        month = apr,
          eid = {arXiv:2504.03571},
        pages = {arXiv:2504.03571},
          doi = {10.48550/arXiv.2504.03571},
archivePrefix = {arXiv},
       eprint = {2504.03571},
 primaryClass = {astro-ph.GA},
       adsurl = {https://ui.adsabs.harvard.edu/abs/2025arXiv250403571A}
}

@ARTICLE{Ferrami2025,
       author = {{Ferrami}, Giovanni and {Adams}, Nathan J. and {Westcott}, Lewi and {Harvey}, Thomas and {Jansen}, Rolf A. and {Diego}, Jose M. and {Estrada-Carpente}, Vince and {Windhorst}, Rogier A. and {Conselice}, Christopher J. and {Koekemoer}, Anton M. and {D'Silva}, Jordan C.~J. and {Willmer}, Christopher and {Wyithe}, J. Stuart B. and {Rutkowski}, Michael J. and {Cohen}, Seth H. and {Frye}, Brenda L. and {Grogin}, Norman A.},
        title = "{Galaxy-scale lens search in the PEARLS NEP TDF and CEERS JWST fields}",
      journal = {arXiv e-prints},
         year = 2025,
        month = may,
          eid = {arXiv:2505.17318},
        pages = {arXiv:2505.17318},
          doi = {10.48550/arXiv.2505.17318},
archivePrefix = {arXiv},
       eprint = {2505.17318},
 primaryClass = {astro-ph.GA},
       adsurl = {https://ui.adsabs.harvard.edu/abs/2025arXiv250517318F}
}

@ARTICLE{Kurtz1998,
       author = {{Kurtz}, Michael J. and {Mink}, Douglas J.},
        title = "{RVSAO 2.0: Digital Redshifts and Radial Velocities}",
      journal = {\pasp},
         year = 1998,
        month = aug,
       volume = {110},
       number = {750},
        pages = {934-977},
          doi = {10.1086/316207},
archivePrefix = {arXiv},
       eprint = {astro-ph/9803252},
 primaryClass = {astro-ph},
       adsurl = {https://ui.adsabs.harvard.edu/abs/1998PASP..110..934K}
}

@ARTICLE{CalistroRivera2016,
       author = {{Calistro Rivera}, Gabriela and {Lusso}, Elisabeta and {Hennawi}, Joseph F. and {Hogg}, David W.},
        title = "{AGNfitter: A Bayesian MCMC Approach to Fitting Spectral Energy Distributions of AGNs}",
      journal = {\apj},
         year = 2016,
        month = dec,
       volume = {833},
       number = {1},
          eid = {98},
        pages = {98},
          doi = {10.3847/1538-4357/833/1/98},
archivePrefix = {arXiv},
       eprint = {1606.05648},
 primaryClass = {astro-ph.GA},
       adsurl = {https://ui.adsabs.harvard.edu/abs/2016ApJ...833...98C}
}

@ARTICLE{Delhaize2017,
       author = {{Delhaize}, J. and {Smol{\v{c}}i{\'c}}, V. and {Delvecchio}, I. and {Novak}, M. and {Sargent}, M. and {Baran}, N. and {Magnelli}, B. and {Zamorani}, G. and {Schinnerer}, E. and {Murphy}, E.~J. and {Aravena}, M. and {Berta}, S. and {Bondi}, M. and {Capak}, P. and {Carilli}, C. and {Ciliegi}, P. and {Civano}, F. and {Ilbert}, O. and {Karim}, A. and {Laigle}, C. and {Le F{\`e}vre}, O. and {Marchesi}, S. and {McCracken}, H.~J. and {Salvato}, M. and {Seymour}, N. and {Tasca}, L.},
        title = "{The VLA-COSMOS 3 GHz Large Project: The infrared-radio correlation of star-forming galaxies and AGN to z {\ensuremath{\lesssim}} 6}",
      journal = {\aap},
         year = 2017,
        month = jun,
       volume = {602},
          eid = {A4},
        pages = {A4},
          doi = {10.1051/0004-6361/201629430},
archivePrefix = {arXiv},
       eprint = {1703.09723},
 primaryClass = {astro-ph.GA},
       adsurl = {https://ui.adsabs.harvard.edu/abs/2017A&A...602A...4D}
}

@ARTICLE{Whitaker2017,
       author = {{Whitaker}, Katherine E. and {Pope}, Alexandra and {Cybulski}, Ryan and {Casey}, Caitlin M. and {Popping}, Gerg{\"o} and {Yun}, Min S.},
        title = "{The Constant Average Relationship between Dust-obscured Star Formation and Stellar Mass from z = 0 to z = 2.5}",
      journal = {\apj},
         year = 2017,
        month = dec,
       volume = {850},
       number = {2},
          eid = {208},
        pages = {208},
          doi = {10.3847/1538-4357/aa94ce},
archivePrefix = {arXiv},
       eprint = {1710.06872},
 primaryClass = {astro-ph.GA},
       adsurl = {https://ui.adsabs.harvard.edu/abs/2017ApJ...850..208W}
}

@ARTICLE{Silver2025,
       author = {{Silver}, Ross and {Civano}, Francesca and {Zhao}, Xiurui and {Creech}, Samantha and {Willmer}, Christopher N.~A. and {Willner}, S.~P. and {Windhorst}, Rogier A. and {Yan}, Haojing and {Koekemoer}, Anton M. and {O'Brien}, Rosalia and {Ortiz}, III, Rafael and {Jansen}, Rolf A. and {Maksym}, W. Peter and {Cappelluti}, Nico and {Fornasini}, Francesca and {Carleton}, Timothy and {Cohen}, Seth H. and {Honor}, Rachel and {Summers}, Jake and {D'Silva}, Jordan C.~J. and {Laha}, Sibasish and {Coe}, Dan and {Conselice}, Christopher J. and {Diego}, Jose M. and {Driver}, Simon P. and {Frye}, Brenda and {Grogin}, Norman A. and {Marshall}, Madeline A. and {Pirzkal}, Nor and {Robotham}, Aaron and {Ryan}, Jr, Russell E.},
        title = "{PEARLS: NuSTAR and XMM-Newton Extragalactic Survey of the JWST North Ecliptic Pole Time-domain Field III}",
      journal = {\apj, in press},
         year = 2025,
        month = oct,
          eid = {arXiv:2510.24858},
        pages = {arXiv:2510.24858},
          doi = {10.48550/arXiv.2510.24858},
archivePrefix = {arXiv},
       eprint = {2510.24858},
 primaryClass = {astro-ph.HE},
       adsurl = {https://ui.adsabs.harvard.edu/abs/2025arXiv251024858S}
}

@ARTICLE{Peng2002,
       author = {{Peng}, Chien Y. and {Ho}, Luis C. and {Impey}, Chris D. and {Rix}, Hans-Walter},
        title = "{Detailed Structural Decomposition of Galaxy Images}",
      journal = {\aj},
         year = 2002,
        month = jul,
       volume = {124},
       number = {1},
        pages = {266-293},
          doi = {10.1086/340952},
archivePrefix = {arXiv},
       eprint = {astro-ph/0204182},
 primaryClass = {astro-ph},
       adsurl = {https://ui.adsabs.harvard.edu/abs/2002AJ....124..266P}
}

@ARTICLE{Peng2010,
       author = {{Peng}, Chien Y. and {Ho}, Luis C. and {Impey}, Chris D. and {Rix}, Hans-Walter},
        title = "{Detailed Decomposition of Galaxy Images. II. Beyond Axisymmetric Models}",
      journal = {\aj},
         year = 2010,
        month = jun,
       volume = {139},
       number = {6},
        pages = {2097-2129},
          doi = {10.1088/0004-6256/139/6/2097},
archivePrefix = {arXiv},
       eprint = {0912.0731},
 primaryClass = {astro-ph.CO},
       adsurl = {https://ui.adsabs.harvard.edu/abs/2010AJ....139.2097P}
}

@ARTICLE{Mineo2014,
       author = {{Mineo}, S. and {Gilfanov}, M. and {Lehmer}, B.~D. and {Morrison}, G.~E. and {Sunyaev}, R.},
        title = "{X-ray emission from star-forming galaxies - III. Calibration of the L$_{X}$-SFR relation up to redshift z {\ensuremath{\approx}} 1.3}",
      journal = {\mnras},
         year = 2014,
        month = jan,
       volume = {437},
       number = {2},
        pages = {1698-1707},
          doi = {10.1093/mnras/stt1999},
archivePrefix = {arXiv},
       eprint = {1207.2157},
 primaryClass = {astro-ph.HE},
       adsurl = {https://ui.adsabs.harvard.edu/abs/2014MNRAS.437.1698M}
}

@ARTICLE{Civano2016,
       author = {{Civano}, F. and {Marchesi}, S. and {Comastri}, A. and {Urry}, M.~C. and {Elvis}, M. and {Cappelluti}, N. and {Puccetti}, S. and {Brusa}, M. and {Zamorani}, G. and {Hasinger}, G. and {Aldcroft}, T. and {Alexander}, D.~M. and {Allevato}, V. and {Brunner}, H. and {Capak}, P. and {Finoguenov}, A. and {Fiore}, F. and {Fruscione}, A. and {Gilli}, R. and {Glotfelty}, K. and {Griffiths}, R.~E. and {Hao}, H. and {Harrison}, F.~A. and {Jahnke}, K. and {Kartaltepe}, J. and {Karim}, A. and {LaMassa}, S.~M. and {Lanzuisi}, G. and {Miyaji}, T. and {Ranalli}, P. and {Salvato}, M. and {Sargent}, M. and {Scoville}, N.~J. and {Schawinski}, K. and {Schinnerer}, E. and {Silverman}, J. and {Smolcic}, V. and {Stern}, D. and {Toft}, S. and {Trakhtenbrot}, B. and {Treister}, E. and {Vignali}, C.},
        title = "{The Chandra Cosmos Legacy Survey: Overview and Point Source Catalog}",
      journal = {\apj},
         year = 2016,
        month = mar,
       volume = {819},
       number = {1},
          eid = {62},
        pages = {62},
          doi = {10.3847/0004-637X/819/1/62},
archivePrefix = {arXiv},
       eprint = {1601.00941},
 primaryClass = {astro-ph.GA},
       adsurl = {https://ui.adsabs.harvard.edu/abs/2016ApJ...819...62C}
}

@ARTICLE{Summers2023,
       author = {{Summers}, Jake and {Windhorst}, Rogier A. and {Cohen}, Seth H. and {Jansen}, Rolf A. and {Carleton}, Timothy and {Kamieneski}, Patrick S. and {Holwerda}, Benne W. and {Conselice}, Christopher J. and {Adams}, Nathan J. and {Frye}, Brenda L. and {Diego}, Jose M. and {Willmer}, Christopher N.~A. and {Ortiz}, Rafael and {Cheng}, Cheng and {Pigarelli}, Alex and {Robotham}, Aaron and {D'Silva}, Jordan C.~J. and {Tompkins}, Scott and {Driver}, Simon P. and {Yan}, Haojing and {Coe}, Dan and {Grogin}, Norman and {Koekemoer}, Anton M. and {Marshall}, Madeline A. and {Pirzkal}, Nor and {Ryan}, Russell E.},
        title = "{Magellanic System Stars Identified in SMACS J0723.3-7327 James Webb Space Telescope Early Release Observations Images}",
      journal = {\apj},
         year = 2023,
        month = dec,
       volume = {958},
       number = {2},
          eid = {108},
        pages = {108},
          doi = {10.3847/1538-4357/acffb9},
archivePrefix = {arXiv},
       eprint = {2306.13037},
 primaryClass = {astro-ph.GA},
       adsurl = {https://ui.adsabs.harvard.edu/abs/2023ApJ...958..108S}
}

@ARTICLE{Hickox2009,
       author = {{Hickox}, Ryan C. and {Jones}, Christine and {Forman}, William R. and {Murray}, Stephen S. and {Kochanek}, Christopher S. and {Eisenstein}, Daniel and {Jannuzi}, Buell T. and {Dey}, Arjun and {Brown}, Michael J.~I. and {Stern}, Daniel and {Eisenhardt}, Peter R. and {Gorjian}, Varoujan and {Brodwin}, Mark and {Narayan}, Ramesh and {Cool}, Richard J. and {Kenter}, Almus and {Caldwell}, Nelson and {Anderson}, Michael E.},
        title = "{Host Galaxies, Clustering, Eddington Ratios, and Evolution of Radio, X-Ray, and Infrared-Selected AGNs}",
      journal = {\apj},
         year = 2009,
        month = may,
       volume = {696},
       number = {1},
        pages = {891-919},
          doi = {10.1088/0004-637X/696/1/891},
archivePrefix = {arXiv},
       eprint = {0901.4121},
 primaryClass = {astro-ph.GA},
       adsurl = {https://ui.adsabs.harvard.edu/abs/2009ApJ...696..891H}
}

@ARTICLE{Bonzini2012,
       author = {{Bonzini}, M. and {Mainieri}, V. and {Padovani}, P. and {Kellermann}, K.~I. and {Miller}, N. and {Rosati}, P. and {Tozzi}, P. and {Vattakunnel}, S. and {Balestra}, I. and {Brandt}, W.~N. and {Luo}, B. and {Xue}, Y.~Q.},
        title = "{The Sub-mJy Radio Population of the E-CDFS: Optical and Infrared Counterpart Identification}",
      journal = {\apjs},
         year = 2012,
        month = nov,
       volume = {203},
       number = {1},
          eid = {15},
        pages = {15},
          doi = {10.1088/0067-0049/203/1/15},
archivePrefix = {arXiv},
       eprint = {1209.4176},
 primaryClass = {astro-ph.CO},
       adsurl = {https://ui.adsabs.harvard.edu/abs/2012ApJS..203...15B}
}

@ARTICLE{Willmer2023,
       author = {{Willmer}, Christopher N.~A. and {Ly}, Chun and {Kikuta}, Satoshi and {Kattner}, S.~A. and {Jansen}, Rolf A. and {Cohen}, Seth H. and {Windhorst}, Rogier A. and {Smail}, Ian and {Tompkins}, Scott and {Beacom}, John F. and {Cheng}, Cheng and {Conselice}, Christopher J. and {Frye}, Brenda L. and {Koekemoer}, Anton M. and {Hathi}, Nimish and {Hyun}, Minhee and {Im}, Myungshin and {Willner}, S.~P. and {Zhao}, X. and {Brisken}, Walter A. and {Civano}, F. and {Cotton}, William and {Hasinger}, G{\"u}nther and {Maksym}, W. Peter and {Rieke}, Marcia J. and {Grogin}, Norman A.},
        title = "{PEARLS: Near-infrared Photometry in the JWST North Ecliptic Pole Time Domain Field}",
      journal = {\apjs},
         year = 2023,
        month = nov,
       volume = {269},
       number = {1},
          eid = {21},
        pages = {21},
          doi = {10.3847/1538-4365/acf57d},
archivePrefix = {arXiv},
       eprint = {2309.00031},
 primaryClass = {astro-ph.GA},
       adsurl = {https://ui.adsabs.harvard.edu/abs/2023ApJS..269...21W}
}
\bibliographystyle{aasjournal}



\renewcommand{\thetable}{\arabic{table}}
\setcounter{table}{0}
\startlongtable
\begin{deluxetable*}{rcccrrrlccrrlc}
\tablecaption{Radio counterpart identifications}
\label{t:ID}
\tabletypesize{\scriptsize}
\tablewidth{0pt}
\tablehead{
\colhead{ID}&
\multicolumn{2}{c}{3~GHz position} & 
\colhead{$S(3~GHz$)}&
\multicolumn{2}{c}{F444W position}&
\colhead{sep}&
\colhead{flags}&
\colhead{$m(\rm F444W)$}&
\colhead{$a$}&
\colhead{$b$}&
\colhead{PA}&
\colhead{best $z$}&
\colhead{$Q$}\\
& \colhead{RA}&
\colhead{Decl}&
\colhead{\mmJy} &
\colhead{RA}&
\colhead{Decl}&
\colhead{\arcsec}& 
&
\colhead{AB}&
\colhead{\arcsec}&
\colhead{\arcsec}&
\multicolumn{1}{c}{\arcdeg}&
}
\startdata
25&260.460836&65.838044&15.0&260.460899&65.838019&0.13&&22.20&0.32&0.24&120&2.28& \\
31&260.473735&65.829625&17.5&260.473841&65.829639&0.16&&22.22&0.17&0.15&57&1.02& \\
38&260.489397&65.835872&26.8&260.489408&65.835885&0.05&&21.61&0.34&0.29&54&1.77& \\
56&260.524069&65.825378&10.6&\no&\no&&i&\no&\no&\no&&\no& \\
60&260.528619&65.833651&18.0&260.528546&65.833640&0.12&ceT&19.56&0.43&0.37&27&1.39& \\
63&260.531782&65.815637&23.1&260.531746&65.815628&0.06&r&19.80&0.33&0.27&119&1.07& \\
64&260.536668&65.805224&11.3&260.536771&65.805282&0.26&cr&20.62&0.31&0.28&34&0.98& \\
74&260.562709&65.810879&14.0&260.562707&65.810898&0.07&&19.44&0.95&0.38&16&0.0796&4 \\
76&260.567194&65.822036&21.2&260.567161&65.822031&0.05&c&16.17&2.82&1.62&113&0.1037&3 \\
81&260.573670&65.850682&12.4&260.573530&65.850727&0.26&b&\no&\no&\no&\no&\no& \\
89&260.579095&65.822131&7.6&260.579092&65.822189&0.21&&22.18&0.31&0.23&28&3.51& \\
99&260.591638&65.844431&10.6&260.591816&65.844351&0.39&bt&\no&\no&\no&\no&0.7115&3 \\
109&260.608503&65.799874&8.8&260.608444&65.799848&0.13&t&20.82&0.37&0.20&54&0.94& \\
112&260.613972&65.805143&10.7&260.614040&65.805183&0.18&T&20.13&0.31&0.27&108&1.2& \\
114&260.616616&65.841301&11.9&260.616475&65.841317&0.22&&20.22&0.31&0.27&91&0.84& \\
115&260.616635&65.820108&12.6&260.616639&65.820138&0.11&&21.17&0.29&0.25&159&1.01& \\
117&260.623537&65.802403&21.6&260.623543&65.802418&0.05&c&20.12&0.43&0.34&92&1.60& \\
118&260.623603&65.837770&18.3&260.623615&65.837779&0.04&&20.48&0.31&0.21&0&1.56& \\
119&260.623828&65.869103&11.7&260.623883&65.869105&0.08&&19.50&0.49&0.39&165&1.07& \\
120&260.624178&65.837675&14.1&260.624156&65.837665&0.05&u&21.24&0.43&0.29&94&1.33& \\
121&260.624289&65.867888&97.7&260.624277&65.867883&0.03&crt&18.73&0.47&0.39&35&1.0172&4 \\
123&260.625902&65.876446&13.5&260.625867&65.876428&0.08&&19.58&0.78&0.55&64&0.8070&3 \\
124&260.626563&65.852201&472.0&260.626563&65.852200&0.00&rv&19.65&0.55&0.46&170&1.00& \\
125&260.626782&65.783912&10.2&260.626793&65.783908&0.02&&21.87&0.38&0.15&79&0.8599&4 \\
127&260.627065&65.791766&18.6&260.627091&65.791766&0.04&u&19.32&0.51&0.29&117&1.01& \\
130&260.630059&65.854053&37.5&260.630038&65.854051&0.03&cu&20.33&0.61&0.43&46&0.7126&4 \\
131&260.631160&65.822183&5.9&260.631081&65.822170&0.12&ej&22.60&0.20&0.14&84&2.996&3 \\
134&260.635045&65.861177&340.0&260.635067&65.861172&0.04&ctu&21.88&0.18&0.17&153&1.78& \\
135&260.635530&65.829730&6.8&260.635528&65.829714&0.06&jt&21.63&0.27&0.17&92&1.875&3 \\
136&260.635817&65.844471&10.0&260.635681&65.844511&0.25&&21.48&0.37&0.27&3&2.37& \\
137&260.636041&65.786557&6.1&260.636039&65.786619&0.22&t&21.23&0.25&0.17&97&1.09& \\
138&260.636382&65.794383&8.9&260.636337&65.794381&0.07&cTx&20.82&0.26&0.21&130&0.82& \\
141&260.639141&65.799377&1071.0&260.639145&65.799392&0.05&crtv&18.79&0.65&0.58&156&1.0175&4 \\
142&260.639584&65.844964&36.2&260.639570&65.844964&0.02&chrtuxz&18.16&0.48&0.41&16&1.1476&4 \\
145&260.639970&65.785139&8.2&260.639935&65.785171&0.13&c&23.40&0.18&0.14&83&2.09& \\
147&260.641242&65.838425&7.8&260.641441&65.838464&0.33&&21.78&0.28&0.19&17&1.46& \\
148&260.641898&65.821400&8.8&260.641922&65.821360&0.15&&20.50&0.31&0.19&165&1.43& \\
150&260.643468&65.855116&112.0&260.643458&65.855119&0.02&tu&20.72&0.54&0.36&58&3.29& \\
153&260.644973&65.864442&11.2&260.644964&65.864457&0.05&&20.90&0.33&0.25&174&0.65& \\
154&260.645290&65.798054&17.0&260.645261&65.798069&0.07&c&20.65&0.38&0.22&40&2.45& \\
156&260.648098&65.887050&12.1&260.648039&65.887046&0.09&c&22.60&0.20&0.15&139&3.48& \\
157&260.648132&65.781641&10.3&260.648195&65.781547&0.35&&23.97&0.18&0.17&161&4.7& \\
158&260.648271&65.906885&22.7&260.648261&65.906895&0.04&t&19.89&0.61&0.30&36&0.76& \\
159&260.648600&65.797384&22.4&260.648570&65.797391&0.05&Tu&19.25&0.79&0.35&62&0.2932&4 \\
160&260.649464&65.842763&16.0&260.649563&65.842771&0.15&&20.68&0.60&0.26&47&1.68& \\
162&260.651753&65.795567&17.6&260.651768&65.795567&0.02&u&20.10&0.66&0.26&142&1.18& \\
163&260.651878&65.820542&6.7&260.651805&65.820561&0.13&c&22.84&0.40&0.24&14&1.27& \\
168&260.656028&65.893823&27.9&260.656011&65.893817&0.03&jt&21.64&0.32&0.19&176&1.730&3 \\
170&260.658132&65.814724&7.7&260.658125&65.814713&0.04&ou&21.17&0.27&0.23&112&1.38& \\
172&260.658604&65.893287&22.5&260.658602&65.893290&0.01&&22.20&0.29&0.22&49&2.49& \\
173&260.658643&65.849813&52.8&260.658674&65.849818&0.05&u&18.41&0.92&0.81&136&0.4972&4 \\
175&260.659842&65.822095&14.2&260.659814&65.822090&0.04&ct&20.51&0.25&0.20&48&1.09& \\
176&260.659965&65.878809&12.5&260.659931&65.878795&0.07&&22.32&0.20&0.16&30&1.55& \\
178&260.660215&65.789579&13.9&260.660373&65.789585&0.23&st&19.95&0.58&0.47&97&1.1288&3 \\
180&260.661146&65.787400&9.4&260.661175&65.787406&0.05&&20.28&0.39&0.24&1&0.84& \\
182&260.662389&65.861950&74.3&260.662400&65.861952&0.02&v&19.53&0.83&0.49&136&0.95& \\
187&260.668045&65.877821&32.9&260.668013&65.877820&0.05&r&20.04&0.34&0.30&177&1.41& \\
188&260.669085&65.887495&11.6&260.669113&65.887485&0.05&&21.16&0.29&0.25&157&1.2& \\
192&260.671320&65.803513&18.2&260.671333&65.803492&0.08&ct&20.00&0.60&0.51&104&0.7105&3 \\
193&260.671558&65.710706&21.4&\no&\no&&io&\no&\no&\no&\no&\no& \\
194&260.671570&65.711591&81.4&260.671561&65.711597&0.03&cksRTxy&17.38&1.06&0.66&62&0.1774&4 \\
195&260.672848&65.871072&7.0&260.672828&65.871095&0.09&&27.64&0.08&0.05&104&0.2& \\
197&260.674708&65.739125&32.5&260.674703&65.739121&0.02&u&21.83&0.37&0.18&173&1.91& \\
200&260.678447&65.837277&20.4&260.678480&65.837261&0.07&chu&20.36&0.40&0.39&1&0.5653&4 \\
201&260.678508&65.768986&50.4&260.678584&65.768945&0.19&etux&19.10&0.61&0.53&151&1.1008&4 \\
202&260.678780&65.878008&25.1&260.678876&65.877970&0.20&o&19.76&0.62&0.43&165&1.1299&3 \\
203&260.679007&65.741979&13.3&260.679011&65.741974&0.02&&25.94&0.06&0.05&139&2.3& \\
204&260.679358&65.773721&13.5&260.679391&65.773724&0.05&et&20.44&0.30&0.21&76&1.62& \\
205&260.679861&65.844657&19.3&260.679858&65.844672&0.05&crt&19.70&0.50&0.42&9&0.89& \\
206&260.680350&65.846506&14.8&260.680224&65.846467&0.23&&20.40&0.46&0.24&29&0.98& \\
207&260.681048&65.869282&64.7&260.681060&65.869290&0.03&rtu&18.97&0.38&0.37&3&3.11& \\
210&260.684936&65.857711&29.9&260.684921&65.857710&0.02&&19.65&0.50&0.32&71&0.7837&3 \\
212&260.686613&65.798259&15.7&260.686600&65.798268&0.04&&23.23&0.25&0.20&152&4.14& \\
213&260.686950&65.722979&27.5&260.686855&65.722985&0.14&o&20.36&0.49&0.42&69&2.2& \\
214&260.687140&65.784347&9.0&260.687112&65.784358&0.06&&20.33&0.34&0.23&161&0.64& \\
215&260.688413&65.869241&13.4&260.688419&65.869229&0.04&j&21.83&0.23&0.17&35&1.954&3 \\
216&260.688836&65.869135&13.0&260.688894&65.869117&0.11&c&22.06&0.41&0.25&36&1.82& \\
218&260.690829&65.738099&29.9&260.690856&65.738083&0.07&crtux&20.10&0.32&0.28&32&1.26& \\
219&260.691058&65.751806&15.2&260.691052&65.751783&0.09&tu&19.59&0.46&0.41&33&1.21& \\
220&260.691565&65.752944&8.2&260.691542&65.752929&0.06&ct&20.59&0.42&0.34&96&2.19& \\
222&260.692870&65.861899&40.8&260.692875&65.861907&0.03&cruz&18.24&1.00&0.39&77&0.2925&4 \\
223&260.695592&65.784323&13.0&260.695567&65.784333&0.05&cRT&20.91&0.28&0.21&13&0.9& \\
225&260.696971&65.906709&20.2&260.696979&65.906725&0.06&&21.53&0.35&0.27&75&3.5& \\
226&260.697570&65.712623&26.9&260.697579&65.712616&0.03&&21.07&0.27&0.20&65&0.8& \\
229&260.700068&65.936547&28.4&260.700063&65.936570&0.08&&21.11&0.28&0.23&166&2.12& \\
231&260.700998&65.896774&12.6&260.700998&65.896782&0.03&c&20.05&0.58&0.43&48&1.0165&4 \\
232&260.701244&65.776170&50.3&260.701253&65.776179&0.04&cRTu&19.91&0.28&0.24&26&1.05& \\
233&260.701674&65.881397&14.2&260.701636&65.881383&0.08&o&21.93&0.24&0.22&17&1.38& \\
234&260.702080&65.866715&21.9&260.702045&65.866719&0.05&&22.58&0.34&0.22&139&4.51& \\
236&260.702588&65.710023&8.3&260.702531&65.710053&0.14&t&20.73&0.26&0.24&139&0.87& \\
237&260.704074&65.900344&50.7&260.704063&65.900346&0.02&&19.65&1.02&0.59&36&0.8210&3 \\
240&260.707135&65.859031&21.4&260.707052&65.859048&0.14&t&20.04&0.44&0.31&75&0.6558&4 \\
241&260.708168&65.732152&8.9&260.708116&65.732174&0.11&c&21.18&0.22&0.20&154&1.11& \\
242&260.709688&65.838421&17.3&260.709713&65.838435&0.06&cT&19.42&0.52&0.29&50&1.45& \\
244&260.711641&65.809166&5.5&260.711710&65.809134&0.15&j&22.45&0.25&0.15&25&2.164&3 \\
245&260.712947&65.785691&7.1&260.712898&65.785704&0.09&&21.09&0.41&0.21&94&0.9174&3 \\
246&260.714532&65.753392&69.0&260.714500&65.753358&0.13&ctu&18.86&0.55&0.44&64&0.5375&4 \\
247&260.715248&65.883939&11.9&260.715095&65.883942&0.23&cjt&20.04&0.39&0.35&28&1.958&3 \\
248&260.715468&65.742532&11.1&260.715289&65.742528&0.26&osu&19.99&0.35&0.29&132&0.9156&3 \\
251&260.716339&65.711697&22.8&260.716345&65.711694&0.01&&21.33&0.29&0.17&46&1.62& \\
253&260.717989&65.937267&25.5&260.718039&65.937266&0.07&&18.86&0.61&0.59&9&0.6338&4 \\
256&260.719510&65.743311&20.2&260.719517&65.743331&0.07&tu&20.30&0.29&0.22&1&1.64& \\
257&260.719954&65.904871&24.3&260.719962&65.904872&0.01&&21.54&0.41&0.26&13&2.68& \\
258&260.719997&65.763175&15.8&260.720021&65.763189&0.06&u&20.51&0.27&0.20&52&0.6558&4 \\
260&260.721446&65.813161&57.4&260.721442&65.813167&0.02&tv&18.98&1.51&0.74&104&0.5445&4 \\
267&260.725968&65.795907&13.2&260.725955&65.795908&0.02&&21.85&0.28&0.16&53&1.55& \\
268&260.726124&65.812468&18.5&260.726281&65.812469&0.23&os&21.25&0.16&0.09&68&0.54& \\
269&260.727932&65.906089&24.7&260.727928&65.906092&0.01&&22.95&0.29&0.19&0&1.48& \\
270&260.728265&65.708595&24.1&260.728218&65.708632&&aemw&\no&0.60&0.20&12&\no& \\
271&260.728365&65.801743&17.1&260.728332&65.801759&0.07&ctx&19.75&0.49&0.41&159&0.4558&4 \\
276&260.731090&65.751261&7.5&260.731112&65.751259&0.03&&23.61&0.17&0.16&15&5.48& \\
278&260.731338&65.810081&6.8&260.731465&65.810040&0.24&&23.38&0.27&0.15&34&1.8& \\
280&260.731923&65.730233&11.0&260.731891&65.730253&0.09&&20.81&0.37&0.18&153&0.97& \\
283&260.733994&65.798503&55.3&260.733993&65.798503&0.00&crtu&18.57&0.67&0.48&26&0.3278&4 \\
284&260.734636&65.849219&6.6&260.734702&65.849225&0.10&j&22.89&0.22&0.15&103&2.270&3 \\
285&260.734793&65.770993&8.9&260.734896&65.770983&0.16&cjT&22.26&0.19&0.17&159&2.174&3 \\
287&260.736498&65.779211&4.7&\no&\no&&i&\no&\no&\no&\no&\no& \\
288&260.737386&65.844827&7.9&260.737354&65.844847&0.09&co&21.31&0.24&0.10&176&1.24& \\
289&260.737916&65.773786&13.5&260.737928&65.773799&0.05&cjtu&21.52&0.38&0.28&146&1.849&3 \\
291&260.739516&65.770729&6.1&260.739391&65.770695&0.22&&20.75&0.48&0.29&77&0.76& \\
292&260.739942&65.925135&26.9&260.739970&65.925099&0.14&r&19.99&0.43&0.25&140&0.7& \\
293&260.739960&65.793233&5.4&260.739947&65.793227&0.03&&24.08&0.17&0.14&20&2.45& \\
295&260.742757&65.906624&12.7&260.742720&65.906661&&ac&\no&0.30&0.28&170&\no& \\
297&260.745008&65.798769&6.9&260.744969&65.798748&0.10&&21.08&0.33&0.26&58&0.3276&3 \\
298&260.745197&65.792766&12.5&260.745023&65.792809&0.30&&21.40&0.38&0.28&36&1.05& \\
299&260.745602&65.842277&12.2&260.745631&65.842307&0.12&rt&19.45&0.49&0.29&170&0.94& \\
300&260.746309&65.784217&18.1&260.746296&65.784221&0.02&cRTu&20.84&0.21&0.18&22&2.51& \\
302&260.747308&65.755821&13.6&260.747339&65.755786&0.13&u&20.99&0.37&0.28&6&2.29& \\
305&260.748780&65.783504&46.7&260.748774&65.783502&0.01&u&21.39&0.23&0.18&30&2.38& \\
306&260.750438&65.770480&17.0&260.750474&65.770475&0.06&cou&22.57&0.14&0.11&78&2.11& \\
307&260.750781&65.721062&10.2&260.750749&65.721097&0.13&&22.18&0.21&0.16&7&2.39& \\
308&260.751498&65.735379&8.1&260.751427&65.735376&0.11&crtz&19.75&0.57&0.28&29&0.1391&4 \\
310&260.752712&65.826804&11.4&260.752727&65.826824&0.08&c&23.05&0.17&0.15&152&4.62& \\
312&260.754829&65.845603&16.8&260.754848&65.845590&0.06&cu&19.67&0.57&0.36&157&0.9167&4 \\
313&260.755657&65.721894&19.8&260.755598&65.721885&0.09&&21.52&0.26&0.16&106&0.98& \\
314&260.756213&65.713171&324.0&260.756207&65.713177&0.02&RTu&20.38&0.23&0.18&162&1.55& \\
315&260.757048&65.817354&9.2&260.757015&65.817366&0.07&ct&20.98&0.34&0.30&74&1.08& \\
317&260.757885&65.861544&34.2&260.757865&65.861567&0.09&ou&20.65&0.67&0.22&92&2.26& \\
318&260.758054&65.860451&9.6&260.758268&65.860462&0.32&cjT&22.61&0.20&0.16&98&1.963&3 \\
319&260.758640&65.800806&103.0&260.758637&65.800812&0.02&chrux&18.11&0.77&0.56&33&0.1787&4 \\
320&260.758650&65.838235&30.8&260.758646&65.838248&0.05&tu&20.35&0.43&0.33&158&1.4267&3 \\
325&260.760204&65.822150&8.3&260.760223&65.822161&0.05&&22.55&0.23&0.14&7&2.04& \\
326&260.760514&65.934074&53.0&260.760532&65.934082&0.04&t&20.79&0.28&0.25&78&1.87& \\
327&260.760675&65.804321&11.7&260.760661&65.804280&0.15&&20.82&0.38&0.19&32&1.32& \\
328&260.761803&65.819778&6.3&260.761840&65.819757&0.09&&21.68&0.25&0.20&3&0.7072&4 \\
329&260.762456&65.815098&5.8&260.762586&65.815154&0.28&&21.09&0.31&0.21&119&0.87& \\
336&260.767063&65.757148&8.1&260.767019&65.757150&0.07&c&22.60&0.17&0.13&9&1.86& \\
337&260.767780&65.816263&33.2&260.767750&65.816265&0.04&cu&21.16&0.39&0.22&66&2.34& \\
338&260.767932&65.779383&19.1&260.767932&65.779389&0.02&ctux&20.61&0.31&0.25&174&1.14& \\
342&260.771767&65.808253&9.9&260.771740&65.808258&0.04&&22.08&0.28&0.23&160&1.9& \\
348&260.776623&65.855291&20.3&260.776598&65.855294&0.04&cjTx&20.50&0.39&0.24&122&0.8861&4 \\
349&260.776777&65.813659&17.2&260.776780&65.813656&0.01&c&20.27&0.45&0.32&123&1.59& \\
352&260.779090&65.844935&8.7&260.779074&65.844984&0.18&&21.82&0.27&0.19&119&0.78& \\
355&260.783195&65.801419&8.1&260.783184&65.801390&0.10&ct&21.04&0.24&0.22&79&1.78& \\
356&260.785638&65.827293&6.9&260.785665&65.827274&0.08&T&20.78&0.25&0.25&115&1.57& \\
363&260.789215&65.802433&17.1&260.789215&65.802429&0.02&chTxz&19.93&0.44&0.39&58&1.6019&4 \\
364&260.790754&65.820649&9.4&260.790783&65.820644&&acux&\no&0.20&0.15&159&\no& \\
372&260.793575&65.801777&14.3&260.793537&65.801792&0.08&tu&21.07&0.57&0.39&69&1.05& \\
373&260.793829&65.795339&19.4&260.793762&65.795351&0.11&ct&19.28&0.58&0.31&162&0.88& \\
377&260.794800&65.801461&9.6&260.794881&65.801450&0.13&jt&21.31&0.35&0.25&54&1.120&3 \\
378&260.794980&65.817308&7.8&260.794958&65.817251&0.21&j&22.84&0.18&0.17&165&2.724&3 \\
382&260.799718&65.837316&100.3&260.799715&65.837317&0.01&v&17.96&1.68&0.97&18&0.3741&4 \\
383&260.802214&65.857679&9.1&260.802273&65.857713&0.15&&20.16&0.52&0.25&23&0.5435&3 \\
387&260.805730&65.853684&9.6&260.805747&65.853694&0.04&&20.94&0.25&0.22&85&1.1028&3 \\
394&260.809765&65.820758&11.1&260.809765&65.820742&0.06&jo&20.86&0.40&0.32&59&1.747&3 \\
400&260.820081&65.845363&23.2&260.820071&65.845341&0.08&&21.74&0.41&0.26&174&1.83& \\
401&260.820084&65.847544&7.1&260.820106&65.847549&0.04&j&21.86&0.23&0.19&163&2.016&3 \\
403&260.820588&65.845965&181.0&260.820591&65.845958&0.03&v&21.41&0.36&0.25&172&0.19& \\
417&260.828859&65.834456&10.0&260.828839&65.834482&0.10&t&20.18&0.52&0.37&167&0.88& \\
420&260.830446&65.835317&14.7&260.830415&65.835298&0.08&c&21.27&0.30&0.23&152&0.66& \\
422&260.831667&65.818214&6.0&260.831664&65.818217&0.01&&20.79&0.53&0.33&173&0.9187&3 \\
426&260.836190&65.826451&6.2&260.836132&65.826464&0.10&&21.48&0.40&0.20&165&1.54& \\
427&260.836595&65.838006&8.7&260.836616&65.838009&0.03&&21.21&0.43&0.24&179&2.4& \\
429&260.838505&65.837650&6.7&260.838517&65.837672&0.08&&20.93&0.40&0.29&98&0.8900&4 \\
431&260.838691&65.819208&11.3&260.838663&65.819222&0.07&T&20.33&0.62&0.19&86&0.2933&4 \\
432&260.838770&65.822131&4.9&260.838670&65.822109&0.17&co&22.22&0.17&0.16&129&1.9& \\
434&260.843775&65.829343&50.5&260.843785&65.829340&0.02&v&20.12&0.66&0.41&86&1.18& \\
445&260.849590&65.825843&19.9&260.849553&65.825853&0.08&mosu&21.93&0.30&0.15&169&3.5& \\
447&260.850789&65.853059&13.7&260.850771&65.853042&&aeT&\no&0.27&0.21&143&\no& \\
458&260.860454&65.815023&5.1&\no&\no&&i&\no&\no&\no&\no&\no& \\
466&260.868850&65.814154&36.1&260.868860&65.814176&0.08&u&21.60&0.22&0.19&103&2.5& \\
470&260.872968&65.813148&32.7&260.872974&65.813145&0.01&cux&21.67&0.24&0.22&88&1.74& \\
472&260.875564&65.843196&7.0&260.875643&65.843188&0.12&t&20.53&0.66&0.42&91&0.81& \\
480&260.883723&65.827424&7.5&260.883655&65.827449&0.13&&20.35&0.41&0.36&147&0.95& \\
489&260.893204&65.824861&15.1&260.893215&65.824841&0.07&&19.90&0.52&0.22&53&0.74& \\
492&260.894275&65.841678&26.2&260.894295&65.841671&0.04&rt&18.88&0.68&0.44&135&0.1395&4 \\
493&260.894310&65.836465&27.8&260.894239&65.836486&0.13&cu&19.01&0.51&0.36&130&0.8504&3 \\
497&260.896708&65.817467&30.3&260.896727&65.817457&0.05&cRTuxz&18.31&0.31&0.26&74&0.7807&4 \\
499&260.897831&65.827950&24.5&260.897839&65.827962&0.05&ctuz&19.70&0.45&0.39&144&0.6739&4 \\
500&260.897952&65.807336&16.0&260.897985&65.807342&&aTz&\no&0.31&0.28&45&0.4453&4 \\
501&260.898331&65.815979&46.6&260.898345&65.815980&0.02&otu&19.80&0.39&0.27&0&0.8262&3 \\
502&260.898508&65.827138&41.8&260.898518&65.827140&0.02&u&19.81&0.58&0.48&73&0.68& \\
505&260.901148&65.846463&57.0&260.901157&65.846467&0.02&u&20.74&0.30&0.27&59&1.6& \\
512&260.908244&65.840239&5.7&\no&\no&&i&\no&\no&\no&\no&\no& \\
515&260.910619&65.839162&7.7&260.910700&65.839166&0.12&&20.76&0.36&0.28&119&0.72& \\
517&260.912756&65.831912&20.7&260.912797&65.831908&0.06&&19.98&0.48&0.35&102&1.41& \\
527&260.918772&65.816556&9.9&260.918820&65.816556&0.07&cu&20.84&0.27&0.20&118&0.86& \\
528&260.919455&65.831340&512.0&260.919471&65.831340&0.02&RTv&19.75&0.37&0.36&47&0.3760&4 \\
531&260.920856&65.805691&16.0&260.920863&65.805713&0.08&&20.54&0.33&0.20&155&1.56& \\
533&260.921434&65.811211&15.4&260.921314&65.811214&0.18&u&19.96&0.49&0.46&143&0.1745&4 \\
542&260.927551&65.826319&54.4&260.927559&65.826317&0.01&tu&23.09&0.15&0.13&82&2.19& \\
544&260.930487&65.842785&24.4&260.930511&65.842795&&aeu&\no&0.52&0.18&16&\no& \\
547&260.933532&65.810916&43.8&260.933566&65.810926&0.06&u&21.35&0.31&0.22&15&2.54& \\
548&260.933814&65.812938&8.0&260.933861&65.812939&0.07&&21.65&0.24&0.20&161&1.5& \\
549&260.935288&65.821329&18.9&260.935323&65.821334&0.05&t&18.47&1.12&0.87&111&0.7544&4 \\
552&260.938186&65.826799&14.4&260.938251&65.826787&0.10&u&20.74&0.32&0.19&148&0.9249&3 \\
559&260.945340&65.819396&14.2&260.945347&65.819393&0.01&&20.05&0.75&0.27&15&0.8& \\
564&260.951667&65.834585&7.2&260.951579&65.834555&0.17&&21.27&0.49&0.19&4&1.59& \\
566&260.954516&65.815456&7.1&260.954501&65.815421&0.13&&20.86&0.33&0.21&119&1.05& \\
567&260.954822&65.821479&16.3&260.954839&65.821476&0.03&&20.38&0.51&0.40&64&1.57& \\
572&260.959373&65.822064&12.1&260.959461&65.822057&0.13&&20.89&0.39&0.27&142&2.69& \\
593&260.981425&65.810748&127.0&260.981447&65.810752&0.04&eu&19.00&0.47&0.30&31&0.8926&4 \\
596&260.985889&65.823470&9.1&260.986060&65.823470&0.25&&20.10&0.54&0.37&13&1.90& \\

\enddata
\tablecomments{Table~\ref{t:ID} is available as a machine-readable
  table (MRT), datafile1.txt in the source directory.   Source sizes $a$ and $b$ are NIRCam semi-major and semi-minor axis dimensions in arcseconds as determined by SExtractor, and PA is in degrees east from north.  The last column ``Q'' is the quality rating for spectroscopic redshifts.  If missing, redshift is photometric based on available VLA, JWST, and HST data. The spectroscopic redshift quality code ($Q$) follows the same criteria adopted by the DEEP2 survey \citep{Newman2013}: $Q=4$ indicates a secure redshift, while $Q=3$ indicates a redshift with a 90\% or higher certainty of being correct.
The flag meanings are as follows:\\
a: the source is outside the LW images (preventing LW photometry) but visible in the SW images. The source position, dimensions, and separation from the radio position are from F200W, not the initial automated search on F444W.\\
b: NIRISS F200W source outside NIRCam FoV: position is NIRISS F200W.\\
c: Chandra X-ray source within 1\arcsec.\\
e: the source is near an image edge, and photometry may be inaccurate.\\
h: spectroscopic redshift from Hectospec.\\
i: no source visible with NIRCam.\\
j: NIRISS redshift or confirmation.\\
k: F356W, F410W, F444W flux densities are lower limits because of saturation.\\
m: position and separation measured manually, source dimensions estimated.\\
o: overlapping or confused sources or diffraction spike may affect photometry in some bands.\\
rR: pointlike nucleus per \citet{Ortiz2024}; capitalized if \galfit\ prefers an unresolved source.\\
tT: pointlike nucleus found independently of \citet{Ortiz2024}; capitalized if \galfit\ prefers an unresolved source.\\
u: source observed by the VLBA but undetected.\\
v: VLBA source detected.\\
w: wrong source found by automated search.\\
x: XMM X-ray source within 2\arcsec.\\
y: NuSTAR X-ray source within 2\arcsec.\\
z: AGN spectrum per \citet{Lamareille2010} criteria or broad emission lines.}
\end{deluxetable*}

\clearpage
\setcounter{table}{2}
\startlongtable
\begin{deluxetable*}{rRRrrrrrcccc@{~}l}
\tablecaption{Galaxy properties derived from SED fits}
\label{t:SEDfits}
\tabletypesize{\small}
\tablewidth{0pt}
\tablehead{
\colhead{ID}&
\multicolumn{1}{c}{$z$}&
\multicolumn{1}{c}{$\delta z$}&
\multicolumn{1}{c}{$\log M_*$}&
\multicolumn{1}{c}{$\delta \log M_*$}&
\multicolumn{1}{c}{$\log \rm SFR$}&
\multicolumn{1}{c}{$\delta \log \rm SFR$}&
\multicolumn{1}{c}{$A_V$}&
\multicolumn{1}{c}{$\delta A_V$}&
\multicolumn{1}{c}{$f_{\rm AGN}$}&
\multicolumn{1}{c}{$\delta f_{\rm AGN}$}&
\multicolumn{2}{l}{SED class}
}
\startdata
025&2.28&0.31&10.16&0.14&1.43&0.38&0.58&0.24&0.44&0.21&RQ&GAL \\
031&1.02&0.36&9.39&0.45&0.59&0.71&2.27&0.72&0.63&0.23&RL&AGN \\
038&1.77&0.16&10.36&0.09&1.01&0.28&0.66&0.29&0.34&0.23&RL&GAL \\
060&1.39&0.17&10.97&0.11&1.67&0.33&0.82&0.20&0.24&0.23&RQ&GAL \\
063&1.07&0.14&10.88&0.14&1.50&0.17&1.47&0.38&0.18&0.19&RQ&GAL \\
064&0.98&0.12&10.64&0.15&0.31&0.77&0.69&0.41&0.26&0.17&RL&GAL \\
074&0.08&\no&9.23&0.03&$-$0.98&0.06&0.14&0.00&0.05&0.05&RQ&GAL \\
076&0.10&\no&10.88&0.06&$-$1.07&0.30&0.14&0.00&0.06&0.05&RL&GAL \\
089&3.51&0.86&10.32&0.16&1.89&0.21&0.55&0.27&0.46&0.24&RQ&AGN \\
109&0.94&0.49&10.04&0.53&1.23&0.62&2.00&0.76&0.38&0.21&RQ&AGN \\
112&1.20&0.32&10.65&0.35&1.36&0.50&1.98&0.77&0.53&0.26&RQ&AGN \\
114&0.84&0.39&10.28&0.59&1.36&0.72&2.52&0.51&0.37&0.21&RQ&AGN \\
115&1.01&0.58&10.11&0.31&1.28&0.66&0.78&0.65&0.45&0.24&RQ&GAL \\
117&1.60&0.08&10.70&0.15&2.20&0.16&1.29&0.19&0.15&0.21&RQ&GAL \\
118&1.56&0.34&10.68&0.20&1.37&1.04&1.34&0.58&0.51&0.25&RQ&GAL \\
119&1.07&0.23&10.95&0.14&1.34&0.33&1.24&0.47&0.33&0.27&RQ&GAL \\
120&1.33&0.52&10.26&0.27&1.31&0.39&1.45&0.71&0.40&0.27&RQ&GAL \\
121&1.02&\no&11.02&0.10&2.08&0.08&1.36&0.05&0.12&0.09&RQ&GAL \\
123&0.81&\no&10.75&0.11&1.29&0.35&0.98&0.45&0.14&0.19&RQ&GAL \\
124&1.00&0.09&11.18&0.09&$-$0.74&1.24&0.27&0.27&0.27&0.16&RL&GAL \\
125&0.86&\no&9.49&0.07&0.79&0.08&0.41&0.06&0.15&0.13&RQ&GAL \\
127&1.01&0.23&11.10&0.14&1.35&0.42&1.41&0.40&0.23&0.23&RQ&GAL \\
130&0.71&\no&10.73&0.07&$-$0.97&0.98&0.14&0.01&0.10&0.02&RL&GAL \\
131&3.00&\no&10.21&0.04&1.40&0.17&0.44&0.10&0.47&0.26&RQ&AGN \\
134&1.78&0.66&10.09&0.68&1.19&2.03&1.55&0.89&0.80&0.17&RL&AGN \\
135&1.87&\no&10.18&0.07&1.10&0.08&0.41&0.02&0.38&0.12&RQ&GAL \\
136&2.37&0.58&10.57&0.15&1.60&0.36&0.72&0.32&0.50&0.27&RQ&GAL \\
137&1.09&0.33&10.20&0.27&1.01&0.37&1.70&0.73&0.29&0.23&RQ&GAL \\
138&0.82&0.14&10.00&0.31&0.90&0.61&1.96&0.87&0.70&0.20&RQ&AGN \\
141&1.02&\no&11.28&0.07&0.12&0.08&0.80&0.12&0.33&0.12&RL&GAL \\
142&1.15&\no&10.93&0.26&1.89&0.30&2.66&0.31&0.63&0.14&RQ&AGN \\
145&2.09&1.39&9.79&0.34&1.08&0.41&1.57&0.93&0.37&0.25&RQ&AGN \\
147&1.46&0.21&9.94&0.16&1.40&0.22&0.69&0.22&0.18&0.22&RQ&GAL \\
148&1.43&0.19&10.51&0.09&1.79&0.25&1.36&0.13&0.14&0.20&RQ&GAL \\
150&3.29&1.10&10.89&0.29&2.42&0.29&0.88&0.20&0.54&0.16&RQ&AGN \\
153&0.65&0.08&9.90&0.09&1.11&0.21&1.34&0.14&0.02&0.07&RQ&GAL \\
154&2.45&1.93&11.05&0.54&2.49&0.70&2.04&0.86&0.48&0.24&RQ&AGN \\
156&3.48&0.36&10.56&0.08&1.34&0.31&0.82&0.11&0.40&0.29&RL&GAL \\
157&4.70&1.87&10.42&0.29&1.78&0.50&2.00&0.76&0.46&0.28&RQ&AGN \\
158&0.76&0.08&10.68&0.08&1.16&0.24&1.27&0.23&0.07&0.12&RQ&GAL \\
159&0.29&\no&10.33&0.13&0.26&0.52&0.54&0.43&0.11&0.10&RQ&GAL \\
160&1.68&0.36&10.38&0.20&1.92&0.27&0.69&0.20&0.32&0.24&RQ&GAL \\
162&1.18&0.14&10.62&0.18&1.90&0.31&1.20&0.33&0.10&0.15&RQ&GAL \\
163&1.27&1.47&9.52&0.71&1.11&0.72&0.48&0.43&0.41&0.17&RQ&AGN \\
168&1.73&\no&10.21&0.18&1.05&0.48&0.76&0.38&0.44&0.19&RL&GAL \\
170&1.38&0.29&10.15&0.18&1.34&0.24&0.93&0.28&0.34&0.27&RQ&GAL \\
172&2.49&0.15&10.54&0.14&1.06&0.54&2.07&0.68&0.38&0.24&RL&GAL \\
173&0.50&\no&11.26&0.06&$-$0.60&0.19&0.16&0.07&0.11&0.04&RL&GAL \\
175&1.09&0.15&10.27&0.28&1.65&0.26&2.61&0.41&0.25&0.23&RQ&AGN \\
176&1.55&0.26&10.11&0.13&0.26&0.79&0.99&0.56&0.48&0.27&RL&GAL \\
178&1.13&\no&10.67&0.04&1.59&0.07&0.82&0.01&0.11&0.16&RQ&GAL \\
180&0.84&0.32&10.31&0.53&1.04&0.47&2.19&0.70&0.41&0.22&RQ&AGN \\
182&0.95&0.10&11.05&0.11&0.24&1.69&0.40&0.41&0.46&0.17&RL&GAL \\
187&1.41&0.40&10.83&0.19&1.45&0.77&1.10&0.70&0.55&0.23&RQ&GAL \\
188&1.20&0.27&10.12&0.24&1.11&1.10&1.06&0.45&0.51&0.25&RQ&GAL \\
192&0.71&\no&10.54&0.07&1.08&0.24&0.83&0.14&0.09&0.12&RQ&GAL \\
194&0.18&\no&10.65&0.14&0.01&0.50&0.63&0.31&0.42&0.14&RQ&AGN \\
195&0.20&0.29&6.54&1.27&$-$2.14&2.79&0.60&0.43&0.18&0.11&RL&AGN \\
197&1.91&0.41&10.01&0.30&1.15&0.75&1.01&0.48&0.71&0.18&RL&AGN \\
200&0.57&\no&9.98&0.12&1.26&0.10&0.78&0.12&0.02&0.05&RQ&GAL \\
201&1.10&\no&11.21&0.14&1.17&0.51&0.56&0.41&0.28&0.19&RQ&GAL \\
202&1.13&\no&10.82&0.17&1.76&0.35&1.02&0.28&0.11&0.13&RQ&GAL \\
203&2.30&1.65&8.93&0.58&0.21&0.82&1.48&0.87&0.62&0.26&RL&AGN \\
204&1.62&0.25&10.55&0.10&1.97&0.17&1.36&0.17&0.24&0.26&RQ&GAL \\
205&0.89&0.14&10.94&0.14&1.13&1.08&0.72&0.52&0.22&0.15&RQ&AGN \\
206&0.98&0.20&10.52&0.12&1.35&0.26&1.41&0.29&0.11&0.15&RQ&GAL \\
207&3.11&1.08&11.44&0.23&3.05&0.22&0.87&0.18&0.53&0.13&RQ&AGN \\
210&0.78&\no&10.65&0.05&1.57&0.06&1.36&0.01&0.04&0.06&RQ&GAL \\
212&4.14&0.30&10.36&0.09&1.63&0.24&0.90&0.22&0.43&0.29&RQ&AGN \\
213&2.20&1.83&11.26&0.76&2.70&0.93&1.69&0.91&0.33&0.29&RQ&GAL \\
214&0.64&0.43&10.25&0.63&0.87&0.59&2.33&0.63&0.38&0.22&RQ&AGN \\
215&1.95&\no&10.33&0.10&1.06&0.21&1.17&0.32&0.49&0.22&RQ&AGN \\
216&1.82&0.20&10.44&0.10&$-$0.27&1.99&1.37&0.38&0.40&0.29&RL&GAL \\
218&1.26&0.33&10.80&0.15&1.52&0.72&0.97&0.46&0.42&0.16&RQ&GAL \\
219&1.21&0.21&10.90&0.16&1.86&0.30&1.18&0.36&0.17&0.21&RQ&GAL \\
220&2.19&0.47&10.99&0.14&1.74&0.63&1.08&0.61&0.31&0.20&RQ&GAL \\
222&0.29&\no&10.61&0.10&0.77&0.33&1.03&0.33&0.07&0.07&RQ&GAL \\
223&0.90&0.20&9.78&0.32&1.14&0.34&1.11&0.38&0.47&0.19&RQ&AGN \\
225&3.50&0.29&11.28&0.10&1.66&0.68&2.67&0.27&0.39&0.25&RQ&GAL \\
226&0.80&0.40&9.83&0.76&0.70&3.17&1.90&0.73&0.63&0.17&RQ&AGN \\
229&2.12&0.35&10.87&0.17&1.43&0.90&2.34&0.62&0.19&0.21&RQ&GAL \\
231&1.02&\no&10.71&0.09&1.51&0.26&0.46&0.14&0.06&0.10&RQ&GAL \\
232&1.05&0.21&10.61&0.10&1.13&0.89&1.00&0.26&0.67&0.19&RQ&AGN \\
233&1.38&0.98&9.93&0.39&1.40&0.56&0.57&0.27&0.48&0.21&RQ&GAL \\
234&4.51&0.84&11.28&0.36&2.65&0.77&2.40&0.62&0.38&0.27&RQ&GAL \\
236&0.87&0.18&10.41&0.17&0.89&0.34&1.42&0.56&0.12&0.16&RQ&AGN \\
237&0.82&\no&10.95&0.09&0.56&0.36&0.79&0.16&0.28&0.16&RL&AGN \\
240&0.66&\no&10.57&0.08&0.87&0.26&0.67&0.22&0.11&0.14&RQ&GAL \\
241&1.11&0.64&9.89&1.00&1.36&1.06&2.62&0.42&0.54&0.21&RQ&AGN \\
242&1.45&0.21&11.02&0.11&2.07&0.32&1.19&0.40&0.27&0.27&RQ&GAL \\
244&2.16&\no&10.25&0.15&0.03&0.63&0.96&0.31&0.60&0.20&RL&GAL \\
245&0.92&\no&10.08&0.08&1.01&0.09&0.84&0.12&0.15&0.13&RQ&GAL \\
246&0.54&\no&10.91&0.04&1.39&0.05&0.82&0.03&0.01&0.03&RQ&GAL \\
247&1.96&\no&11.08&0.14&1.49&0.31&0.96&0.47&0.44&0.24&RQ&GAL \\
248&0.92&\no&10.61&0.06&1.53&0.08&1.35&0.09&0.03&0.06&RQ&GAL \\
251&1.62&0.55&10.30&0.23&1.26&0.26&0.95&0.31&0.38&0.22&RQ&GAL \\
253&0.63&\no&11.04&0.09&1.08&0.51&0.55&0.33&0.14&0.12&RQ&GAL \\
256&1.64&0.14&10.97&0.12&1.06&2.66&1.41&0.43&0.49&0.29&RQ&GAL \\
257&2.68&0.09&10.94&0.04&$-$0.18&0.50&1.36&0.05&0.46&0.28&RL&GAL \\
258&0.66&\no&10.16&0.04&1.10&0.04&0.82&0.01&0.02&0.04&RQ&GAL \\
260&0.54&\no&11.21&0.07&$-$1.70&1.33&0.16&0.07&0.10&0.04&RL&GAL \\
267&1.55&0.45&10.29&0.24&0.81&3.49&1.08&0.69&0.60&0.23&RL&GAL \\
268&0.54&0.08&9.43&0.13&0.99&0.18&1.33&0.13&0.01&0.03&RQ&GAL \\
269&1.48&1.49&9.70&0.75&1.12&0.78&1.20&0.92&0.69&0.22&RQ&AGN \\
271&0.46&\no&10.23&0.09&0.92&0.17&0.57&0.20&0.03&0.05&RQ&GAL \\
276&5.48&0.70&10.15&0.11&1.78&0.12&0.43&0.14&0.39&0.28&RQ&AGN \\
278&1.80&0.98&9.61&0.77&0.85&1.81&2.08&0.86&0.72&0.23&RQ&AGN \\
280&0.97&0.11&10.47&0.12&0.73&0.45&1.02&0.46&0.28&0.21&RQ&GAL \\
283&0.33&\no&10.53&0.19&1.01&0.34&0.87&0.50&0.09&0.12&RQ&GAL \\
284&2.27&\no&9.89&0.09&1.12&0.09&0.81&0.09&0.45&0.25&RQ&GAL \\
285&2.17&\no&10.53&0.05&0.25&0.09&0.41&0.04&0.40&0.31&RL&GAL \\
288&1.24&0.78&10.01&0.84&1.54&1.09&1.27&0.36&0.32&0.26&RQ&GAL \\
289&1.85&\no&10.42&0.08&1.14&0.13&0.79&0.13&0.38&0.18&RQ&GAL \\
291&0.76&0.13&10.30&0.14&0.82&0.33&0.96&0.36&0.08&0.12&RQ&AGN \\
292&0.70&0.08&10.66&0.13&1.10&0.26&1.40&0.46&0.08&0.11&RQ&GAL \\
293&2.45&0.75&9.34&0.18&0.82&0.20&0.20&0.23&0.36&0.28&RL&AGN \\
297&0.33&\no&9.33&0.07&0.18&0.17&0.41&0.07&0.01&0.03&RQ&GAL \\
298&1.05&0.29&9.98&0.24&0.56&2.51&0.49&0.42&0.54&0.20&RQ&GAL \\
299&0.94&0.12&10.98&0.10&1.41&0.38&1.12&0.28&0.13&0.18&RQ&GAL \\
300&2.51&0.54&10.90&0.13&2.09&0.39&1.19&0.47&0.50&0.32&RQ&GAL \\
302&2.29&1.25&10.58&0.31&2.12&0.41&0.73&0.36&0.49&0.30&RQ&AGN \\
305&2.38&0.29&10.79&0.10&1.55&0.59&1.29&0.31&0.32&0.18&RL&GAL \\
306&2.11&1.14&10.24&0.47&1.33&1.36&1.54&0.83&0.50&0.27&RQ&AGN \\
307&2.39&0.52&10.54&0.12&0.88&0.56&1.34&0.47&0.45&0.26&RL&GAL \\
308&0.14&\no&9.55&0.02&$-$0.72&0.02&0.41&0.00&0.00&0.00&RQ&GAL \\
310&4.62&1.10&10.78&0.18&2.13&0.35&1.40&0.38&0.42&0.24&RQ&AGN \\
312&0.92&\no&10.66&0.06&1.64&0.14&1.36&0.07&0.03&0.07&RQ&GAL \\
313&0.98&0.38&9.95&0.29&0.57&0.66&1.38&0.61&0.56&0.19&RL&AGN \\
314&1.55&0.15&10.51&0.14&1.39&0.52&0.78&0.30&0.53&0.24&RL&GAL \\
315&1.08&0.42&10.31&0.35&1.07&0.43&1.94&0.84&0.39&0.23&RQ&AGN \\
317&2.26&1.60&10.91&1.20&2.51&1.33&2.15&0.68&0.67&0.20&RQ&AGN \\
318&1.96&\no&9.84&0.07&1.19&0.10&0.79&0.10&0.43&0.22&RQ&GAL \\
319&0.18&\no&9.99&0.06&0.87&0.10&0.39&0.07&0.04&0.05&RQ&GAL \\
320&1.43&\no&10.47&0.06&1.87&0.11&0.78&0.12&0.25&0.17&RQ&GAL \\
325&2.04&0.98&10.22&0.45&1.69&0.74&2.66&0.30&0.46&0.31&RQ&AGN \\
326&1.87&1.02&10.80&0.34&1.83&0.44&1.44&0.64&0.49&0.29&RQ&AGN \\
327&1.32&0.69&10.32&0.45&1.93&0.56&1.20&0.28&0.15&0.18&RQ&GAL \\
328&0.71&\no&9.65&0.09&0.81&0.15&0.43&0.13&0.02&0.05&RQ&GAL \\
329&0.87&0.13&10.31&0.13&0.62&0.36&1.15&0.50&0.13&0.17&RQ&GAL \\
336&1.86&0.18&10.09&0.10&0.40&0.39&0.45&0.34&0.49&0.29&RL&GAL \\
337&2.34&0.18&10.94&0.10&0.35&2.55&1.27&0.29&0.54&0.26&RL&GAL \\
338&1.14&0.19&10.58&0.17&0.54&0.99&0.54&0.45&0.46&0.25&RL&GAL \\
342&1.90&0.79&10.21&0.22&1.48&0.69&1.56&0.69&0.54&0.26&RQ&AGN \\
348&0.89&\no&10.04&0.17&0.88&0.33&1.00&0.46&0.72&0.14&RQ&AGN \\
349&1.59&1.24&10.77&0.93&2.35&1.20&2.47&0.61&0.40&0.30&RQ&GAL \\
352&0.78&0.88&9.60&1.28&1.05&1.61&1.55&0.81&0.79&0.20&RQ&AGN \\
355&1.78&0.40&10.29&0.15&1.73&0.15&0.44&0.12&0.36&0.28&RQ&AGN \\
356&1.57&0.25&10.44&0.21&1.86&0.22&1.23&0.31&0.15&0.21&RQ&GAL \\
363&1.60&\no&10.58&0.14&1.82&0.22&0.70&0.85&0.43&0.10&RQ&AGN \\
372&1.05&0.20&9.98&0.17&1.43&0.28&1.20&0.35&0.22&0.23&RQ&GAL \\
373&0.88&0.10&11.01&0.10&1.49&0.19&1.33&0.19&0.07&0.12&RQ&AGN \\
377&1.12&\no&10.10&0.08&0.97&0.16&1.34&0.14&0.28&0.24&RQ&GAL \\
378&2.72&\no&9.86&0.06&1.51&0.06&0.42&0.09&0.41&0.24&RQ&AGN \\
382&0.37&\no&11.19&0.05&0.08&1.78&0.19&0.24&0.10&0.03&RL&GAL \\
383&0.54&\no&10.36&0.10&0.50&0.45&1.04&0.30&0.10&0.06&RQ&GAL \\
387&1.10&\no&10.24&0.07&1.21&0.11&0.84&0.11&0.23&0.19&RQ&GAL \\
394&1.75&\no&10.37&0.18&1.77&0.18&1.26&0.22&0.39&0.28&RQ&GAL \\
400&1.83&1.01&10.27&0.36&1.68&0.42&1.49&0.57&0.66&0.26&RQ&AGN \\
401&2.02&\no&10.06&0.32&1.31&0.13&1.36&0.02&0.56&0.20&RQ&AGN \\
403&0.19&0.46&9.05&2.49&0.34&2.87&1.52&0.87&0.74&0.20&RQ&AGN \\
417&0.88&0.06&10.65&0.09&1.08&0.17&0.75&0.17&0.10&0.14&RQ&GAL \\
420&0.66&0.31&9.39&0.69&0.73&0.73&2.64&0.34&0.57&0.18&RQ&AGN \\
422&0.92&\no&10.33&0.11&1.04&0.20&0.74&0.26&0.09&0.14&RQ&GAL \\
426&1.54&0.14&10.00&0.10&1.46&0.15&0.81&0.06&0.33&0.27&RQ&GAL \\
427&2.40&0.15&10.98&0.06&1.12&0.67&2.73&0.02&0.37&0.23&RQ&GAL \\
429&0.89&\no&10.08&0.06&1.21&0.05&0.40&0.05&0.04&0.08&RQ&GAL \\
431&0.29&\no&10.00&0.10&$-$0.63&1.22&0.28&0.35&0.09&0.04&RL&GAL \\
432&1.90&0.20&9.94&0.09&1.61&0.08&1.36&0.13&0.15&0.22&RQ&GAL \\
434&1.18&0.25&10.67&0.15&0.45&1.47&0.95&0.51&0.58&0.20&RL&GAL \\
445&3.50&0.22&10.01&0.14&1.63&0.09&2.72&0.10&0.52&0.33&RQ&AGN \\
466&2.50&0.63&10.58&0.11&1.60&0.39&0.80&0.29&0.42&0.22&RQ&GAL \\
470&1.74&0.17&10.43&0.08&0.86&0.36&0.41&0.19&0.36&0.25&RL&GAL \\
472&0.81&0.13&10.43&0.13&0.96&0.38&0.79&0.26&0.10&0.14&RQ&GAL \\
480&0.95&0.17&10.49&0.11&1.30&0.24&0.82&0.17&0.10&0.14&RQ&GAL \\
489&0.74&0.16&10.73&0.12&0.98&0.41&1.40&0.34&0.09&0.12&RQ&GAL \\
492&0.14&\no&9.98&0.09&$-$0.88&0.38&0.23&0.17&0.10&0.05&RL&GAL \\
493&0.85&\no&11.00&0.08&1.74&0.04&1.36&0.01&0.08&0.10&RQ&AGN \\
497&0.78&\no&10.99&0.07&1.29&0.27&1.13&0.27&0.70&0.14&RQ&AGN \\
499&0.67&\no&10.68&0.06&1.13&0.10&0.79&0.11&0.05&0.08&RQ&GAL \\
501&0.83&\no&10.65&0.15&1.33&0.40&1.16&0.34&0.23&0.20&RQ&GAL \\
502&0.68&0.09&10.45&0.10&1.45&0.19&0.86&0.18&0.04&0.07&RQ&GAL \\
505&1.60&0.40&10.45&0.31&1.56&0.38&1.20&0.42&0.46&0.20&RQ&GAL \\
515&0.72&0.12&10.19&0.15&1.00&0.31&1.18&0.33&0.05&0.10&RQ&AGN \\
517&1.41&0.13&10.63&0.20&2.17&0.13&1.27&0.23&0.09&0.16&RQ&GAL \\
527&0.86&0.10&10.37&0.11&0.25&0.59&0.41&0.23&0.37&0.16&RL&GAL \\
528&0.38&\no&10.05&0.11&$-$1.24&0.64&0.66&0.29&0.30&0.04&RL&GAL \\
531&1.56&0.88&10.54&0.83&2.04&1.18&2.42&0.68&0.45&0.31&RQ&GAL \\
533&0.17&\no&9.44&0.04&0.10&0.03&0.41&0.00&0.00&0.01&RQ&GAL \\
542&2.19&0.33&9.79&0.11&0.99&0.20&0.40&0.09&0.37&0.25&RL&AGN \\
547&2.54&0.23&10.95&0.14&1.28&0.40&2.25&0.65&0.37&0.24&RL&GAL \\
548&1.50&0.49&10.14&0.29&1.21&0.46&1.49&0.54&0.58&0.27&RQ&GAL \\
549&0.75&\no&11.28&0.10&0.37&0.62&0.74&0.22&0.06&0.11&RL&GAL \\
552&0.92&\no&10.32&0.08&1.16&0.09&1.36&0.05&0.09&0.12&RQ&GAL \\
559&0.80&0.12&10.59&0.09&1.28&0.28&1.36&0.13&0.07&0.10&RQ&GAL \\
564&1.59&0.18&9.94&0.16&1.55&0.10&0.82&0.07&0.37&0.23&RQ&GAL \\
566&1.05&0.21&10.31&0.16&1.18&0.34&0.90&0.25&0.17&0.19&RQ&GAL \\
567&1.57&0.32&10.80&0.15&1.80&0.22&1.40&0.42&0.22&0.23&RQ&AGN \\
572&2.69&0.64&10.82&0.16&2.13&0.26&0.81&0.17&0.43&0.29&RQ&GAL \\
593&0.89&\no&10.89&0.23&1.88&0.35&1.23&0.59&0.28&0.20&RQ&GAL \\
596&1.90&0.18&10.96&0.08&1.72&0.21&0.83&0.13&0.25&0.28&RQ&GAL \\

\enddata
\tablecomments{Table~\ref{t:SEDfits} is available as a machine-readable table (MRT), datafile3.txt in the source directory.  Data are based on fitting the HST and JWST data with CIGALE.  For each parameter, CIGALE fits its derived probability density function with a Gaussian, and the table reports the mean and standard deviation of the Gaussian.  For quantities tabulated as logs, $\delta=(\sigma/\mu)/\log(10)$, where $\mu$ and $\sigma$ are respectively the mean and standard deviation from CIGALE. Column~2 gives the spectroscopic redshift (Table~\ref{t:ID}) when available, and values of  other parameters were calculated with the redshift fixed at the spectroscopic value.  For sources without a spectroscopic redshift, column~2 and~3 give the mean and uncertainty from CIGALE, and other parameters were marginalized over the redshift probability density function.}
\end{deluxetable*}

\clearpage
\startlongtable
\begin{deluxetable}{@{\hspace{1em}}r@{\hspace{3em}}l@{\hspace{3em}}l@{\hspace{3em}}R@{\hspace{3em}}R}
\tablecaption{Redshift Comparison}
\label{t:z}
\tabletypesize{\small}
\tablewidth{0pt}
\tablehead{
 \colhead{\hspace{-1.5em}ID}&
 \colhead{\hspace{-2.8em}\zsp}&
 \colhead{\hspace{-2.8em}$Q$}&
 \colhead{\hspace{-2.8em}\zph}&
 \colhead{$\delta$\zph}
}
\startdata
74 & 0.0796&  4&0.06&0.03 \\
76 & 0.1037&  3&0.15&0.01 \\
89 &  3.766&  2&3.51&0.86 \\
112&  1.810&  2&1.20&0.32 \\
117&  1.802&  2&1.60&0.08 \\
121& 1.0172&  4&1.05&0.22 \\
123& 0.8070&  3&0.83&0.09 \\
125& 0.8599&  4&0.91&0.21 \\
130& 0.7126&  4&0.64&0.05 \\
131&  2.996&  3&2.38&1.26 \\
135&  1.875&  3&2.65&0.33 \\
138&  1.832&  2&0.82&0.14 \\
141& 1.0175&  4&0.91&0.06 \\
142& 1.1476&  4&1.51&0.41 \\
145&  1.011&  2&2.09&1.39 \\
154&  0.928&  2&2.45&1.93 \\
159& 0.2932&  4&0.23&0.08 \\
168&  1.730&  3&1.73&0.25 \\
173& 0.4972&  4&0.46&0.10 \\
178& 1.1288&  3&1.05&0.17 \\
182&  1.258&  2&0.95&0.10 \\
192& 0.7105&  3&0.81&0.12 \\
194& 0.1774&  4&0.23&0.14 \\
200& 0.5653&  4&0.57&0.09 \\
201& 1.1008&  4&0.93&0.11 \\
202& 1.1299&  3&0.89&0.10 \\
210& 0.7837&  3&0.81&0.09 \\
215&  1.954&  3&0.79&0.63 \\
216&  2.036&  2&1.82&0.20 \\
222& 0.2925&  4&0.34&0.06 \\
231& 1.0165&  4&0.87&0.20 \\
237& 0.8210&  3&0.82&0.13 \\
240& 0.6558&  4&0.67&0.06 \\
244&  2.164&  3&2.99&0.42 \\
245& 0.9174&  3&1.03&0.18 \\
246& 0.5375&  4&0.51&0.08 \\
247&  1.958&  3&2.63&0.48 \\
248& 0.9156&  3&0.80&0.12 \\
253& 0.6338&  4&0.57&0.11 \\
258& 0.6558&  4&0.64&0.11 \\
260& 0.5445&  4&0.48&0.09 \\
267&  1.827&  2&1.55&0.45 \\
271& 0.4558&  4&0.47&0.08 \\
283& 0.3278&  4&0.32&0.08 \\
284&  2.270&  3&2.50&0.50 \\
285&  2.174&  3&2.08&0.12 \\
288& 1.3654&  2&1.24&0.78 \\
289&  1.849&  3&1.75&0.19 \\
292& 0.7144&  2&0.70&0.08 \\
297& 0.3276&  3&0.29&0.09 \\
308& 0.1391&  4&0.09&0.03 \\
312& 0.9167&  4&0.88&0.09 \\
318&  1.963&  3&1.51&0.29 \\
319& 0.1787&  4&0.18&0.07 \\
320& 1.4267&  3&1.42&0.28 \\
325&  2.406&  2&2.04&0.98 \\
328& 0.7072&  4&0.66&0.10 \\
336&  1.871&  2&1.86&0.18 \\
342&  2.053&  2&1.90&0.79 \\
348& 0.8861&  4&0.99&0.19 \\
363& 1.6019&  4&0.84&0.56 \\
377&  1.120&  3&1.19&0.28 \\
378&  2.724&  3&2.26&0.63 \\
382& 0.3741&  4&0.32&0.09 \\
383& 0.5435&  3&0.50&0.15 \\
387& 1.1028&  3&1.06&0.32 \\
394&  1.746&  3&1.52&0.43 \\
401&  2.016&  3&1.67&0.68 \\
422& 0.9187&  3&0.85&0.10 \\
429& 0.8900&  4&0.90&0.12 \\
431& 0.2933&  4&0.23&0.06 \\
492& 0.1395&  4&0.06&0.05 \\
493& 0.8504&  3&0.90&0.22 \\
497& 0.7807&  4&0.92&0.14 \\
499& 0.6739&  4&0.68&0.05 \\
501& 0.8262&  3&0.95&0.19 \\
528& 0.3760&  4&0.26&0.06 \\
533& 0.1745&  4&0.18&0.05 \\
549& 0.7544&  4&0.68&0.10 \\
552& 0.9249&  3&1.06&0.29 \\
593& 0.8926&  4&0.94&0.35 \\

\enddata
\tablecomments{Table~\ref{t:z} is available as a machine-readable table (MRT), datafile4.txt in the source directory.   Spectroscopic redshifts shown with four decimal digits are from Binospec or Hectospec, and those with three decimal digits are from NIRISS. 
The spectroscopic redshift quality code Q follows the same
criteria adopted by the DEEP2 survey \citep{Newman2013}: $Q = 4$
indicates a secure redshift, while $Q = 3$ indicates a redshift with a
90\% or higher certainty of being correct, and $Q=2$ means may
potentially yield redshift information but needs more analysis or
re-reduction.  $Q=2$ redshifts were not used for analysis but are given
here for comparison with spectroscopic redshifts.
Values of \zph\ and $\delta$\zph\ are respectively the mean and standard deviation of CIGALE's Gaussian fit to the derived photometric-redshift probability density function.}
\end{deluxetable}

\end{document}